\documentclass[11pt,letterpaper]{article}
\pdfoutput=1
\usepackage{jheppub}
\usepackage{graphicx,epsfig,psfrag,bm,amssymb}
\usepackage{dcolumn}
\usepackage{bm}
\usepackage{color}
\usepackage{mathrsfs,amsfonts,hepunits, color}
\usepackage[utf8]{inputenc}

\newcommand{\be}{\begin{equation}}
\newcommand{\ee}{\end{equation}}
\newcommand{\din}{d_{\rm in}}

\newcommand{\fenc}{f^{\rm enc}}
\newcommand{\fdec}{f^{\rm dec}}
\newcommand{\x}{\mathbf{x}}
\newcommand{\Ybar}{\overline{Y}}

\begin{document}
\title{Topological Obstructions to Autoencoding}
\author[1]{Joshua Batson,}
\affiliation[1]{The Public Health Company, Calle Real, Goleta, CA USA}
\author[2]{C. Grace Haaf,}
\affiliation[2]{Department of Business and Finance, New York University Shanghai, Century Avenue, Shanghai, China}
\author[3,4]{Yonatan Kahn,}
\affiliation[3]{Department of Physics, University of Illinois at Urbana-Champaign, Green Street,
Urbana, IL USA}
\affiliation[4]{Center for Artificial Intelligence Innovation, National Center for Supercomputing Applications, \\University of Illinois at Urbana-Champaign, Clark Street, Urbana, IL USA}
\author[5,6,7]{Daniel A. Roberts}
\affiliation[5]{Center for Theoretical Physics and Department of Physics, Massachusetts
  Institute of Technology, Massachusetts Avenue, Cambridge, MA USA}
  \affiliation[6]{The NSF AI Institute for Artificial Intelligence and Fundamental Interactions}
\affiliation[7]{Salesforce}
\emailAdd{joshua.batson@gmail.com}
\emailAdd{cgh@nyu.edu}
\emailAdd{yfkahn@illinois.edu}
\emailAdd{drob@mit.edu}

\date{\today}

\abstract{
Autoencoders have been proposed as a powerful tool for model-independent anomaly detection in high-energy physics. The operating principle is that events which do not belong to the space of training data will be reconstructed poorly, thus flagging them as anomalies. We point out that in a variety of examples of interest, the connection between large reconstruction error and anomalies is not so clear. In particular, for data sets with nontrivial topology, there will always be points that erroneously seem anomalous due to global issues. Conversely, neural networks typically have an inductive bias or prior to locally interpolate such that undersampled or rare events may be reconstructed with small error, despite actually being the desired anomalies. Taken together, these facts are in tension with the simple picture of the autoencoder as an anomaly detector. Using a series of illustrative low-dimensional examples, we show explicitly how the intrinsic and extrinsic topology of the dataset affects the behavior of an autoencoder and how this topology is manifested in the latent space representation during training. We ground this analysis in the discussion of a mock ``bump hunt'' in which the autoencoder fails to identify an anomalous ``signal'' for reasons tied to the intrinsic topology of $n$-particle phase space.
}

\preprint{MIT-CTP/5264}

\maketitle

\pagebreak

\section{Introduction}

Data of interest in the physical sciences often consists of features of low intrinsic dimensionality packaged in a high-dimensional space. For example, the variants of a gene might be embedded in the much larger space of base pair sequences, or a single fundamental particle might manifest itself as an $N \times N$-pixel ($N \gg 1$) jet image \cite{Cogan:2014oua,deOliveira:2015xxd} in a particle detector at a high-energy physics experiment. A common task is to detect outliers, or ``anomalies,'' in a large data set; a common tool to perform this task is a neural network autoencoder \cite{rumelhart1985learning,pimentel2014review}.\footnote{Anomaly detection in high-energy physics using machine learning is a rich and growing field: some other model-independent strategies include weakly supervised learning (such as classification without labels), density estimation, and likelihood-free anomaly detection. See \cite{Nachman:2020ccu,Feickert:2021ajf} for a review of these and other strategies, and also \cite{Kasieczka:2021xcg} for a summary of some of these techniques as applied to simulated data.} The autoencoder architecture is quite simple: input data is processed and passed through a feed-forward network to a latent layer of smaller width than the input. The output of the latent layer is then processed and unpacked to an output layer of the same size as the input layer. The intuition is that the data is being compressed in the smaller latent layer, and uncompressed on its way out to the output layer. An autoencoder trained on a large data sample is attempting to learn a compressed representation of the data, and a network successful in this task should have small reconstruction error, measured for example by taking the loss function to be the mean squared error between the output and the input.

Nearly all data sets of practical relevance in high-energy physics descend from the manifold of Lorentz-invariant phase space. This manifold, which describes the energies and momenta of particles produced in relativistic collisions, has dimension $3n-4$ for $n$ final-state particles, and has a natural embedding in $\mathbb{R}^{4n}$ whose coordinates comprise the $n$ final-state 4-vectors. Training data for a machine learning task derived from these 4-vectors, whether low-level \cite{Cogan:2014oua,deOliveira:2015xxd,Baldi:2016fql,Barnard:2016qma,Komiske:2016rsd,ATLAS:2017dfg,Kasieczka:2017nvn,Bhimji:2017qvb,Macaluso:2018tck,Guo:2018hbv,Guest:2016iqz,Louppe:2017ipp,Cheng:2017rdo,Egan:2017ojy,Fraser:2018ieu,Almeida:2015jua,Pearkes:2017hku,Roxlo:2018adx} or high-level \cite{Aguilar-Saavedra:2017rzt,Luo:2017ncs,Moore:2018lsr,Komiske:2017aww,Komiske:2018cqr,Komiske:2019asc,Kasieczka:2020nyd}, must still at some level inherit the geometry and topology of phase space \cite{Datta:2017rhs,Butter:2017cot,Datta:2017lxt,Dreyer:2018nbf,Komiske:2019fks,Larkoski:2019nwj,Cesarotti:2020hwb,Komiske:2020qhg,Lai:2020byl,Cai:2020vzx}. In this context, we can make the notion of ``anomaly'' more precise. If background events are drawn from a manifold of fixed particle number $n$, events may be anomalous if they contain more than $n$ particle-like features, in other words if they lie on the $m$-particle phase space manifold with $m > n$. This situation describes some jet substructure observables, specifically $n$-subjettiness \cite{Thaler:2010tr}. Autoencoders (and their generalizations, such as variational autoencoders \cite{kingma2013auto}), have already shown some success in performing this kind of anomaly detection \cite{Farina:2018fyg,Heimel:2018mkt,Cerri:2018anq,Hajer:2018kqm,Roy:2019jae,Blance:2019ibf,Cheng:2020dal,Park:2020pak,Nachman:2020ccu,Romao:2020ocr}. The geometric intuition is that anomalous events lie \emph{off the background manifold}, and thus the autoencoder will fail to reconstruct these events because it is attempting to perform an extrapolation, a task on which neural-network autoencoders tend to perform poorly.

On the other hand, some particles, such as leptons, may be well-characterized by their 4-vectors rather than the more complicated jets characteristic of hadrons. A ``bump hunt'' search for a new particle in events containing leptons will feature anomalous events drawn from the \emph{same} manifold as the background events, but localized to a \emph{submanifold}. For example, in the search for the Higgs in the 4-lepton ``golden channel'' $H \to Z Z^* \to 4\mu$ \cite{Chatrchyan:2013mxa,Aad:2014eva}, the background events have 4 muons in the final state with a broad distribution of invariant masses, and the ``anomalous'' Higgs decay events are distinguished by lying on the submanifold of 4-particle phase space where the invariant mass of all four muons is equal to $m_H^2$. In this case, an autoencoder trained on a sideband data set of background events excluding invariant masses of $m_H^2$ may attempt to perform an interpolation task when run on a Higgs decay event. Such interpolation tasks are generally ``easy'' for neural networks, and thus might be expected to lead to low autoencoder loss for the signal Higgs events, which is the opposite of the desired behavior.

In this paper, we will investigate how the topology of data manifolds may pose a number of important obstructions to autoencoder performance on the second type of anomaly-detection task, where anomalous events lie on a distinguished submanifold of the manifold of background events. Consider an autoencoder trained on a set of 4-vectors sampled from $n$-particle phase space. Since $4n > 3n-4$, the embedding space is clearly redundant, and one might expect that after sufficient training, an autoencoder can achieve essentially zero reconstruction error on the training set for latent dimension $d_l$ equal to the intrinsic data dimension, $3n-4$. However, as we will show, this is impossible because phase space does not have the trivial topology of $\mathbb{R}^{3n-4}$, but rather that of a sphere $S^{3n-4}$.  A generic neural network autoencoder is a composition of continuous maps, so the nontrivial topology makes unavoidable the existence of nearby points on the data manifold which are mapped to distant points in the latent space, exactly as a Mercator projection distorts the poles of the 2-sphere when mapped into $\mathbb{R}^2$.\footnote{One can also consider neural networks with discontinuous activation functions, like a perceptron, though such activations are typically no longer used in practice.} 

The easiest context in which to visualize this topological obstruction is the unit circle, which we will study extensively in order to gain intuition for the breakdown of these maps to the latent space. Points on the circle may be labeled by a single number, an angle $\phi$, but since $\phi$ and $\phi + 2\pi$ represent the same point, an autoencoder which attempts to compress points on the circle to their angular coordinate $\phi$ will rip apart nearby points in the data manifold during the compression. More precisely, in the language of differential topology, the latent space is a single \emph{chart} on the data manifold, which can accurately capture the local geometry but not the global topology, which requires additional charts with transition functions between them.

The failure of the latent representation will imprint spurious features on the data. This has two important and related consequences:
\begin{itemize}
\item If the data manifold has nontrivial topology, there will always be points or regions in the training set with poor reconstruction error, even when the latent dimension is equal to the intrinsic dimension of the data. These regions are not the desired anomalies, but instead avatars of the topological obstruction to mapping the data manifold into a topologically-trivial latent space. 
\item If anomalous events live on a submanifold (as in the Higgs example above), the autoencoder may learn to interpolate smoothly across the submanifold even if the training distribution had no support there, causing the would-be anomalous events to have the same error distribution as background events. 
\end{itemize} 
These observations present obstacles to using autoencoders as practical anomaly detectors. A necessary condition for a successful autoencoder is near-perfect (or at least uniform-loss) reconstruction on the training set -- otherwise the compression of the data is not faithful -- but the topology of the data manifold can render that impossible without additional priors on the network. In addition, the background distribution itself may introduce additional topological or geometrical features; in the physics context, a matrix element governing the background process with poles or zeros at certain values of the kinematic variables may concentrate the events with large loss away from the desired submanifold.

This paper is organized as follows. In Sec.~\ref{sec:Setup} we define our basic autoencoder architecture, where in particular we take the latent dimension $d_l$ equal to the dimension $d$ of the data. We then introduce a specific example in Sec.~\ref{sec:bumphunt} of an autoencoder failing to perform a bump hunt in 3-particle phase space. The remainder of the paper is devoted to understanding the features of that failure by studying a series of low-dimensional examples, motivated by the fact that phase space has the topology of a sphere. We start in Sec.~\ref{sec:circle} with the simplest example, the circle $S^1$ embedded in $\mathbb{R}^2$, and show how the periodicity of the angular coordinate on the circle poses an obstruction to training an autoencoder with a latent layer $\mathbb{R}^1$. Moving to the 2-sphere in Sec.~\ref{sec:dim2}, we construct an easily-visualized analogue of the anomalous submanifold $S^1 \subset S^2$, and examine the interplay between topology, extrinsic geometry, and sampling distributions with a double cone. We confirm that these features persist in higher dimensions in Sec.~\ref{sec:dimn}. Armed with this intuition, we return to the example of 3-particle phase space in Sec.~\ref{sec:phasespace}. We briefly summarize the effects of taking $d_l > d$ in Sec.~\ref{sec:ChangeD}, arguing that this does not cure the issues we have identified, and conclude in Sec.~\ref{sec:conclusion}. Additional details are provided in the Appendices: App.~\ref{app:hyperparams} describes our hyperparameter choices, App.~\ref{app:MoreCircle} studies the $S^1$ example in depth including an analytic investigation of the trained network dynamics, and App.~\ref{app:MoreExamples} describes our studies of spaces with topological obstructions even for $d_l > d$.

Our goal in this work is not to claim that autoencoders are doomed to fail in the high-energy physics context, but rather to make the point that the topology of phase space and the inductive bias of autoencoders toward interpolation are important pieces of prior knowledge which should be considered before attempting a black-box solution to generalized anomaly detection.\footnote{The recent LHC Olympics exercise \cite{Kasieczka:2021xcg} featured a data set with a new particle decaying via two decay modes. Tellingly, this anomaly was not detected by any of the machine-learning strategies proposed by the participants in the exercise. Prior to unblinding, no autoencoder architectures detected true anomalies and several found spurious anomalies in a background-only data set.} In fact, it is somewhat surprising that autoencoders appear to perform worse on the nominally easier task of a bump hunt in leptons than on the superficially much more complicated task of jet image recognition and classification, since leptons live on a phase space of fixed dimension. The increasing prominence of ``physics-inspired neural networks'' -- where networks with important symmetry principles (such as gauge equivariance and Lorentz symmetry) hard-coded into the network architecture perform better than networks which are forced to learn these principles from scratch \cite{Bogatskiy:2020tje,Kanwar:2020xzo,Boyda:2020hsi} -- suggests that knowledge of the topology may in fact be necessary to appropriately interpret the autoencoder performance. We illustrate this point with the low-dimensional examples described above, and speculate on how these principles might be applied in the context of phase space.\footnote{We note that similar considerations have been investigated in \cite{olah_2014,korman2018autoencoding,moor2020topological,hajij2020topology}, though not in the context of physics. In particular, \cite{korman2018autoencoding} notes that nontrivial topological structure in the input data can require an autoencoder with latent dimension larger than the intrinsic dimension of the data, \cite{moor2020topological} considers adding a term to the loss function to force the latent layer to preserve topological structures of the data, and \cite{hajij2020topology} performs an in-depth study of the observations of \cite{olah_2014} to understand how topology is transformed at each layer of a feed-forward network.}

\section{Autoencoder architecture}
\label{sec:Setup}

In this paper, we will implement an autoencoder as a multilayer neural network.
Our baseline network architecture will be as follows: a 5-layer, fully-connected network with layer widths $(\din, d_w, d_l, d_w, \din)$ and loss function $L = ||\mathbf{y} - \mathbf{x}||^2$, where $\mathbf{x}$ is the input and $\mathbf{y}$ is the output. The second and fourth layers have $d_w \gg d_l, \din$ to ensure that for low-dimensional examples we are not artificially penalizing ourselves by using a network with too few parameters to accurately approximate the embedding of the data manifold to $\mathbb{R}^{d_l}$. To verify that the number of network parameters is not the limiting factor in autoencoder performance, we will sometimes add a second layer of width $d_w$ to both the encoder and decoder, so that the full network has 7 layers. 

We will be primarily concerned with autoencoders with $d_l < \din$ but $d_l = d$, so that the latent representation has the same number of degrees of freedom as the manifold from which the data is sampled. We will refer to the map $\mathbb{R}^{\din} \to \mathbb{R}^{d_l}$ as the encoder or latent representation and the map $\mathbb{R}^{d_l} \to \mathbb{R}^{\din}$ as the decoder, each of which is a 1-hidden-layer neural network. We will refer to the full autoencoder map $ \mathbb{R}^{\din} \to  \mathbb{R}^{\din}$ as the model. Our default width will be $d_w = 64$; this is small by the standards of networks used for \textit{e.g.}\ (jet) image recognition, but it is much larger than the $\din \leq 12$ we will be considering in this paper. In each example we train the network with stochastic gradient descent (SGD) for 20,000 epochs using a training set of size $N_{\rm train}$ and a test set of size $N_{\rm test}$, both sampled from the same distribution. Our batch size and learning rate hyperparameters for each example are given in Tab.~\ref{tab:hyperparameters} in App.~\ref{app:hyperparams}; we have checked that our conclusions are robust to changes in these hyperparameters, because in essentially all examples the networks will be trained to convergence but do not overfit the training data.

To visualize the output of the autoencoder, especially in low-dimensional examples, we will plot both the test set data and the predictions of the model on the test set. Occasionally, it will be convenient to present these on the same plot, to see both the density of data points and the density of their images as predicted by the autoencoder. Since our loss function is Euclidean distance, plotting them together can show how large-loss points will be mapped far away from their true locations.

An important feature of an autoencoder is that for $d_l \geq \din$, the global minimum of the loss is always the identity function on $\mathbb{R}^{\din}$. However, this is not a generalizable minimum, since the loss would be zero on any input whatsoever, even one having nothing to do with the data distribution which is a particular submanifold of $\mathbb{R}^{\din}$. Since the goal of an autoencoder is to learn a useful low-dimensional representation of the data -- an encoding -- a successful autoencoder will find its way to a local minimum, which is zero (or close to it) on the training and test sets, but is nonzero on other data. The function of the latent layer with $d_l < \din$ is to prevent the network from finding the trivial global minimum.\footnote{Noise injected into the training data may serve the same purpose, though we do not consider that strategy in this work.} Regardless, the existence of a global minimum for the family of autoencoder architectures with different $d_l$ which is not the desired loss minimum means that initialization and gradient descent algorithms may be an important component of the analysis, and suggests a loss landscape for the autoencoder with a rich structure; we discuss a number of examples in Appendices~\ref{app:MoreCircle} and \ref{app:MoreExamples}. 

\section{Failure of a bump hunt}
\label{sec:bumphunt}

The manifold $\mathcal{M}_{n = 3}$ of 3-particle phase space is defined by imposing energy/momentum conservation (4 constraints) and putting the three particles on mass shell (3 constraints), which imposes 7 algebraic constraints on the three 4-vectors (12 parameters) and yields a 5-dimensional manifold embedded in $\mathbb{R}^{12}$.\footnote{See Ref.~\cite{Larkoski:2020thc} for a detailed study of the geometry of phase space as a Riemannian manifold.} As we will explain in Sec.~\ref{sec:phasespace}, $\mathcal{M}_{n = 3}$ has the topology of the 5-sphere $S^5$, which has important implications for autoencoder behavior.

\begin{figure}[t!]
	\includegraphics[width = 0.27\textwidth]{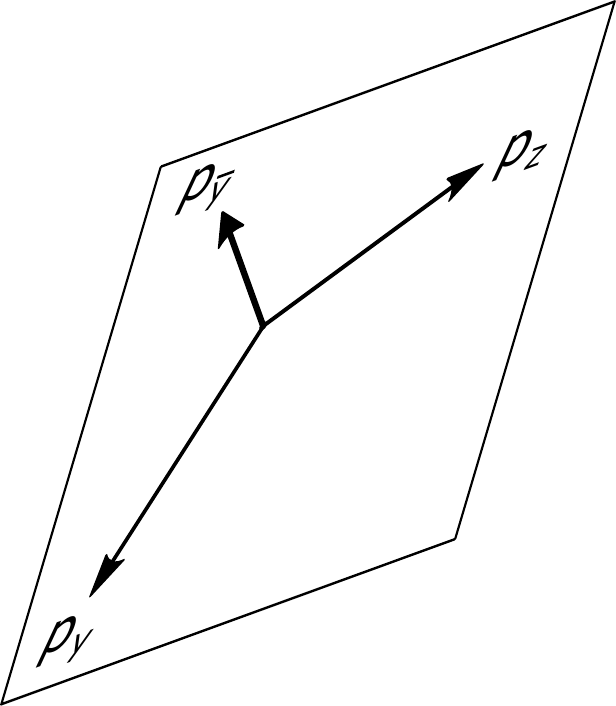} \qquad
	\includegraphics[width = 0.3\textwidth]{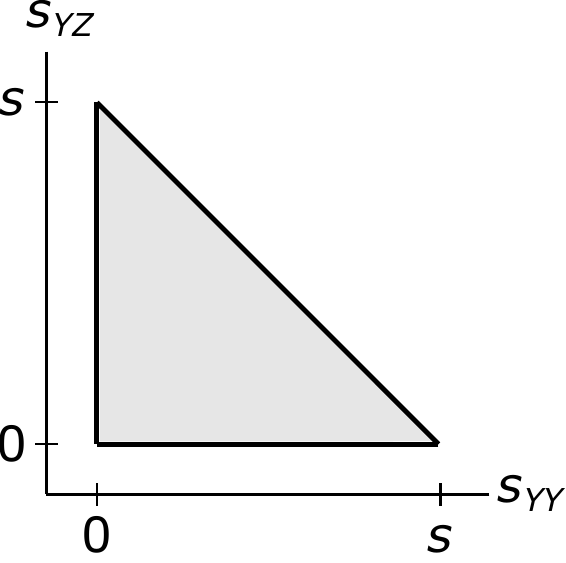} \qquad
	\includegraphics[width = 0.3\textwidth]{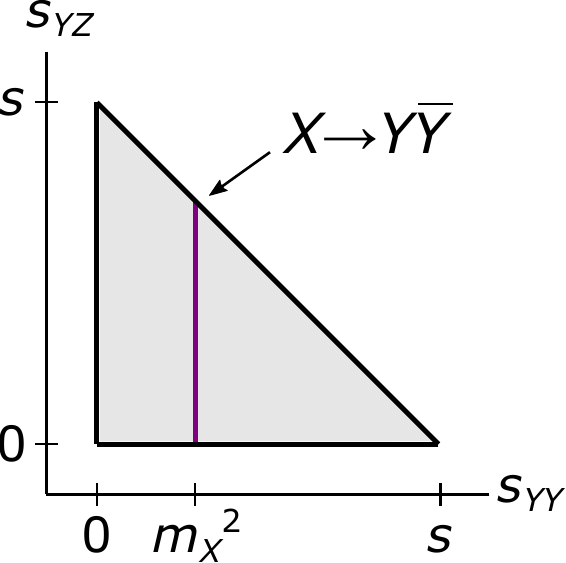} \
	\caption{\textbf{Left:} cartoon of the geometry of the 3-vectors of particles $Y$, $\Ybar$, and $Z$ sampled from 3-particle phase space. In the center-of-momentum frame, the particles are coplanar. \textbf{Center:} example of a Dalitz plot for uniform sampling from 3-particle phase space, which uniformly populates a right isosceles triangle in the $s_{YY}-s_{YZ}$ plane. \textbf{Right:} if the matrix element for the process contains a resonance, say at $s_{YY} = m_{X}^2$ from the intermediate decay $X \to Y \Ybar$, there will be an oversampled ``stripe'' (purple) in the Dalitz plot.}
		\label{fig:Dalitz}
\end{figure}

To approximate the situation typically encountered at colliders (and also to simplify the analysis), we will consider the final state $Y + \Ybar + Z$ where all final-state particles are distinguishable and massless -- the example we have in mind is a bump hunt in leptons, where (say) $Y$ is a muon and $Z$ is a photon, and the collision energy is large enough that the muon is approximately massless. In the center-of-momentum (COM) frame, the three particles are coplanar (Fig.~\ref{fig:Dalitz}, left). The natural measure on phase space is the Lorentz-invariant measure, which for 3-body phase space takes a particularly simple form \cite{PDG}:
\be
\label{eq:dPhi3}
\int d\Phi_3 = \frac{1}{128 \pi^3 s}\int_{\mathcal{R}} ds_{YY} \, ds_{YZ},
\ee
where $\sqrt{s}$ is the collision energy in the COM frame, and $s_{YY} = (p_Y + p_{\Ybar})^2$ and $s_{YZ} = (p_Y + p_Z)^2$ are the invariant squared masses of the $Y\Ybar$ and $YZ$ pairs, respectively. The shape of the region $\mathcal{R}$ depends on the masses of the final-state particles and is conveniently visualized in a Dalitz plot. Events which are sampled uniformly from phase space will uniformly populate $\mathcal{R}$ in the $s_{YY}-s_{YZ}$ plane, which for the massless case is the right isosceles triangle defined by $s_{YY}, s_{YZ} > 0$ and $s_{YY} + s_{YZ} \leq s$ (Fig.~\ref{fig:Dalitz}, center). The remaining three phase space coordinates are Euler angles defining an element of SO(3) which orients the event, and have been integrated over in Eq.~(\ref{eq:dPhi3}), making the Dalitz plot a particularly convenient 2-dimensional projection of the 5-dimensional manifold $\mathcal{M}_{n=3}$. The boundaries of $\mathcal{R}$ correspond to events where two particles are collinear, and the corners of the Dalitz triangle for massless particles correspond to a soft particle whose energy goes to zero; for finite final-state masses, these corners are rounded off. Note that the measure is uniform in any pair of invariant masses, and for massless particles $\mathcal{R}$ is the same triangle for all three such pairs. In real particle physics events, the matrix element for the desired process introduces a non-uniform distribution on $\mathcal{R}$: for example, if a resonance $X$ of mass $m_X$ can decay to $Y\Ybar$, an oversampled ``stripe'' will appear in the Dalitz plot at $s_{YY} = m_X^2$ (Fig.~\ref{fig:Dalitz}, right). In this work we will focus on the intrinsic topology of phase space and only sample uniformly according to Eq.~(\ref{eq:dPhi3}), but we will comment throughout on the role of the sampling distribution, which may itself be incorporated into the geometry of the phase space manifold \cite{Larkoski:2020thc}.

\begin{figure}[t!]
	\includegraphics[width = 0.55\textwidth]{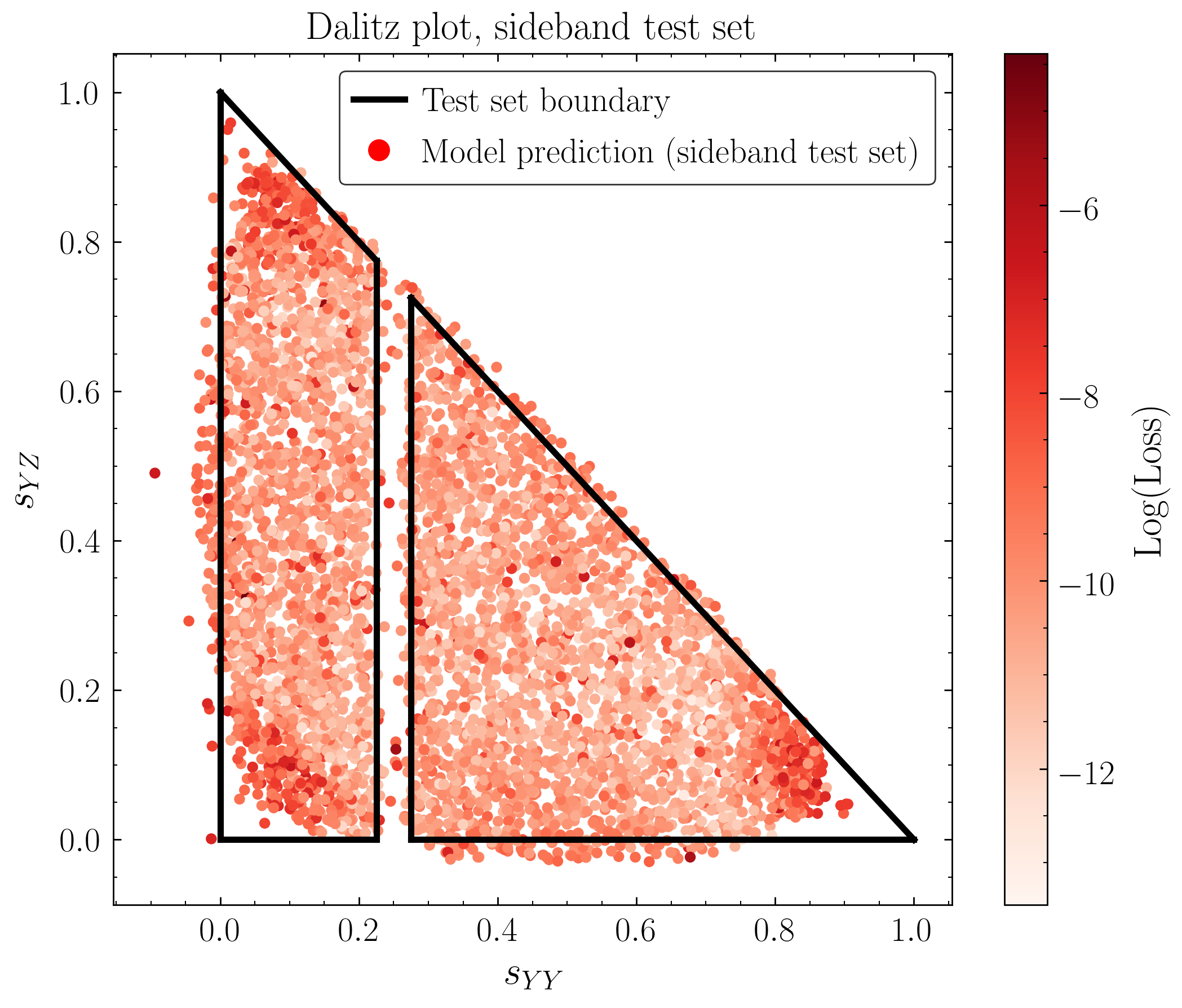}
	\includegraphics[width = 0.45\textwidth]{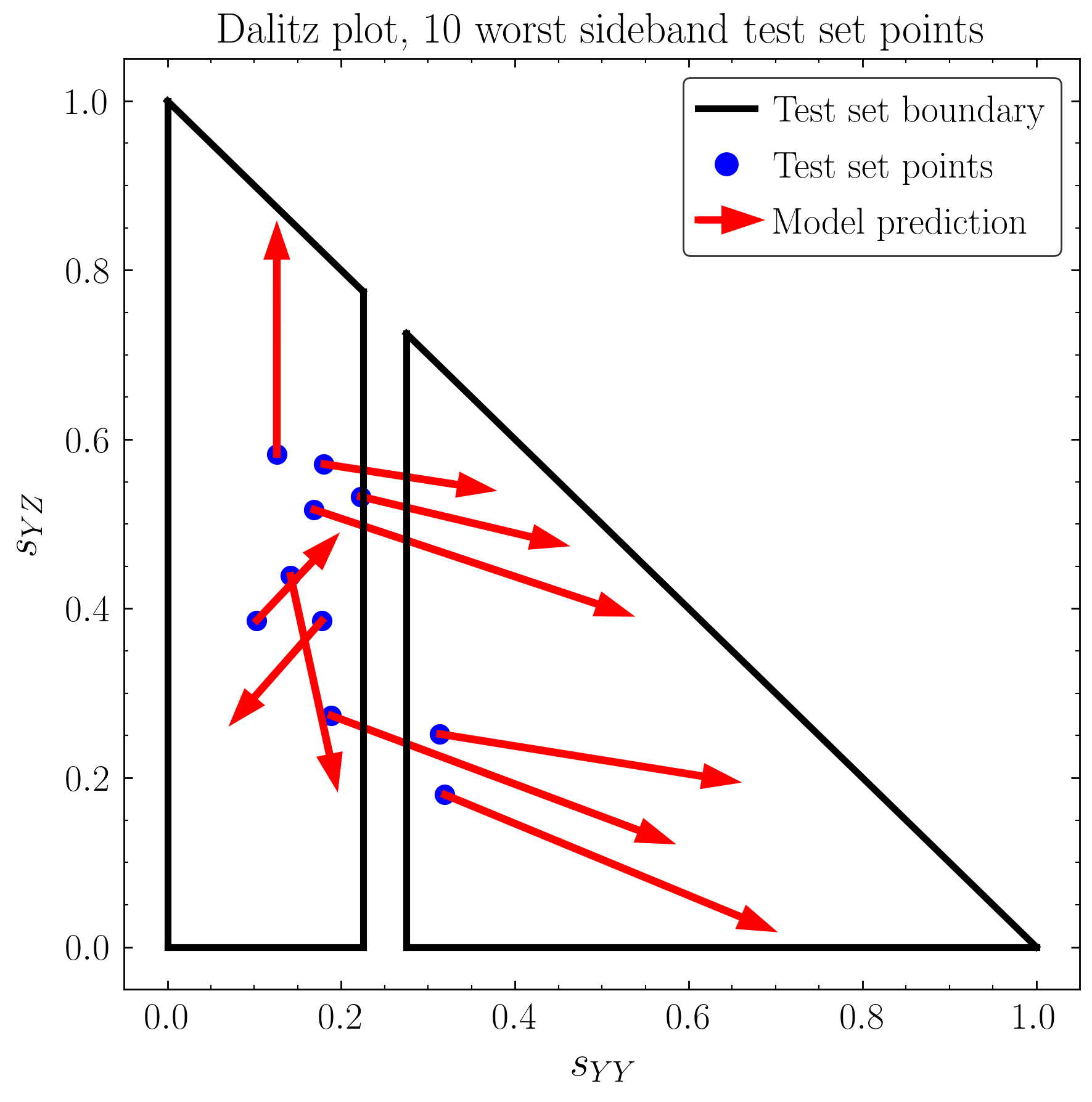}
	\caption{\textbf{Left:} Dalitz plot of model predictions for a test set uniformly sampled on phase space excising a region around $s_{YY} = 0.25$ to mimic a sideband analysis. The boundary of the sampled region is shown in black, and only 5\% of the test set is shown for clarity. \textbf{Right:} The 10 worst-loss points in the sideband test set (blue) and their model predictions (red arrows). The worst loss is located at isolated points near the boundary of the excised interval, and these points are mapped far away in the Dalitz plane. However, the excised region is reproduced fairly well. The corners of the Dalitz triangle are also reproduced poorly, but note that the density of points mapped to the corners is large while the loss of any individual point there is considerably smaller than the worst-loss points.}
	\label{fig:SidebandDalitz}
\end{figure}

We perform a mock ``bump hunt'' by normalizing our units such that $\sqrt{s} = 1$ and choosing a desired invariant mass, say $s_{YY} = 0.25$, corresponding to a heavy unstable particle $X$ of mass $m_X = 0.5$ ($m_X^2 = 0.25$) which decays to $Y\Ybar$. We then train a 7-layer autoencoder with $\din = 12$ and $d_l = 5$ on ``sideband'' data sampled from the distribution (\ref{eq:dPhi3}) excluding the region $0.9\ m_X^2 < s_{YY} < 1.1  m_X^2$; we use this deeper network rather than the 5-layer autoencoder to ensure that there are no issues with network capacity that would inhibit learning the full geometric structure of the phase space manifold. Our setup mimics the standard procedure of fitting a background model to sidebands before examining the signal region. The choice of latent dimension is determined by the dimension of phase space: since the sideband data is drawn from a 5-dimensional manifold, $d_l < 5$ would fail to capture the full geometry of the background data and would result in large losses across the whole data distribution, while $d_l > 5$ would be a redundant parameterization of the data. To distinguish signal from background, we generate two test sets: a sideband test set sampled from the same distribution as the training data, and a signal test set with  $s_{YY} = 0.25$ but otherwise sampled uniformly in phase space.

The boundary of the sampling region for the training and sideband test sets, along with the autoencoder output on the test set colored by loss, is shown in Fig.~\ref{fig:SidebandDalitz} (left). Note that the autoencoder does a fairly good job of identifying the boundaries of the sideband region around $s_{YY} = 0.25$, but has trouble at the corners of the Dalitz triangle which correspond to kinematic endpoints where the energy of one particle goes to zero. While it is true that the autoencoder task is only to minimize the Euclidean distance between the model and the data point-by-point in phase space, the spurious features in the model output imply correlations which will be imprinted on the loss distribution, which is the desired diagnostic for anomaly detection. Furthermore, the largest-loss points (blue), with loss $L \simeq 0.05$, are located near the boundary of the excised interval near $s_{YY} = 0.25$, as shown in  Fig.~\ref{fig:SidebandDalitz} (right). While these large-loss points are mapped far away by the trained model, most points near the excised interval are low-loss and are mapped close to their true locations. Indeed, we will show in the remainder of this paper that the existence of a neighborhood of large-loss points (\textit{i.e.} those whose predictions are far away from the true value, and thus have large loss as measured by Euclidean distance) is a direct consequence of phase space having the topology of a sphere.

\begin{figure}[t!]
	\includegraphics[width = 0.52\textwidth]{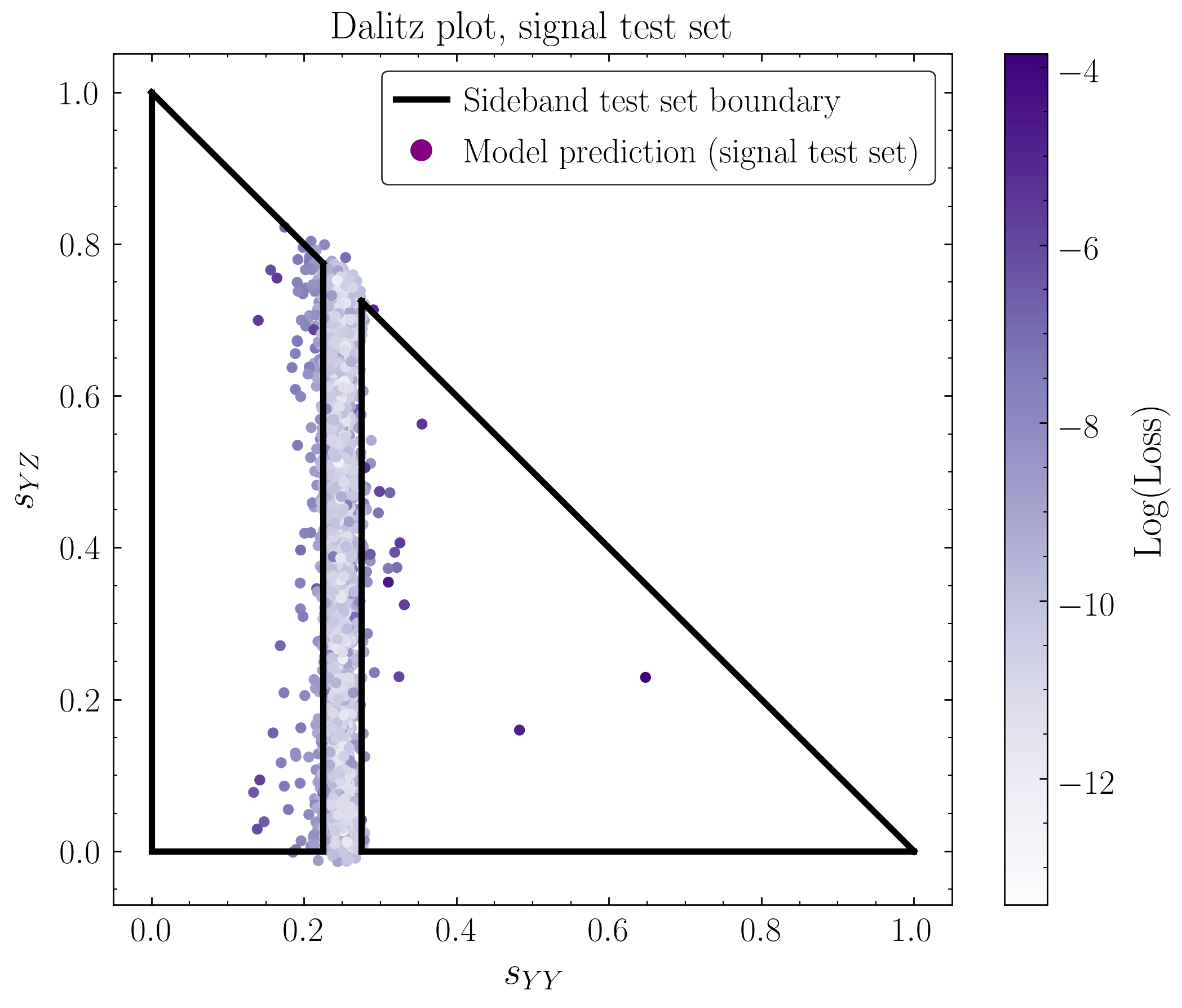}
	\includegraphics[width = 0.46\textwidth]{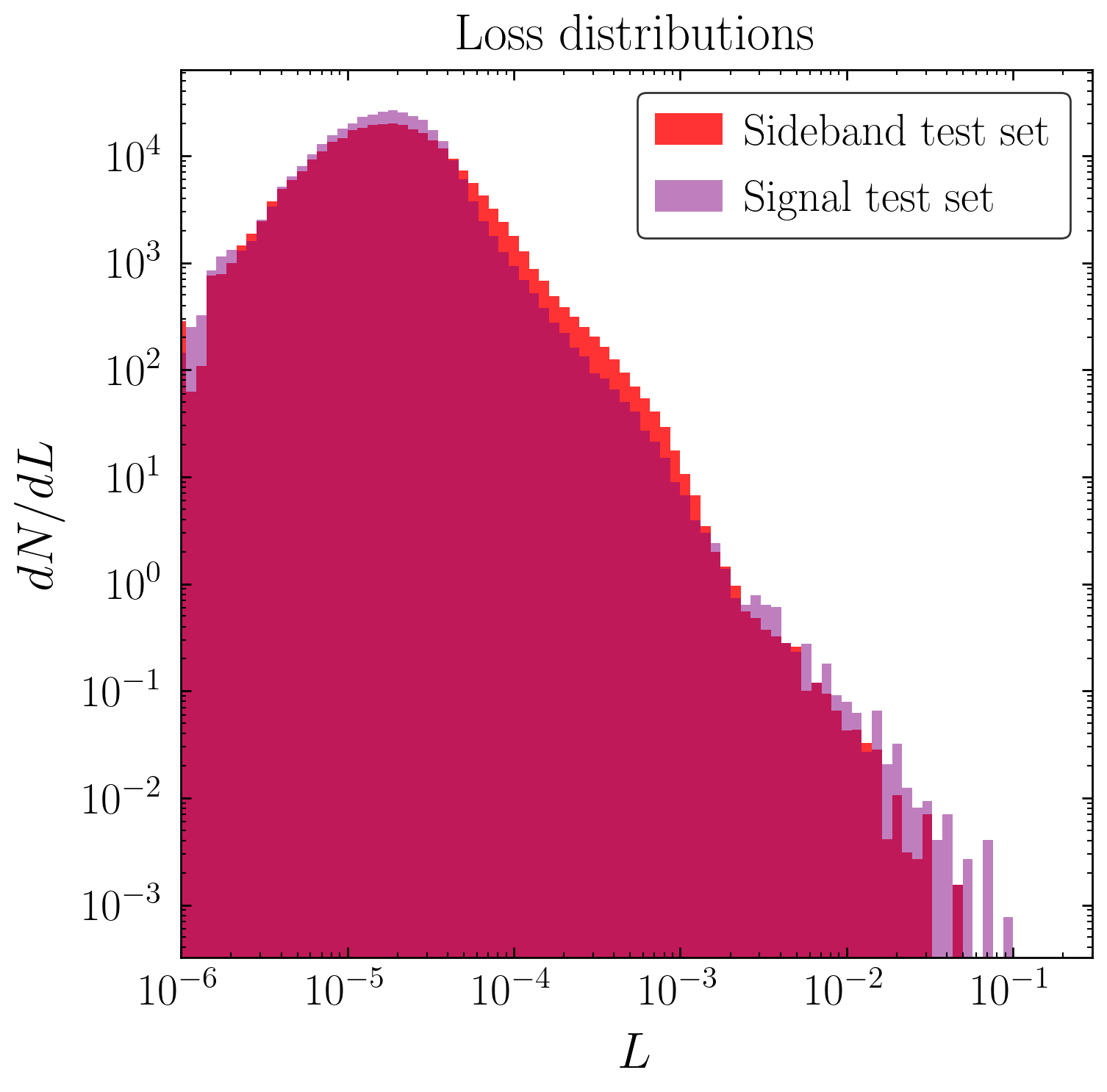}
	\caption{\textbf{Left:} Dalitz plot for a ``signal'' test set with $s_{YY} = 0.25$ but uniformly sampled in phase space otherwise. 5\% of the data is plotted for clarity. The network trained on the sideband distribution learns to interpolate through the signal region, such that the signal events are not flagged as anomalous. \textbf{Right:} Normalized loss distributions for the sideband test set (Fig.~\ref{fig:SidebandDalitz}) and signal test set. Remarkably, the loss distributions are almost identical for the two data sets.}
	\label{fig:SignalDalitz}
\end{figure}

Next, we run the trained autoencoder on the signal test set. The Dalitz plot is shown in Fig.~\ref{fig:SignalDalitz}, left. In the Dalitz plane, the signal data lives on a vertical line (the purple line in Fig.~\ref{fig:Dalitz}, right), and we see that despite the autoencoder never having seen points in this region before, it can smoothly interpolate across it, reconstructing points in the signal region with low loss except for a few isolated points. These points are not any more or less anomalous than the rest of the signal data, but are simply the neighbors of the large-loss points in the sideband test set which get mapped far away by the model. The loss distributions of the two test sets (Fig.~\ref{fig:SignalDalitz}, right) are essentially identical. In particular, there is no obvious large-loss tail for the signal events which would flag them as anomalies, despite events with $s_{YY} = 0.25$ being \emph{entirely absent from the training set}. There is no reasonable decision boundary that one could draw to separate these two distributions. Cutting on whatever small large-loss tail does exist, at say $L = 10^{-2}$, would give a signal efficiency of $\epsilon_S = 7.4 \times 10^{-4}$ and background rejection power $1/\epsilon_B = 3 \times 10^3 \simeq 2/\epsilon_S$ , making this autoencoder an extremely poor anomaly detector for rare events and no better than a random classifier at larger signal efficiency. This simple example should be compared with e.g.\ Ref.~\cite{Farina:2018fyg} where QCD and non-QCD jets have largely non-overlapping loss distributions for a similar autoencoder architecture and reasonable ROC curves which can achieve the same background rejection with $\epsilon_S \simeq 0.1$. 

We have checked that varying $d_l$ does not change this conclusion. For $d_l > 5$ the loss distributions for both background and signal have no large-loss tail (indicating near-perfect reconstruction) and whatever tail exists for the background events exceeds the signal events, so the signal events would be classified as ``less anomalous.'' For $d_l < 5$, the loss tails for both the background and signal distributions are large and nearly identical, and moreover the network fails to identify the boundaries of the sideband intervals, so no information is gained by further reducing the latent dimension. Similarly, changing the length of training does not change our conclusions: the large-loss points persist for both shorter and longer training, and the loss tails for the sideband and signal sets do not separate. We have also verified that the results are identical including both 1\% Gaussian smearing on all 4-momenta coordinates and sampling the signal test set from a Breit-Wigner distribution with $0.5$\% width, both of which resemble typical detector effects and matrix element structures for realistic applications.

This result would seem to preclude using a standard neural network autoencoder to perform a bump hunt in leptons, where the lack of soft and collinear radiation makes the particle 4-vector a decent proxy for what is actually measured at a collider detector -- in other words, parton-level observables are nearly equivalent to detector-level observables -- unless additional features were incorporated into the autoencoder architecture. Given that simple feed-forward autoencoders have already found some success in anomaly detection in jet images \cite{Farina:2018fyg}, it is a priori somewhat surprising that the same network architecture fails at what would naively seem to be an easier problem.\footnote{That said, in \cite{Farina:2018fyg} the jet image autoencoders were only trained on ``cropped'' images containing only individual jets from a (presumably) multi-jet event. Our analysis suggests that similar issues might be encountered if the event were considered as a whole, since (for example) two QCD jets which were occasionally almost collinear (which one would reasonably expect to be part of the background distribution whenever the event in question had three or more partons) would be difficult to distinguish from a single fat jet.} As we will see in the remainder of this paper, because the training data for the lepton bump hunt is sampled directly from $n$-particle phase space, one can better understand these results from the perspective of topology: the network uses its inductive bias toward interpolation to trivialize the topology of $S^{3n-4}$ in the most minimal way possible, which involves localizing all large losses near a single point in phase space and interpolating everywhere else.

\section{Dimension 1: intrinsic topology and the unit circle}
\label{sec:circle}

To elucidate our observations about phase space, we will explore a series of low-dimensional examples which encapsulate particular topological features and are more easily visualized. In dimension 1, all closed manifolds without boundary have the same intrinsic topology as the circle.\footnote{Depending on the embedding $S^1 \to \mathbb{R}^n$, the embedded curve may also have extrinsic topology, depending on whether a projection into a plane $\mathbb{R}^2$ can yield a curve without self-intersections; the canonical example of such nontrivial extrinsic topology is a knot, which we explore further in App.~\ref{app:knot}.} Our examples in this section illustrate two key points:
\begin{enumerate}
\item If the manifold has nontrivial intrinsic topology then there will be some data points on the manifold reconstructed with large loss despite not being anomalies of any kind;
\item The sampling distribution used during training can influence the location of those badly reconstructed points.
\end{enumerate}

To start, we consider the simple example of a training set of points $(x,y)$ equidistantly spaced on the unit circle $S^1$. Since $S^1$ has dimension 1, every point on the data manifold can be represented by a single number, an angle $\phi \in (-\pi, \pi]$ such that $(x,y) = (\cos \phi, \sin \phi)$. Thus, a latent layer with $d_l = 1$ should be able to fully capture the \emph{local} geometric features of this manifold. However, the periodicity of $\phi$ is a topological obstruction to learning the \emph{global} structure of the data manifold. In the language of differential geometry, a choice of coordinates is a \emph{chart} $S^1 \to \mathbb{R}$, but the nontrivial topology of $S^1$ means that it must be covered by at least two charts; without additional structure in the autoencoder network, the latent representation can only provide a single chart. Since $\phi$ is periodic, with $\phi$ and $\phi + 2\pi$ corresponding to the same point in the training set, the latent layer's encoding function $(x,y) \mapsto \fenc(\phi)$ will cover its range at least twice, with one arc of the circle mapping onto the interval $[\min(\fenc),\max(\fenc)]$ and its complementary arc mapping onto the same interval. The reconstruction can be accurate on at most one of those arcs, so we expect the autoencoder to make one of those arcs as large as possible and the other as small as possible.

\begin{figure}[t!]
	\includegraphics[width = 0.44\textwidth]{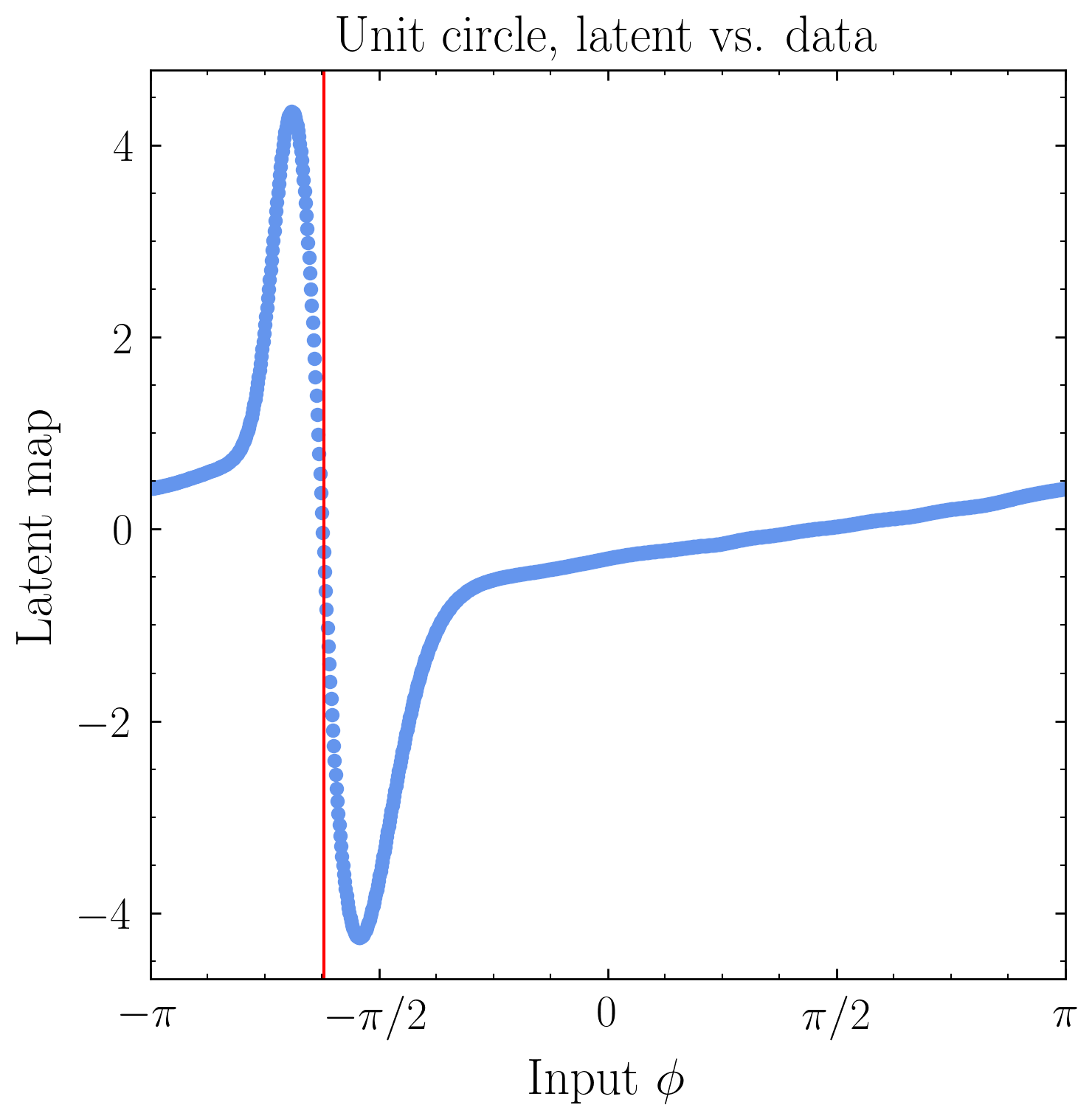} \qquad
	\includegraphics[width = 0.45\textwidth]{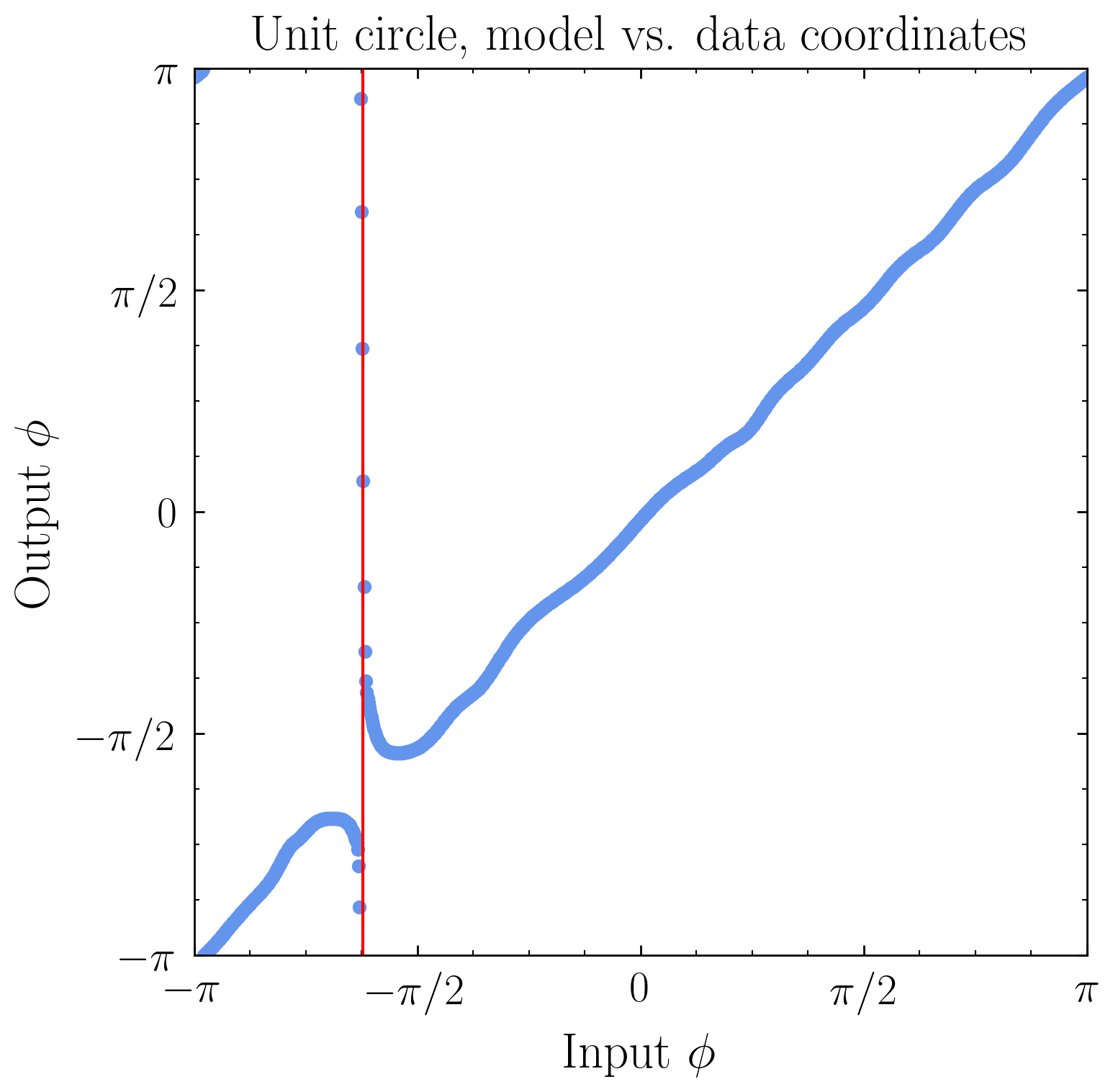}  \\
	\hspace*{-0.2 cm}\includegraphics[width = 0.45\textwidth]{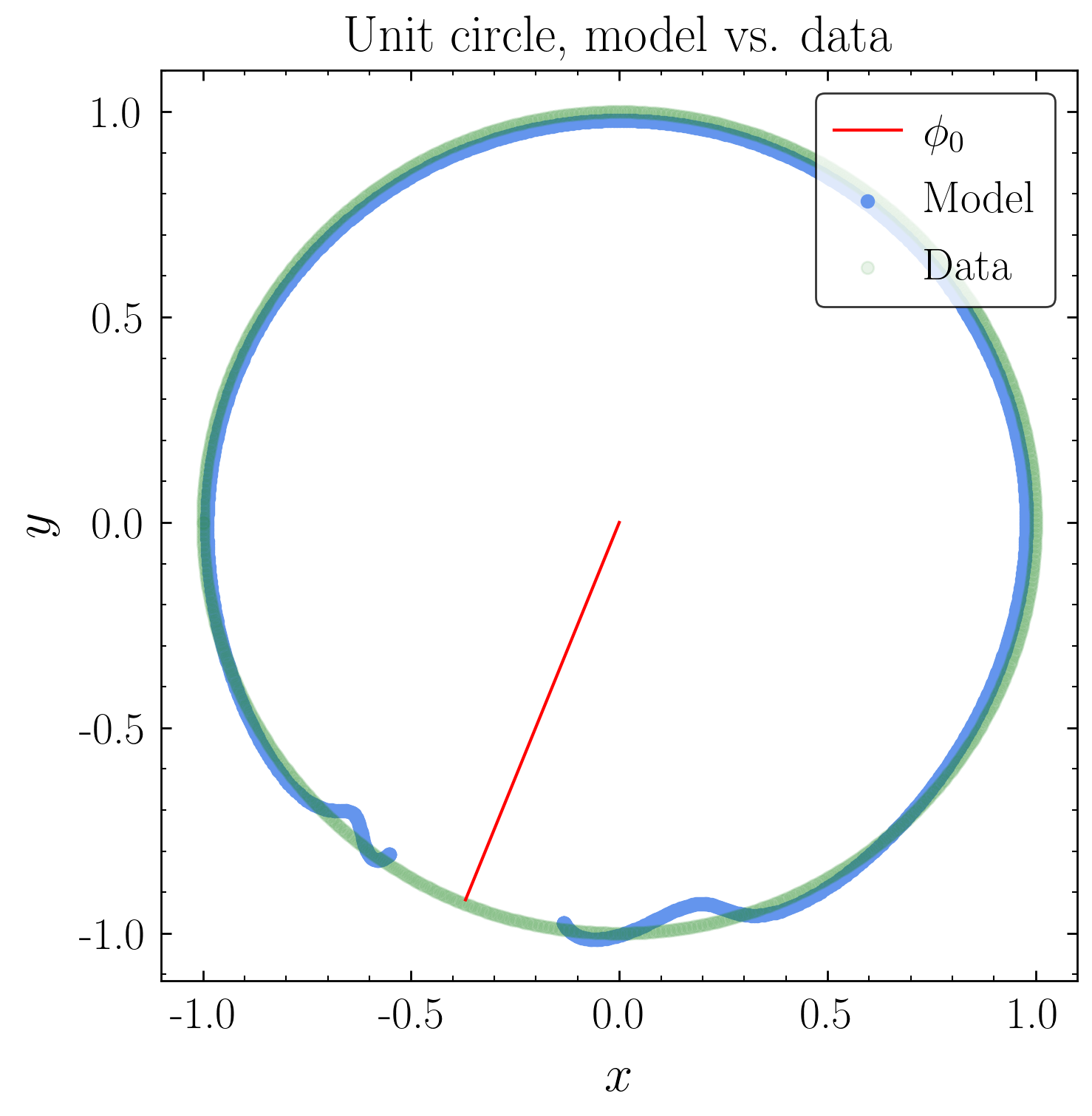} \qquad \ 
	\includegraphics[width = 0.45\textwidth]{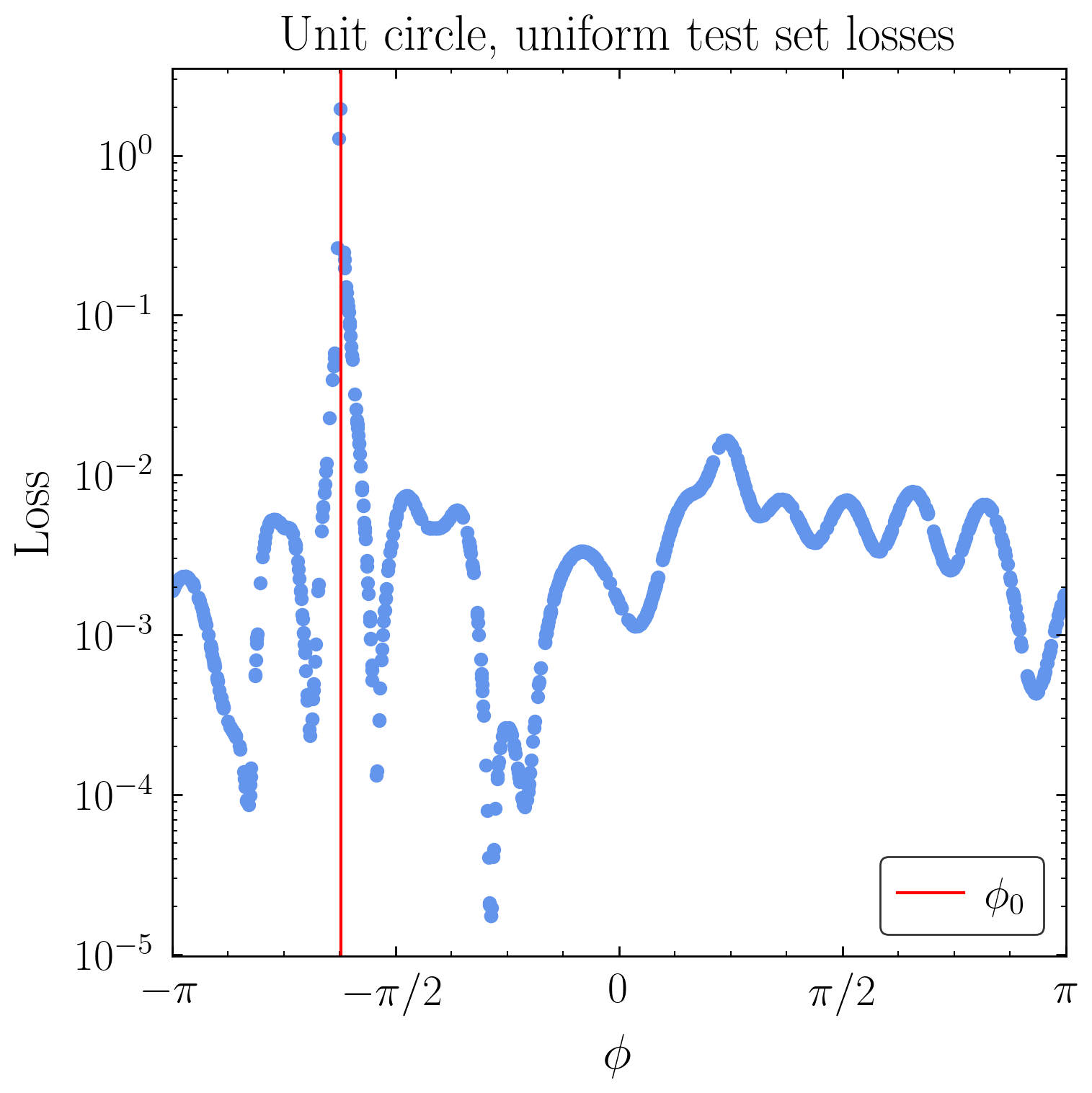} 
	\caption{Performance of an $S^1$ autoencoder with latent dimension $d_l = 1$. The break point $\phi_0$ is shown in red in all plots. \textbf{Top left:} Latent representation as a function of input $\phi$. \textbf{Top right:} Model $\phi$ as a function of input $\phi$. \textbf{Bottom left:} Model points $(x,y)$ compared to data $(x,y)$. \textbf{Bottom right:} Loss as a function of input $\phi$ for a uniformly-sampled test set.}
	\label{fig:S1Uniform}
\end{figure}

\begin{figure}[t!]
	\includegraphics[width = 0.45\textwidth]{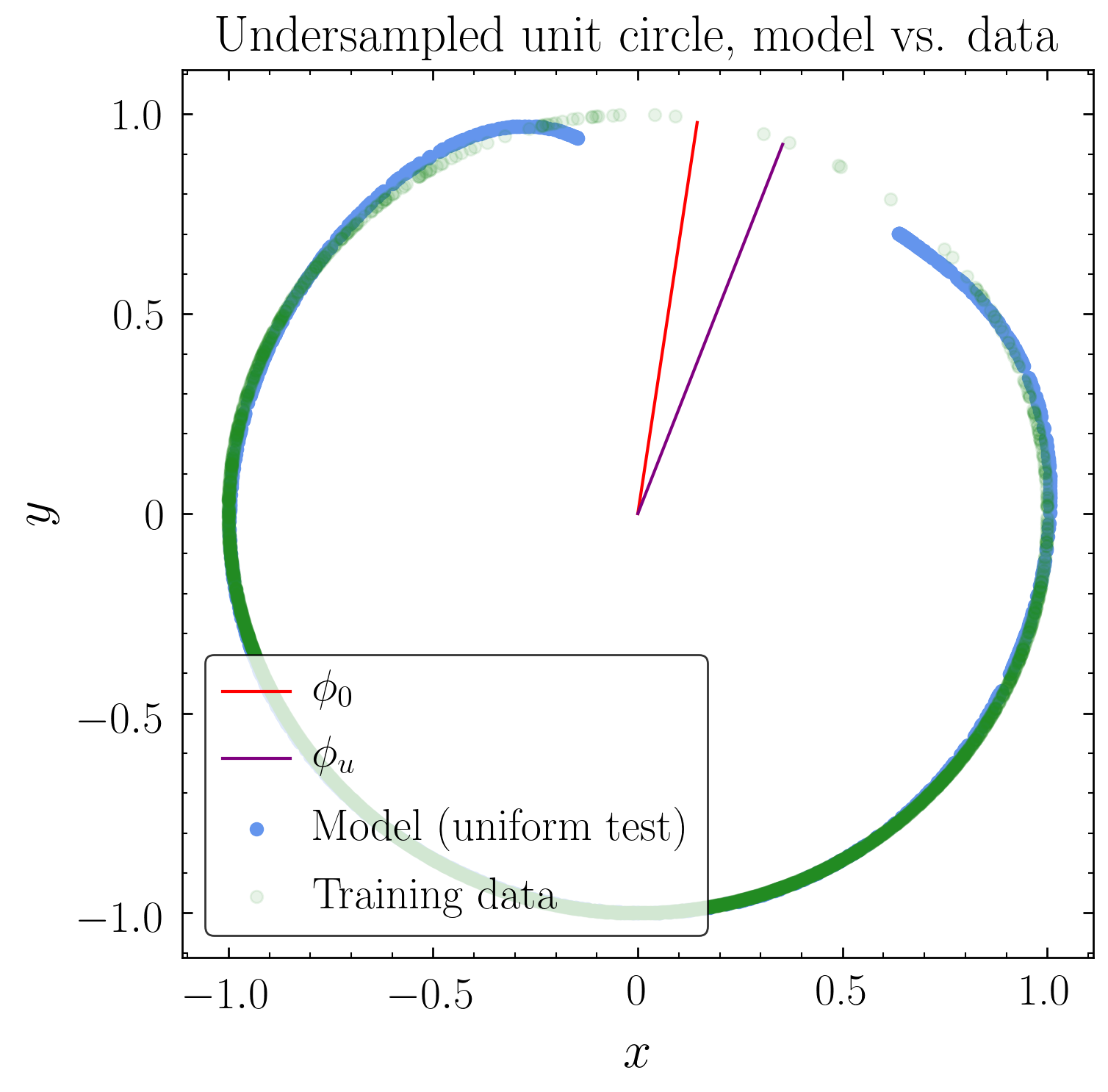} \qquad
	\includegraphics[width = 0.45\textwidth]{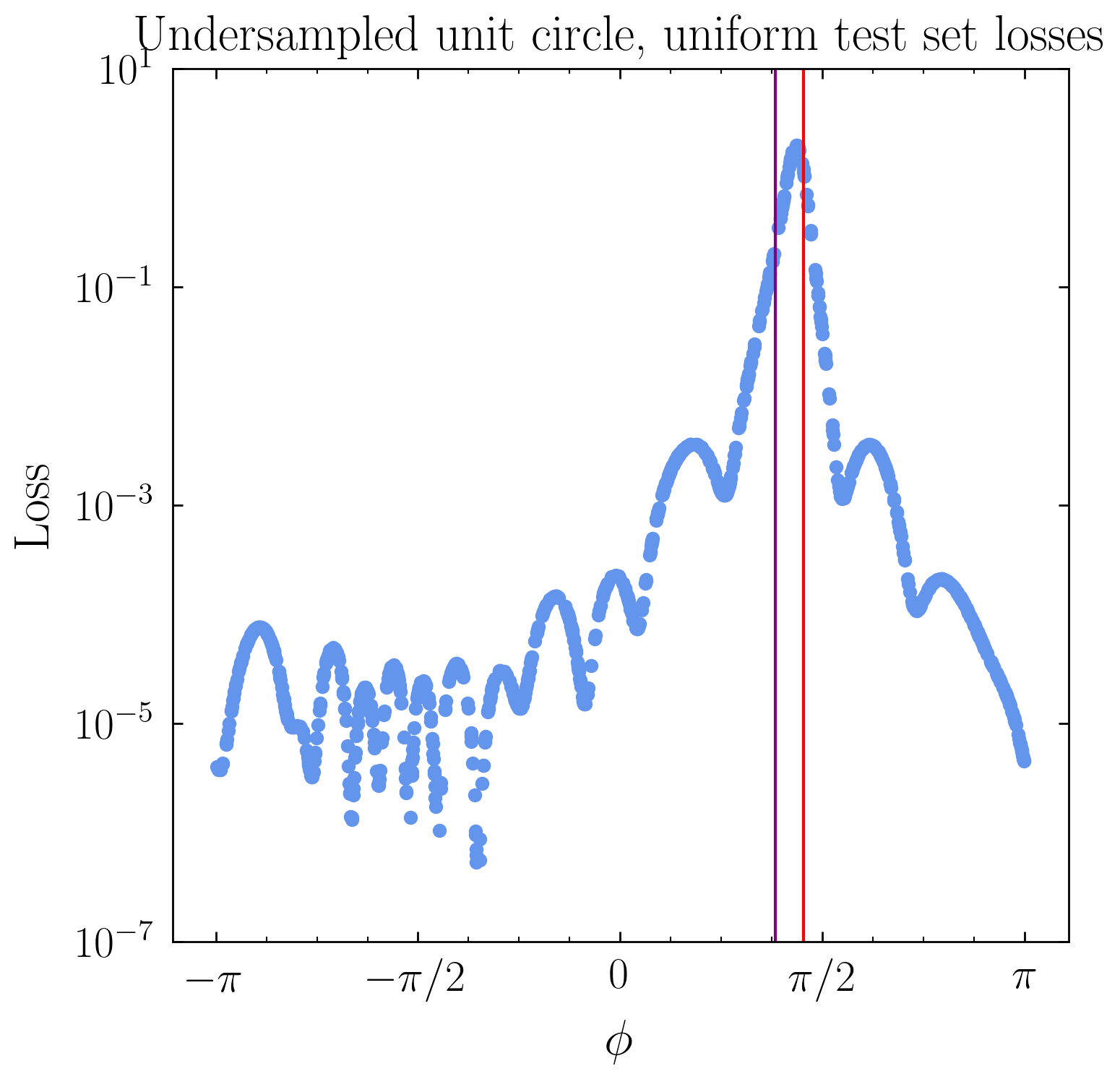}
	\caption{Same as Fig.~\ref{fig:S1Uniform} (bottom row) but for a training set undersampled around $\phi_u$ (purple). The break point $\phi_0$ is shown in red.}
	\label{fig:S1Undersampled}
\end{figure}

Indeed, this is exactly what happens.  Fig~\ref{fig:S1Uniform} (top left) shows the latent representation as a function of the input $\phi$ after training the 5-layer network on a training set composed of equidistant points on the unit circle. The representation fails quite obviously at a particular angle $\phi_0$ (marked in red) which we refer to as the break point; this result is consistent with the fact that the circle with a point excised is topologically equivalent to $\mathbb{R}$.\footnote{Given that the training set is exactly uniform (to machine precision), the appearance of a preferred angle $\phi_0$ is a textbook example of spontaneous symmetry breaking; as the size of the training set increases, a continuous family of identical loss minima parameterized by $\phi_0$ emerges. In future work we plan to investigate how this symmetry breaking is realized on the loss landscape of the autoencoder, given that $\phi_0$ is determined by the stochastic dynamics of network initialization and training.}
As can be seen in the plot of the model output (Fig.~\ref{fig:S1Uniform}, top right, where we define the output $\phi$ as $\tan^{-1}(y/x)$), the autoencoder maps points near $\phi_0$ all over the circle, with output values of $\phi$ ranging from $-\pi$ to $\pi$. This is also easily visualized by plotting the model as points in $\mathbb{R}^2$ (Fig.~\ref{fig:S1Uniform}, bottom left). This leads to large reconstruction error in the neighborhood of $\phi_0$: Fig.~\ref{fig:S1Uniform} (bottom right) shows the losses as a function of $\phi$ on a uniformly-sampled test set. Losses at $\phi_0$ are on the order of $10^3$ times the loss of a generic test set point, despite the fact that the break point is not an anomalous point of any kind but rather just another generic point on the circle. In Appendix~\ref{app:MoreCircle}, we solve the network dynamics for a generic activation function and SGD training and demonstrate why a finite-sized break region around $\phi_0$ persists even after long amounts of training. The size of this break region is roughly independent of the network width and depth for a fixed training length, shrinks very slowly (perhaps logarithmically) with training, and depends primarily on the particular form of the activation function and the training algorithm. We plan to return to the rich interplay between topology and network dynamics illustrated by this simple example in future work.

In fact, the region around the break point persists even for the absurdly small training set of 20 equidistant points on the unit circle. In that case, the loss at the worst point in the training set is only $\sim$10 times the loss for a generic point after 100,000 epochs of training. However, the network has not simply memorized the training data because the output map fails to reconstruct at least one of the training set points. Indeed, a bad point seems to occur as long as the density of the training set is high enough that the break region size exceeds the spacing between data points (for the hyperparameters given in Appendix~\ref{app:hyperparams}, this occurs for training sets containing 15 or more equidistant points on the unit circle).\footnote{We can also formulate this observation in terms of persistent homology, a method of identifying topological features of datasets \cite{carlsson2009topology}: if the data has persistent first homology $H^1$ at the length scale defined by the size of the break region, then it will behave like a topological circle and have a bad point in its reconstruction.} We provide more details in Appendix~\ref{app:MoreCircle}, including a number of other checks showing that the behavior we see persists with different training algorithms and activations, and that near-perfect reconstruction error can be achieved if $d_l = 2$ since the autoencoder finds the trivial global minimum.\footnote{In this case, since with $d_l = 2$ we have $d_l = \din$, the network is learning the identity map on all of $\mathbb{R}^2$. More complicated topologies in higher dimensions may provide obstructions when $d < d_l < \din$, and we study such an example in App.~\ref{sec:torus}.}

These observations are related to general considerations about the performance of neural networks on interpolation and extrapolation tasks. To see this, consider a training set which is undersampled near a randomly chosen point on the unit circle given by $\phi_u$. Fig.~\ref{fig:S1Undersampled} shows the result of training an autoencoder on points sampled from a normal distibution with mean $\phi_u + \pi$ and standard deviation $\pi/3$. The break point now lies in the undersampled region (with $\phi_u$ shown in purple), but all other aspects of the autoencoder behavior are similar to the equidistant training set. In effect, we are asking the autoencoder to perform an interpolation task -- on which neural networks typically have excellent performance -- but the nontrivial topology of the circle makes this task impossible. In this 1-dimensional example, points absent from the training set in the neighborhood of $\phi_u$ are indeed reconstructed with large loss, but this is not because these points are anomalous per se, but rather because the topology forces large loss to occur somewhere and the overall loss is minimized by placing the break point in the region where the fewest training points exist. Indeed, we can choose the location of the break point by changing the sampling distribution. We emphasize again that the reconstruction error for an undersampled topologically-trivial curve is \emph{not} enhanced in the undersampled region; an autoencoder has no trouble learning a distribution on an interval, except near the endpoints where the reconstruction task changes from interpolation to extrapolation. Thus, topology precludes a simple 1-to-1 mapping between autoencoder loss and typicality of data. This behavior persists in higher dimensions, as we discuss further in Sec.~\ref{sec:sphere} below.

\section{Dimension 2}
\label{sec:dim2}

As we begin to investigate higher-dimensional data sets, visualizing both the latent representation of the data and the data manifold itself will become more difficult. Visualization is still manageable in $d = 2$, but to prepare for higher-dimensional examples, we will introduce a useful tool, the loss-versus-distance plot. This is a scatter plot of the autoencoder loss on points in the test set versus their Euclidean distance from the point of largest loss. The intuition is that manifolds which suffer poor reconstruction error in the neighborhood of a single point will show losses anti-correlated with distance from that break point, as in the case of the circle. Indeed, since the $n$-sphere $S^n$ with a single point excised can be covered with a single chart, all autoencoders trained on spheres should exhibit this behavior, regardless of dimension (we will see this explicitly in Sec.~\ref{sec:dimn}). If, on the other hand, the loss appears to be uncorrelated with distance, then the manifold may have more complex topology, requiring tearing along a submanifold (instead of just puncturing) to fit in $\mathbb{R}^n$; we study such examples in App.~\ref{app:MoreExamples}.

Our examples here will illustrate that the issue of intrinsic topology we identified in 1-dimensional data sets persists in $d = 2$. However, in dimension 2 we can have the qualitatively different situation of undersampling our data distribution along a 1-dimensional submanifold, as opposed to dimension 1 where submanifolds are just isolated points. We will see that, depending on the topology of the data manifold, most of the submanifold may be reconstructed with small loss, despite being absent from the training set.

\subsection{The 2-sphere and the paraboloid: interpolation and extrapolation}
\label{sec:sphere}

\begin{figure}[t!]
\begin{center}
	\includegraphics[width = 0.27\textwidth]{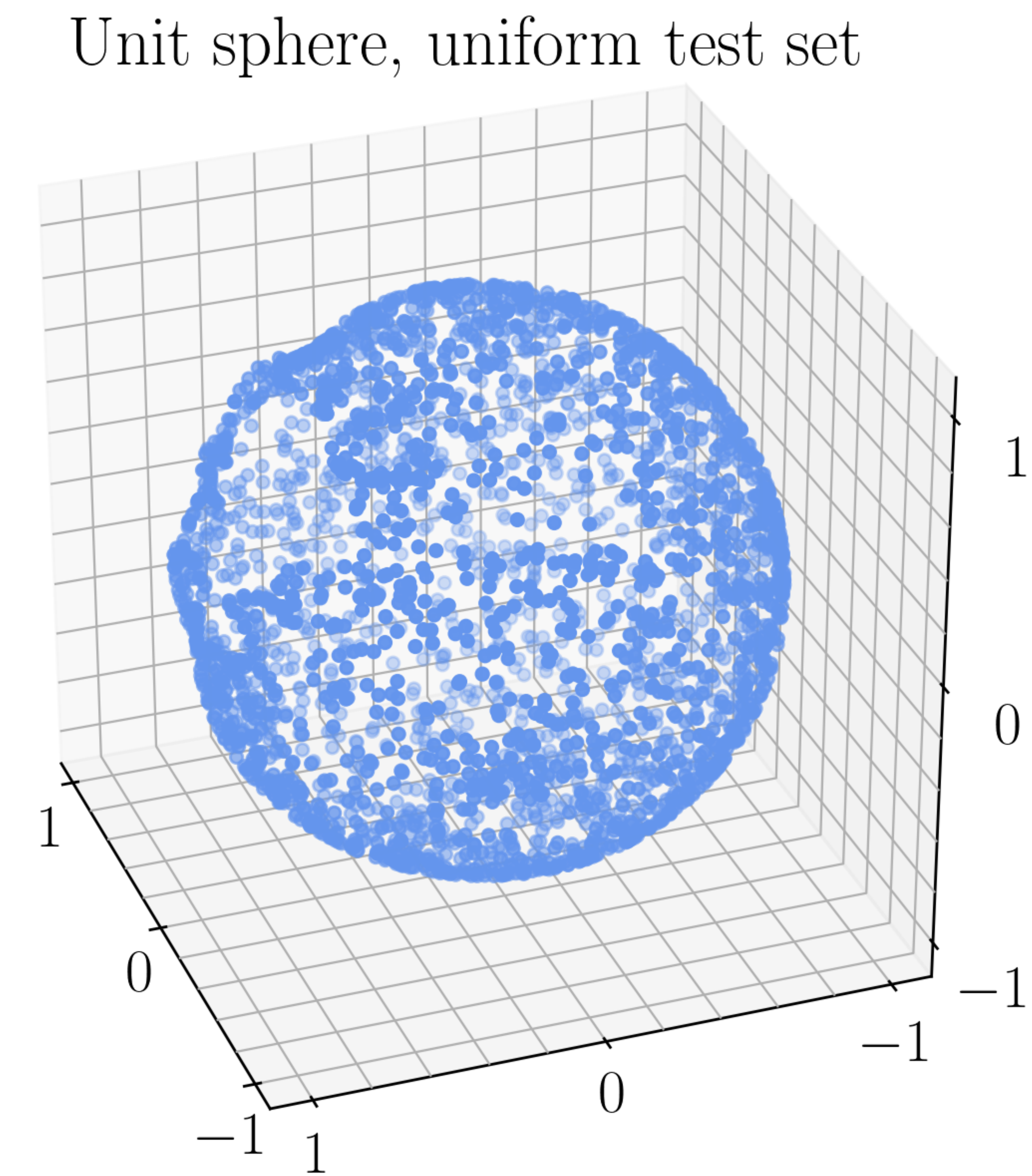}
	\includegraphics[width = 0.35\textwidth]{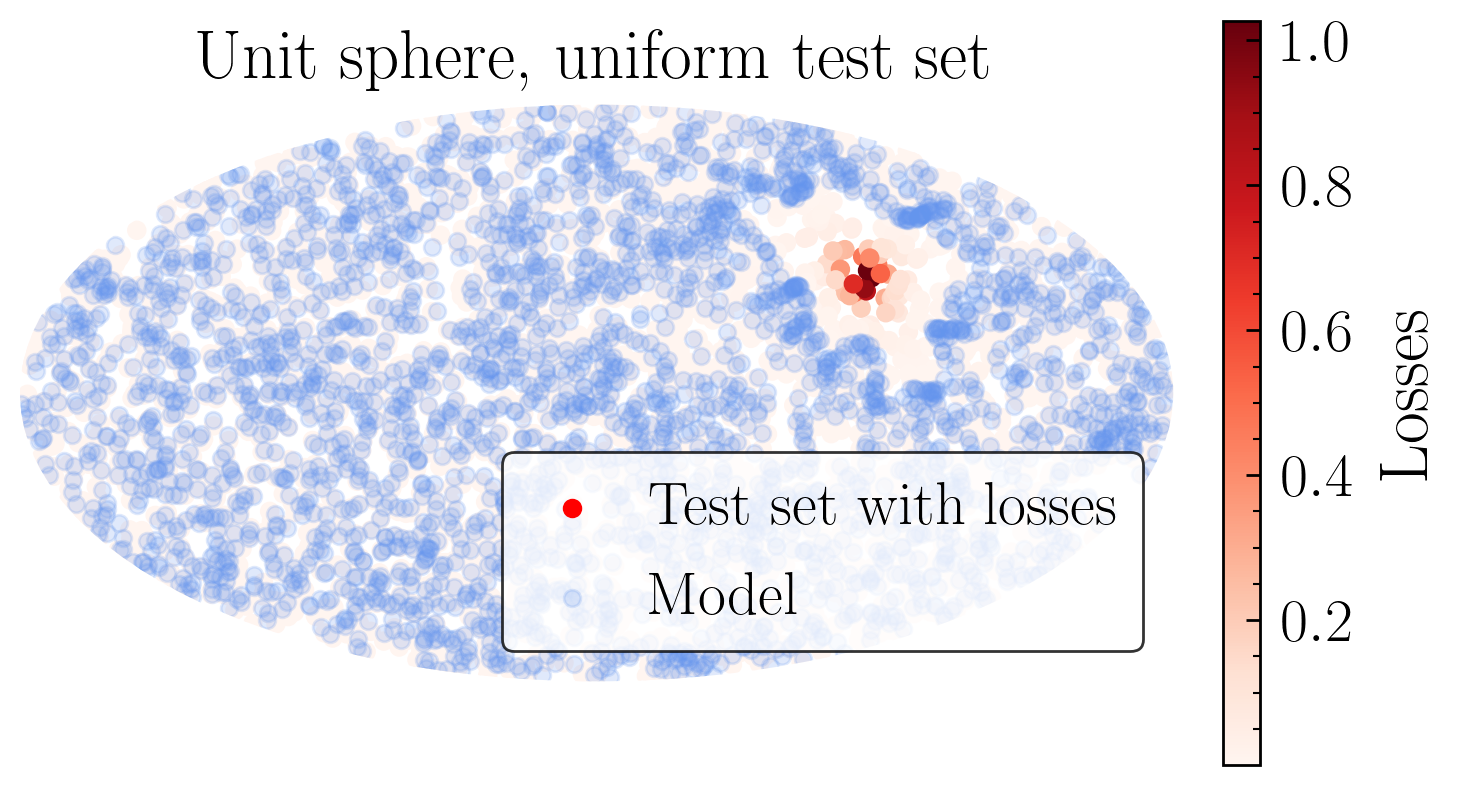}
	\includegraphics[width = 0.27\textwidth]{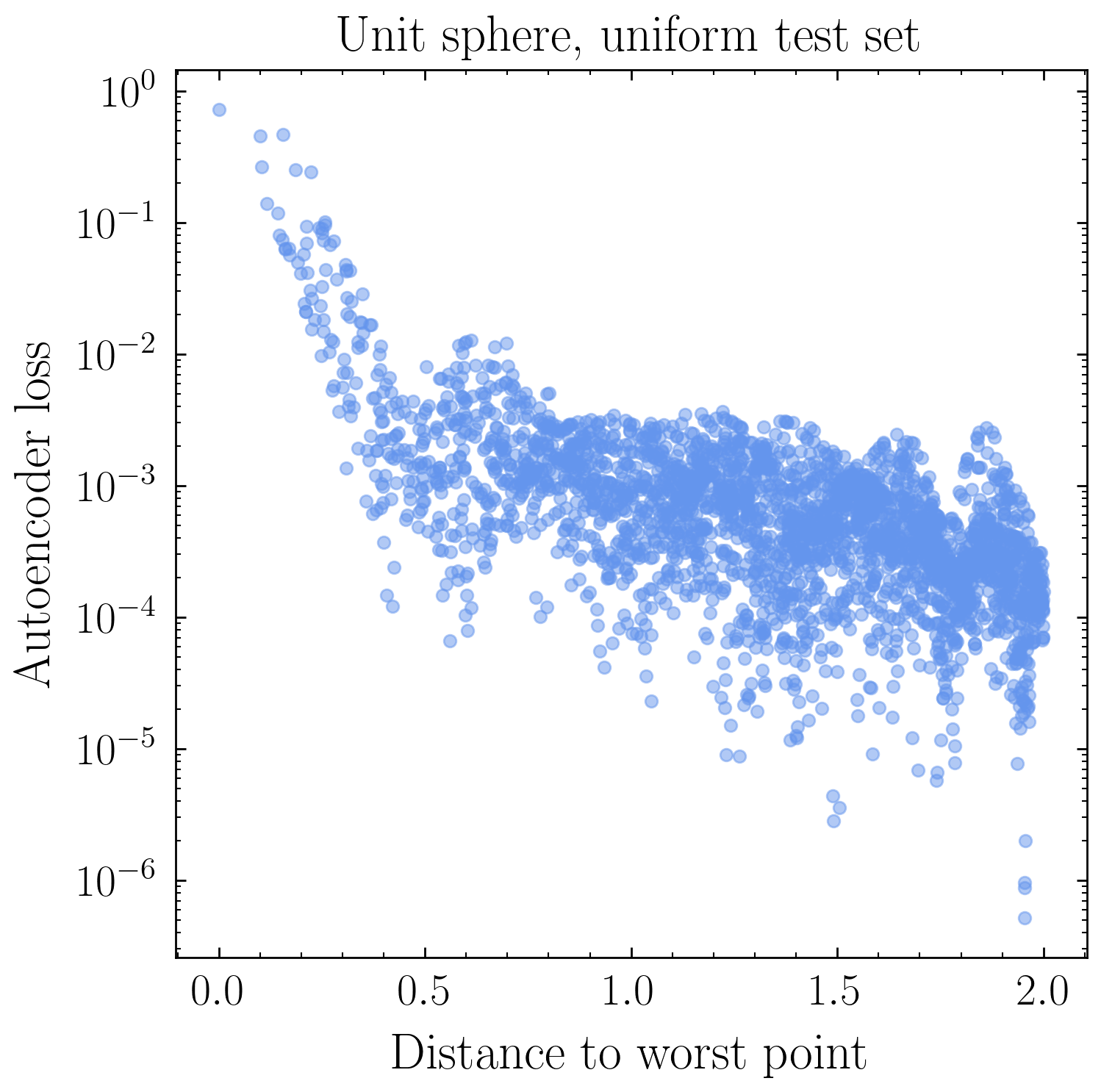}
	\caption{$S^2$ autoencoder. \textbf{Left:} autoencoder output for a uniform test set showing a ``hole'' analogous to the break point for the $S^1$ autoencoder (here at upper left of the figure). \textbf{Center:} Mollweide plot showing the autoencoder output in blue, along with the test set colored by loss, showing that large losses are localized to the hole. \textbf{Right:} loss-versus-distance plot showing that the loss falls monotonically with distance, indicating a localized tearing in the latent map. This visualization more straightforwardly generalizes to higher dimensions.}
	\label{fig:S2Uniform}
\end{center}
\end{figure}

As we did with the unit circle, we consider training an autoencoder on a uniformly-sampled unit sphere $S^2$ (in this example we use uniform sampling rather than equidistant points for the training set, but this makes no material difference for the examples to follow), defined by $x^2 + y^2 + z^2 = 1$ in $\mathbb{R}^3$. Using the same training scheme as with the circle, but now using an autoencoder with $d_l = 2$, we find the results shown in Fig~\ref{fig:S2Uniform}: the loss is localized near a single point on the sphere (as with the circle, this point is randomly chosen by initialization and stochastic dynamics), the autoencoder model punctures the sphere in a region around that point, and the loss is $\sim 10^3$ worse in this region than at a generic point in the test set.\footnote{Unsurprisingly, for $d_l = 1$, the reconstruction is poor everywhere except on a randomly-chosen curve on the sphere, which must have a break region because of the analysis of Sec.~\ref{sec:circle}.} 

\begin{figure}[t!]
\begin{center}
	\includegraphics[width = 0.28\textwidth]{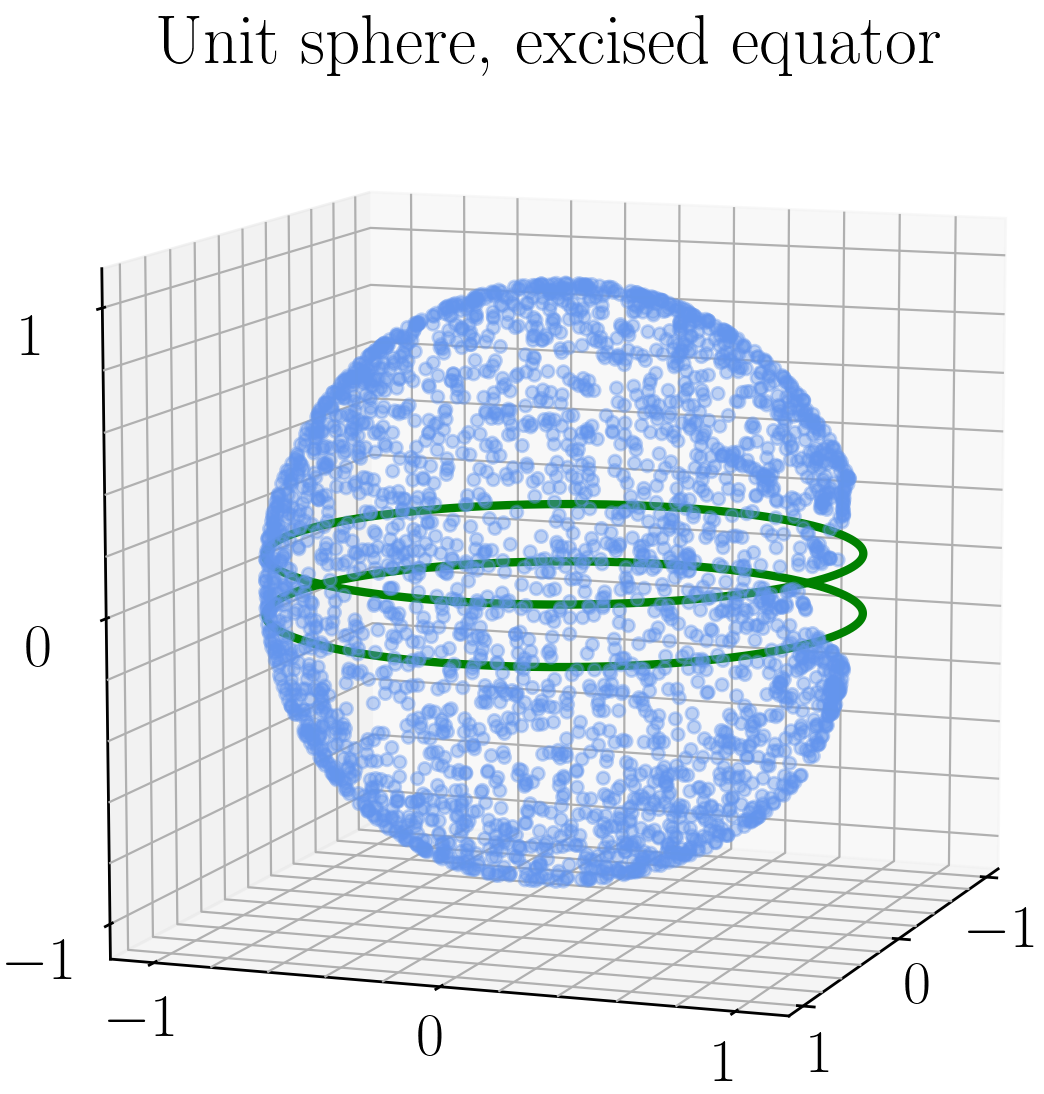}
	\includegraphics[width = 0.4\textwidth]{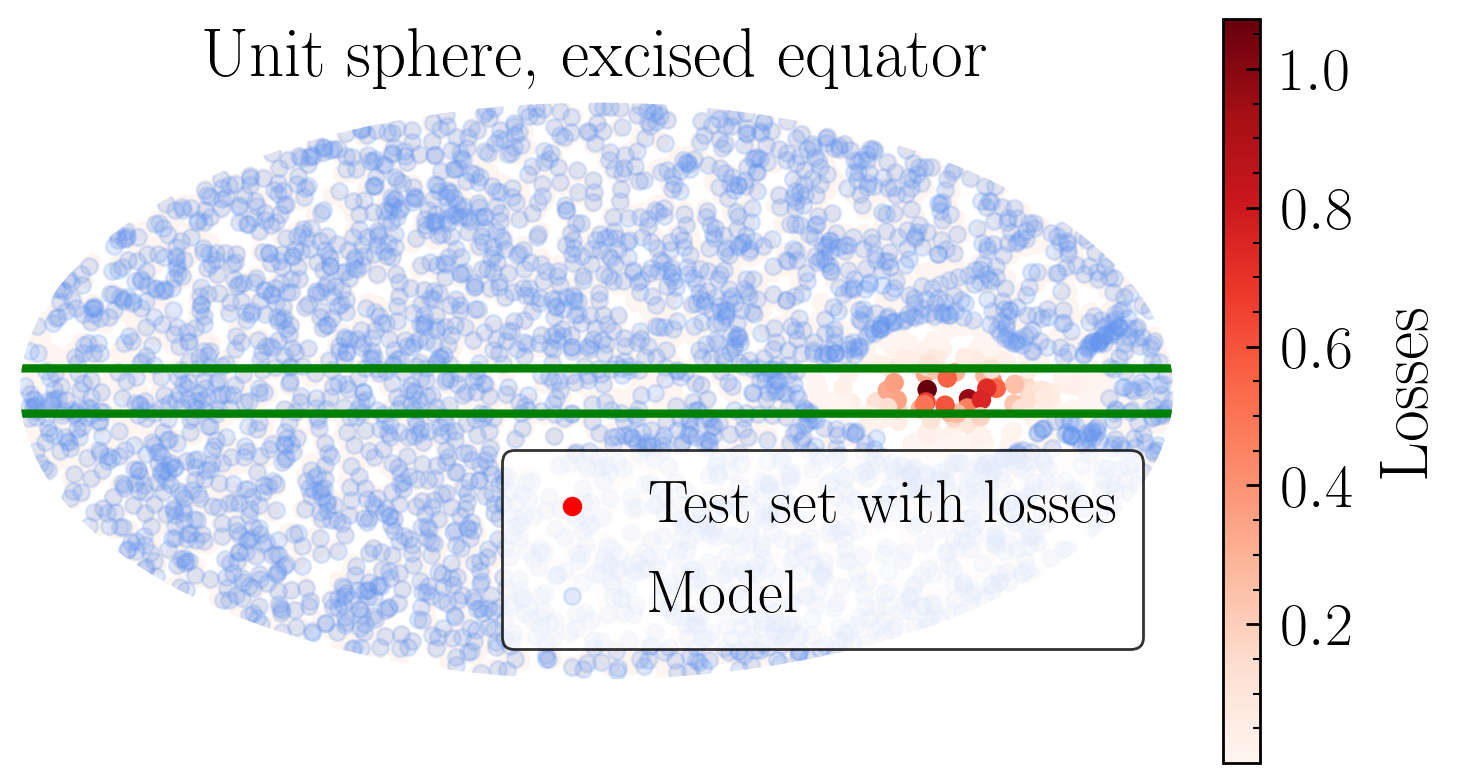}
	\includegraphics[width = 0.28\textwidth]{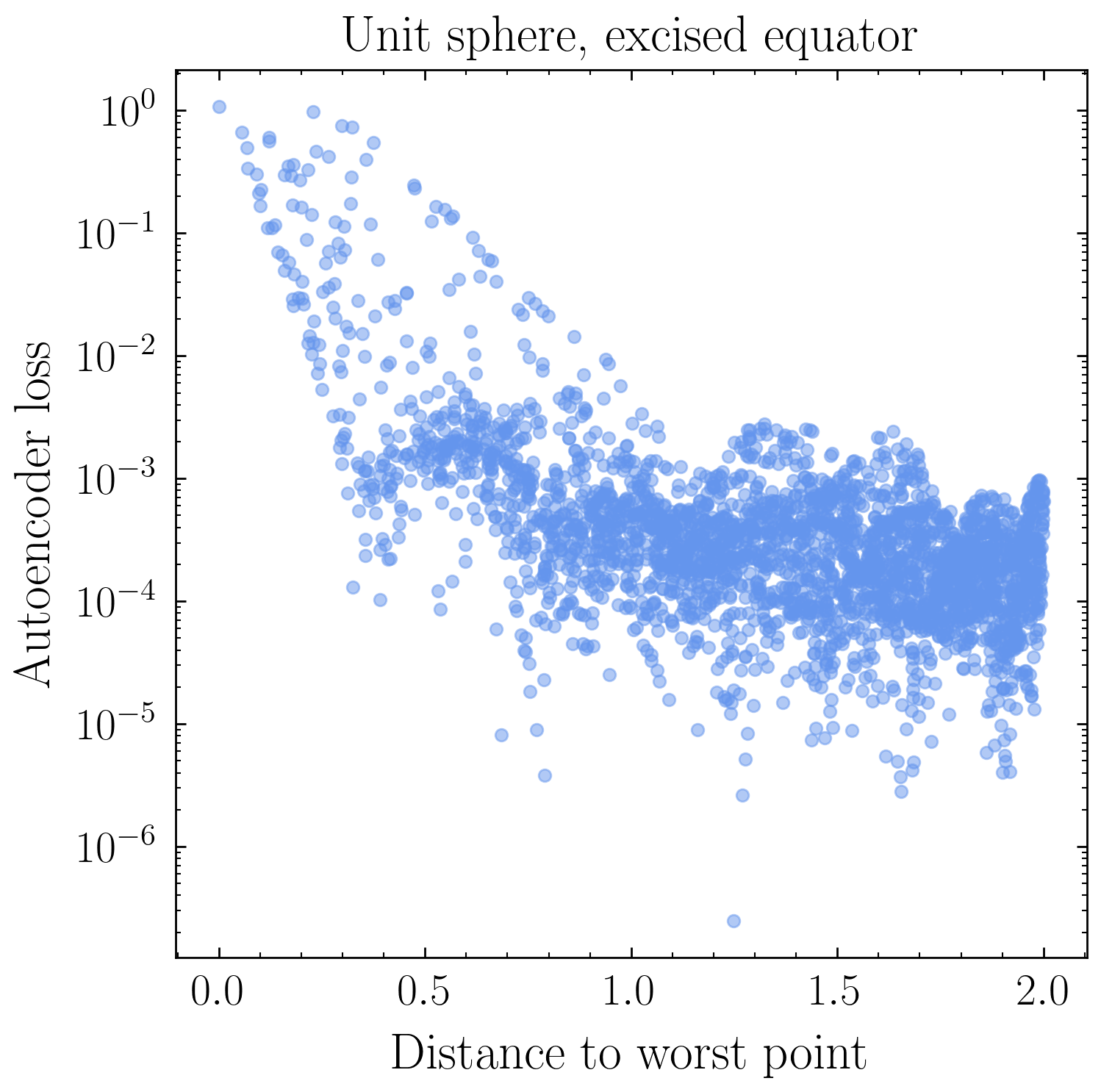}
	\caption{Same as Fig.~\ref{fig:S2Uniform} for a training set with the region at the equator between the green curves excised. The model map can interpolate most of the equator with low loss but breaks at a point along the equator.}
	\label{fig:S2Split}
\end{center}
\end{figure}

As with the circle, we can undersample the sphere at a point, and just as with the circle, the break point of the model map falls in this undersampled region. However, because the sphere is two-dimensional, we now have the opportunity to undersample along an entire 1-dimensional submanifold, for instance the great circle along the equator. This situation is a closer analogy to our bump hunt example, where rare events tend to lie on submanifolds, rather than at isolated points, of phase space. Since another way to trivialize the topology of the sphere is to excise an entire great circle, yielding the topology of two disks, $D^2 \oplus D^2$, we might expect that this will also be a local minimum of the autoencoder loss. However, after training an autoencoder on a uniform distribution on the sphere but with the region around the equator with $|z| < 0.1$ excised entirely, the model map (Fig.~\ref{fig:S2Split} left and center) typically breaks at a random point along the equator; it has no trouble interpolating the rest of the equator (which was absent from the training set entirely) because there is no topological obstruction to doing so. The trained network does occasionally yield the output with the $D^2 \oplus D^2$ topology; however, over many network realizations, the local minimum with a single break point is much more common, and moreover has lower overall loss. The situation we have described is thus complementary to the 1-dimensional case of the unit circle. The best local minimum for the autoencoder is the one which distorts the data manifold at the fewest number of points; since this can be done by removing a single point on $S^2$ (i.e. a submanifold of dimension 0), ``anomalous'' (i.e.\ undersampled) submanifolds of dimension 1 will be interpolated with low loss except perhaps at an isolated point. Without any additional way of influencing the latent representation, this behavior would seem to preclude using autoencoders to learn this family of distributions on the 2-sphere.

\begin{figure}[t!]
\begin{center}
	\includegraphics[width = 0.5\textwidth]{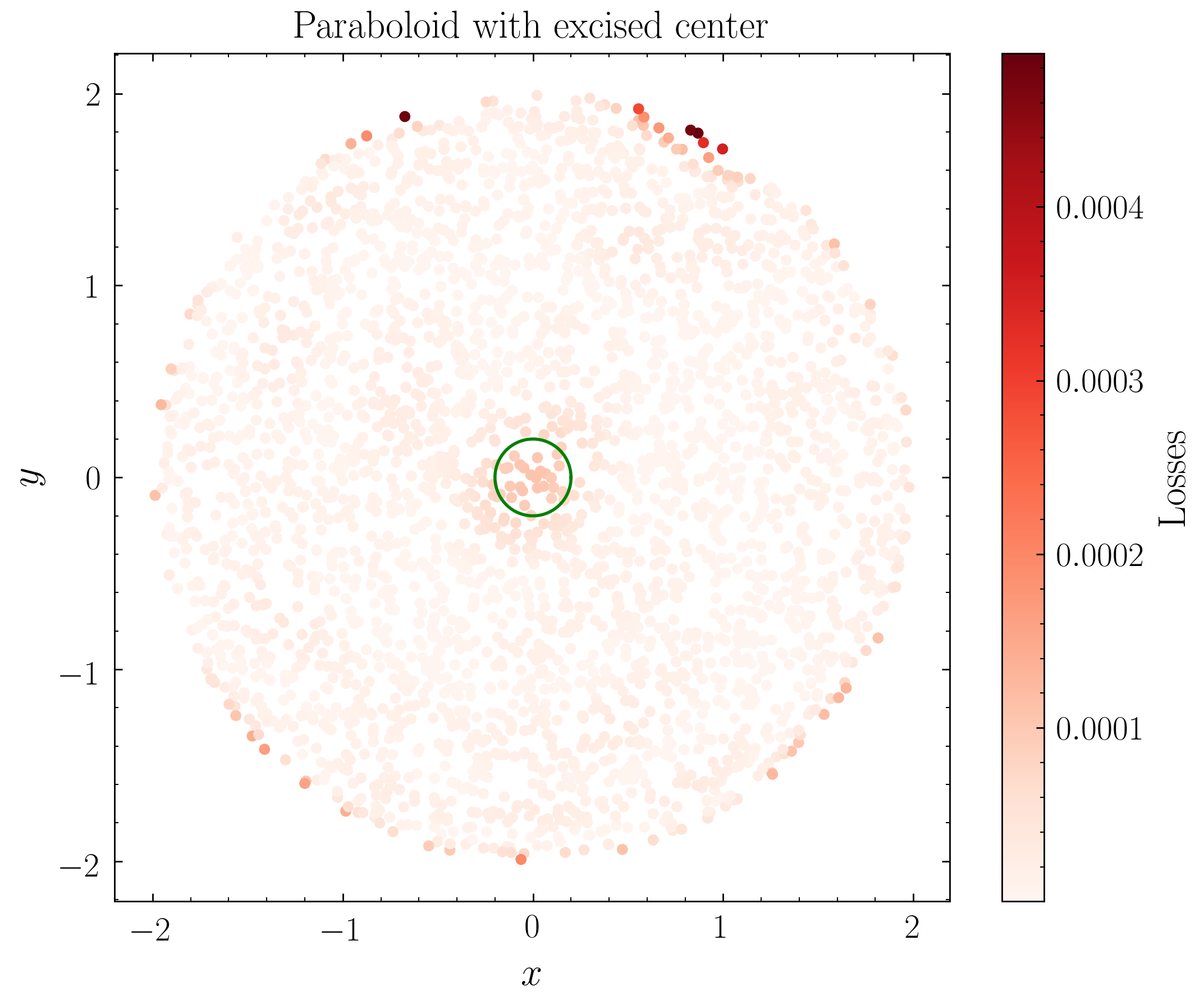}
	\caption{An autoencoder trained on a paraboloid $z = x^2 + y^2$, $z \leq 4$, with the region $z < 0.2$ (green circle) excised, can interpolate the excised region better than it can extrapolate the boundary.}
	\label{fig:Paraboloid}
\end{center}
\end{figure}

To demonstrate that this interpolation is a generic feature of autoencoders, we train a network with the same architecture and hyperparameters as for the sphere example on a topologically trivial surface, the paraboloid $z = x^2 + y^2$ with the region $z < 0.2$ excised. The test set is sampled uniformly in $x$ and $y$ up to $x^2 + y^2 = 4$. Fig.~\ref{fig:Paraboloid} shows the losses on a test set sampled from the full paraboloid with $0 \leq z \leq 4$. The center region is interpolated with much smaller loss than the largest-loss points, which are localized on the boundary. Indeed, the finite extent of the training set implies the topology of a manifold with boundary, and reconstructing the boundary accurately is an extrapolation task, which is generally more difficult for neural networks than the interpolation task of filling in the center. Despite this, the worst loss is more than two orders of magnitude smaller than the worst loss for the excised sphere example, because (neglecting the boundary at $z = 4$) the paraboloid has the same topology as $\mathbb{R}^2$.

\subsection{The double cone: extrinsic geometry and non-uniform sampling}
\label{sec:cone}

\begin{figure}[t!]
\begin{center}
	\includegraphics[width = 0.3\textwidth]{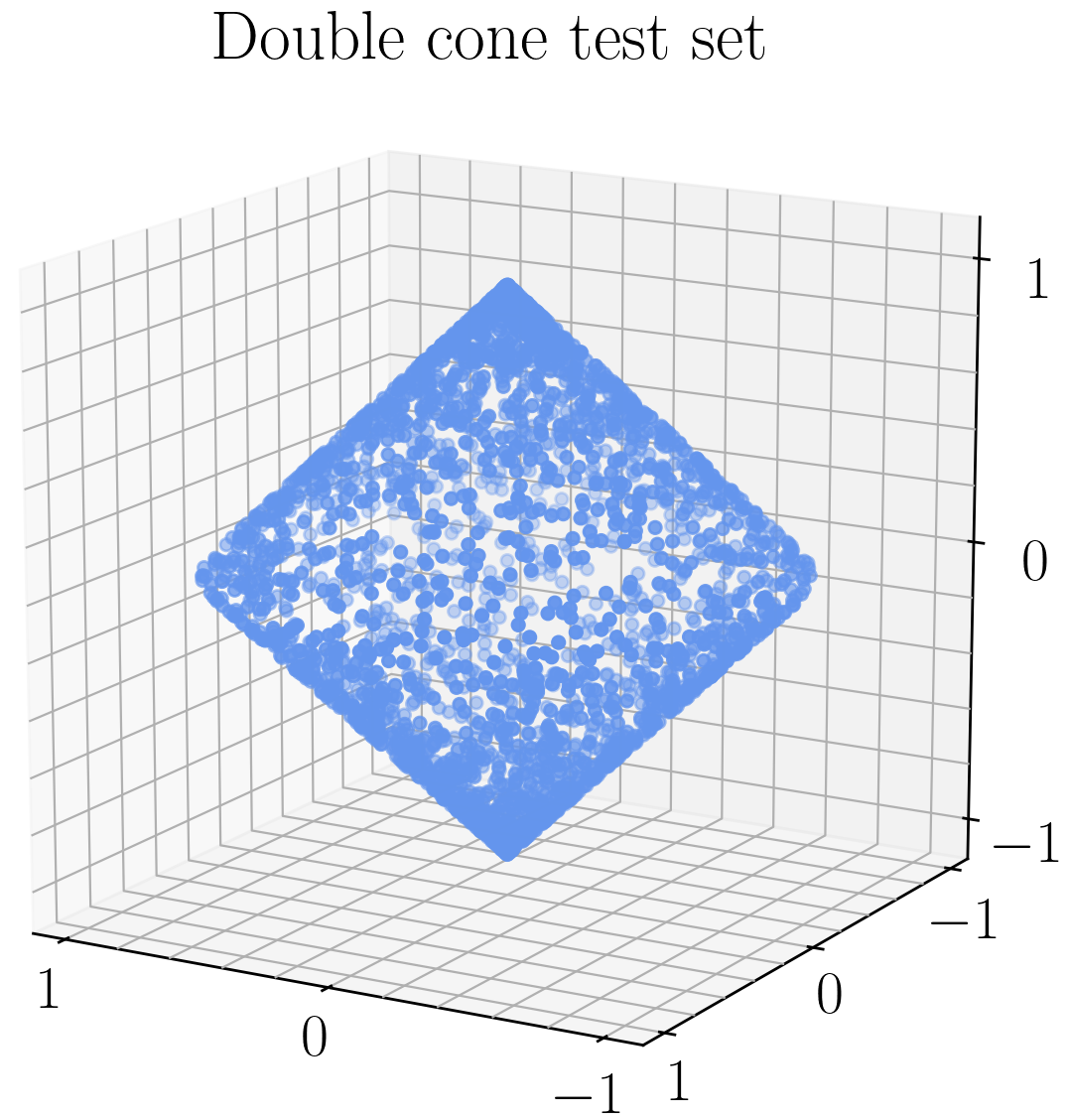} \ \ 
	\includegraphics[width = 0.3\textwidth]{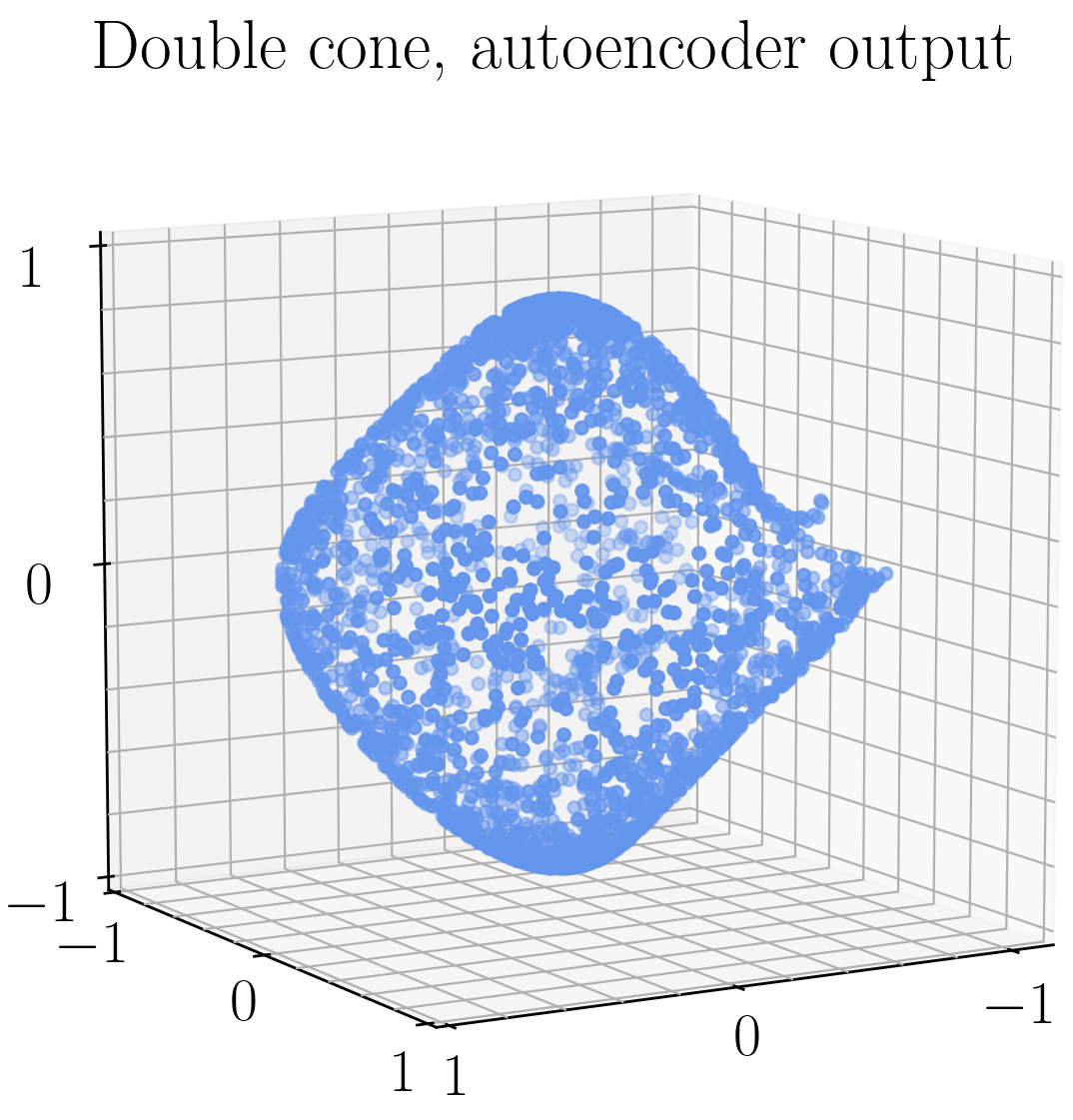} \ \ 
	\includegraphics[width = 0.3\textwidth]{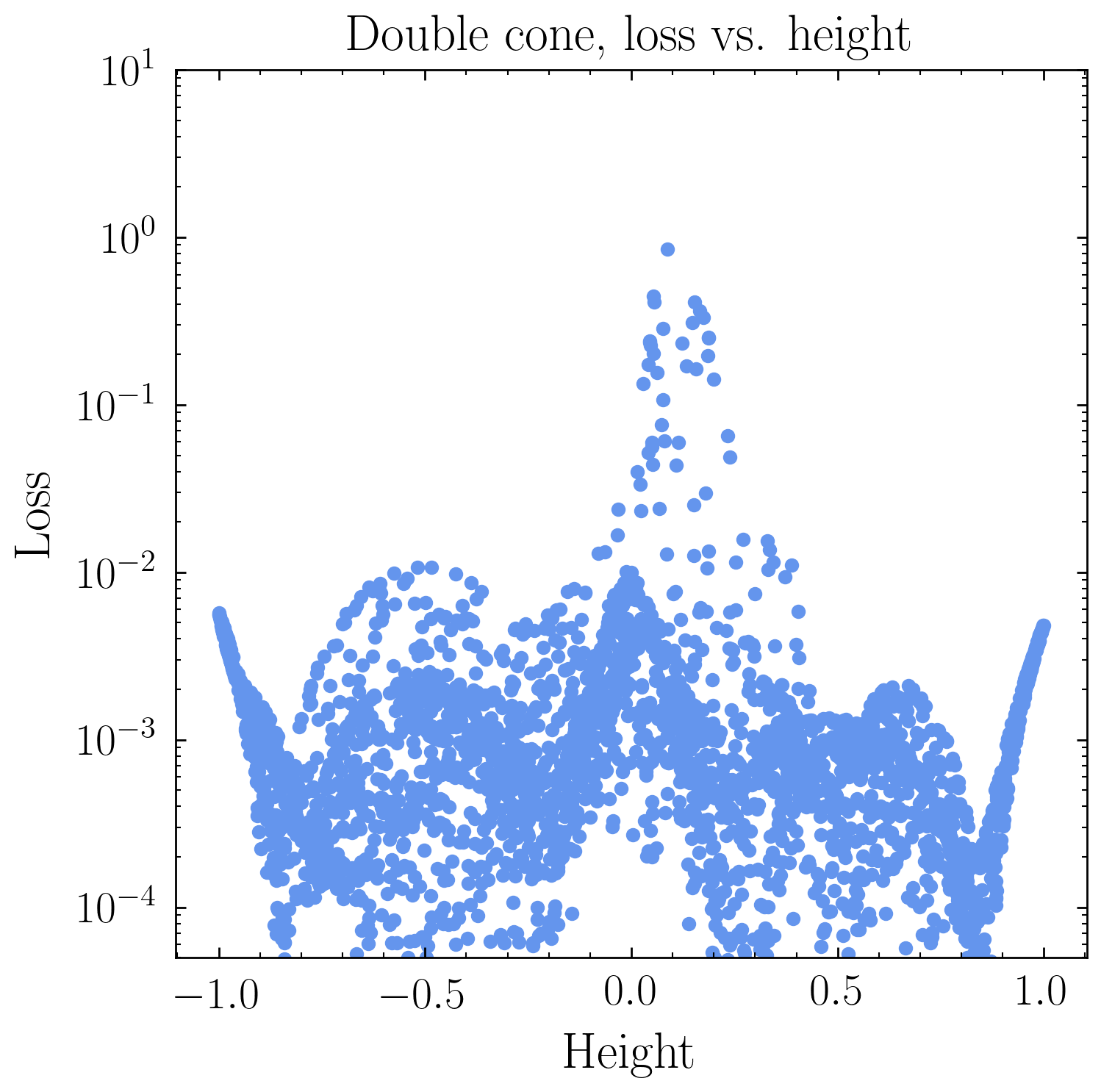}
	\caption{Double cone autoencoder. \textbf{Left:} Training set sampled uniformly by height, thus oversampled at the tips. \textbf{Center:} Autoencoder output, showing a break point as for the $S^2$ autoencoder, as well as somewhat poor reconstruction near the tips. \textbf{Right:} Loss as a function of height, showing the global maximum at the break point but local maxima at the tips.}
	\label{fig:cone}
\end{center}
\end{figure}

Any 2-manifold without boundary or handles is topologically equivalent to the 2-sphere $S^2$. However, the embedding in $\mathbb{R}^3$ can introduce an extrinsic geometry different than that of the round metric on the sphere. For example, a double cone has two distinguished points (the tips of the cones) where the embedding is not differentiable and the extrinsic curvature diverges. As we will see in Sec.~\ref{sec:phasespace} below, this is a decent low-dimensional cartoon of the geometry of massless phase space, where the corners of the Dalitz plot represent the non-differentiable embedding of $\mathcal{M}_{n= 3}$ in $\mathbb{R}^{12}$ at points where the energy of a massless particle goes to zero. In anticipation of that analogy, we will also consider sampling the double cone uniformly in height (analogous to sampling uniformly in the Dalitz triangle), which effectively oversamples near the tips. Fig.~\ref{fig:cone} (left) shows an example of a training set drawn from this distribution, for a right circular cone of height $h = 2$ and equatorial radius $r = 1$. As expected, the density of points near the tips is greater than at the equator.

After training an autoencoder with $d_l = 2$ on the double cone sampled uniformly in height, Fig.~\ref{fig:cone} (center) shows the output of the model on a test set drawn from the same uniform-height distribution. Since the double cone has the topology of $S^2$, there must be a break point, and as with the example of $S^2$ with an excised equator, the break point is located in the ``bulk'' of the cone since the average loss is minimized by placing the break point in the undersampled region. However, the large extrinsic curvature at the tips is an obstruction to reconstructing them well by a smooth function, as can be seen visually from the plot of the model output. In Fig.~\ref{fig:cone} (right) we plot the loss on the test set as a function of the true height of the test set point. The global maximum is at the break point, but there are also local maxima at the tips. The same result is obtained when the equator of the double cone is excised entirely from the training set: the break point now lies on the equator, but the remainder of the equator is interpolated with low loss, and the local maxima at the tips persist. As we will show, the equator in this toy example is analogous to the signal submanifold of fixed 2-particle invariant mass in 3-particle phase space, and our results will be more or less equivalent to Fig.~\ref{fig:cone} (right).

\section{Higher-dimensional spheres}
\label{sec:dimn}

\begin{figure}[t!]
	\includegraphics[width = 0.32\textwidth]{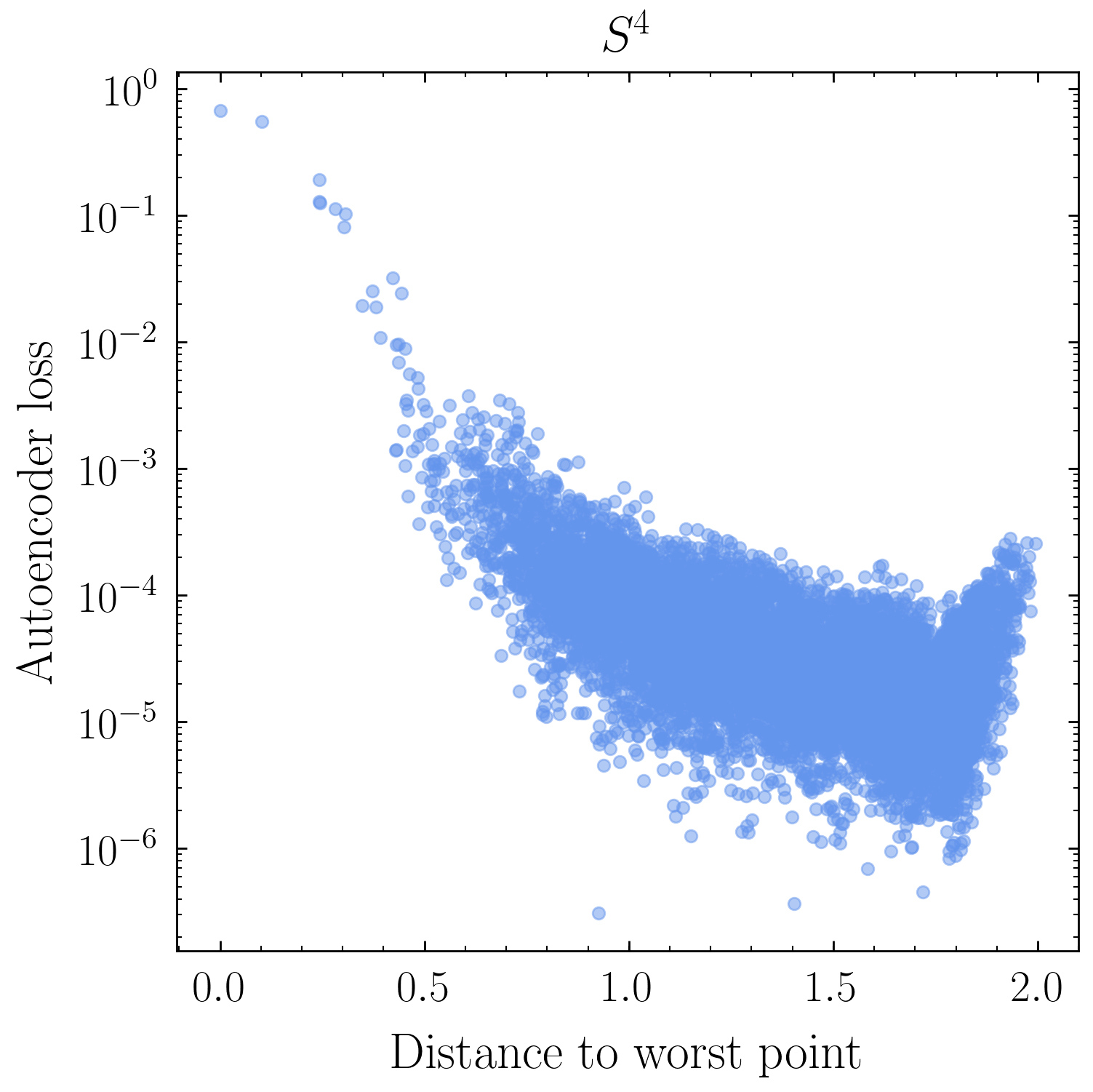}
	\includegraphics[width = 0.32\textwidth]{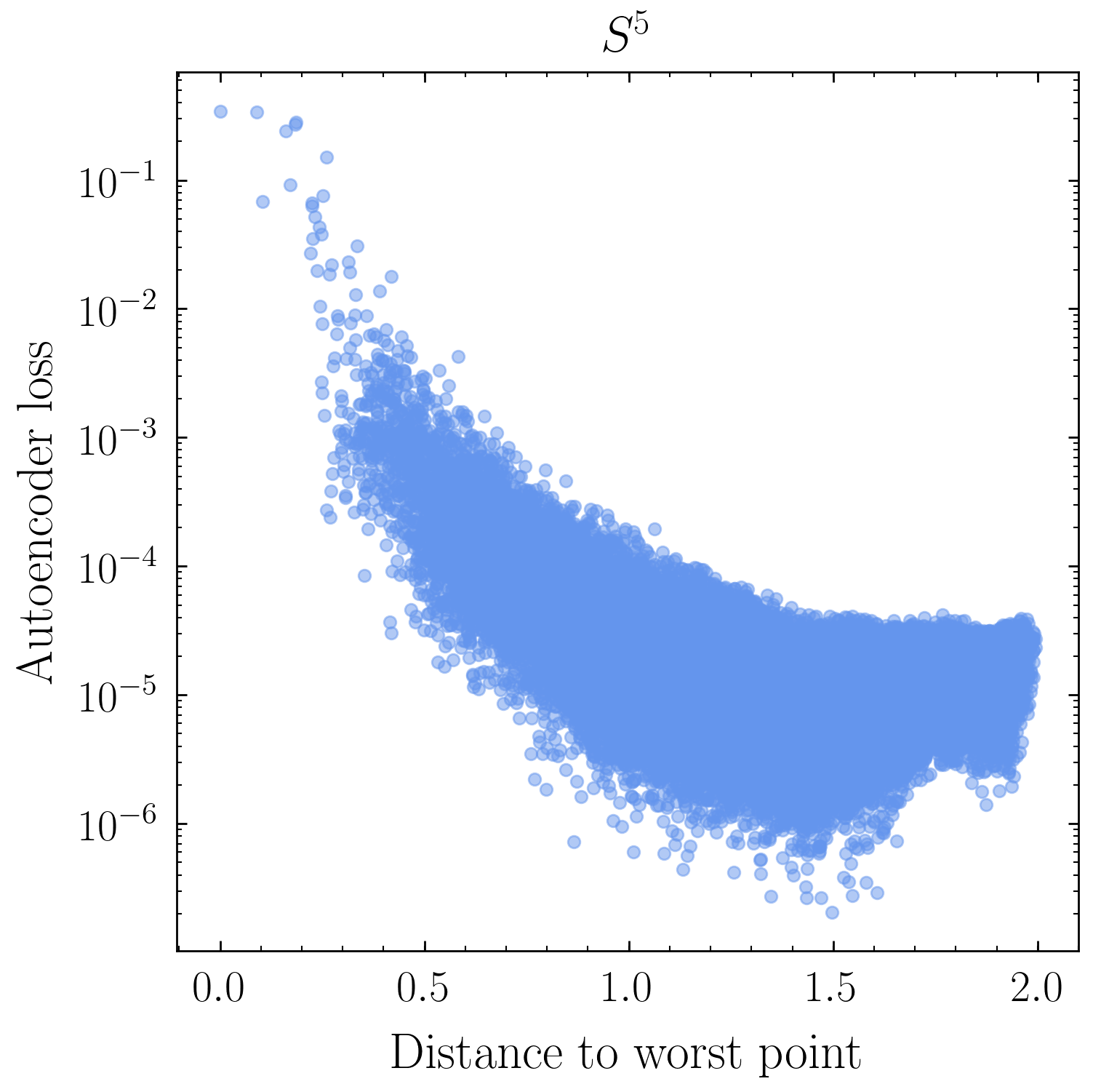}
	\includegraphics[width = 0.32\textwidth]{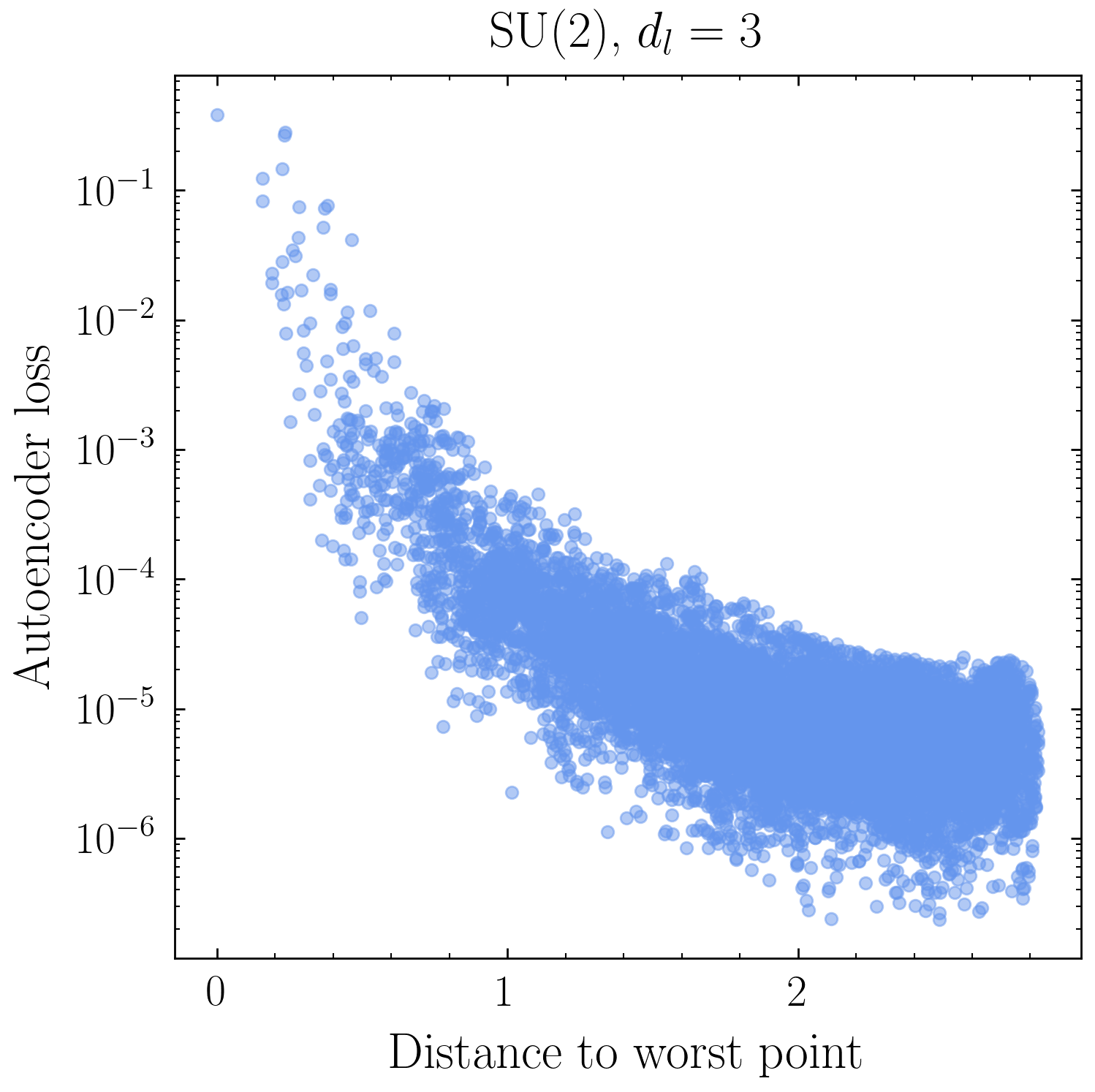}
	\caption{Loss-versus-distance plots for autoencoders trained on $n$-dimensional spheres. For $n = 4,5$, the autoencoder finds the latent map which approximates stereographic projection, with only a single break point. The same is true for SU(2), which is diffeomorphic to the 3-sphere embedded in $\mathbb{R}^8$.}
	\label{fig:Sn}
\end{figure}

Just like the case of $S^2$, the $n$-dimensional sphere $S^n$ can be mapped into $R^n$ everywhere except a single point, in a higher-dimensional analogue of stereographic projection. We can see this explicitly in Fig.~\ref{fig:Sn}, where we have trained the deep 7-layer network with $d_l = n$ on uniformly-sampled training points from the standard embeddings of round spheres, $S^4 \subset \mathbb{R}^5$ and $S^5 \subset \mathbb{R}^6$. We also consider an example of a sphere embedded in higher-dimensional space. The group SU(2), the set of complex $2 \times 2$ matrices $U$ satisfying $U^\dagger U = I_{2 \times 2}$, can be parameterized by a triplet of Euler angles $(\alpha, \beta, \gamma)$ and is diffeomorphic to $S^3$. An element of SU(2) can be mapped into a vector of 8 real numbers, the real and imaginary parts of the matrix entries, and thus embedded in $\mathbb{R}^8$. As shown in Fig.~\ref{fig:Sn}, the SU(2) autoencoder with $\din = 8$ and $d_l = 3$ shows almost identical behavior to the spheres in other dimensions. These examples confirm that the behavior we have been finding -- in particular the utility of the loss-versus-distance plot to visualize the effect of data topology on the autoencoder reconstruction -- persists to higher dimensions. Note that the magnitude of the loss at the break point compared to a generic point on the training manifold, about 5 orders of magnitude, is also robust with respect to dimension with the other network hyperparameters fixed. 

\section{3-body phase space}
\label{sec:phasespace}

Armed with the intuition from the previous lower-dimensional examples, we return to 3-particle phase space. As discussed in Sec.~\ref{sec:bumphunt}, this 5-dimensional manifold $\mathcal{M}_{n=3}$ has a natural embedding in $\mathbb{R}^{12}$; here, we will show that it has the topology of the 5-sphere $S^5$. Intuitively, the mass-shell conditions and the conservation of spatial momenta are topologically trivial, as they can be formulated by saying one variable is a single-valued function of the others. Only the conservation of energy creates topology, and the level sets of the energy function turn out to be spheres. More precisely, suppose the particles have masses $m_i$, energies $E_i$, and spatial momenta $\vec{p}_i$ ($i = 1,2,3$). Then the mass-shell conditions are
$E_i = \sqrt{|\vec{p}_i|^2 + m_i^2}$ for $i=1,2,3$. Since each $E_i$ is determined algebraically by the $p_i$, dropping the $E_i$ coordinates preserves the topology. Let the total initial-state 4-momentum $P = p_1 + p_2 + p_3$ have 4-vector components $P = (E_0, \vec{P}_0)$. Since each $E_i$ is a convex function of the $p_i$, the inequality $E_0 \leq E_1 + E_2 + E_3$ defines a convex origin-symmetric ball in $\mathbb{R}^9$. Conservation of energy says that phase space lies on the boundary of that ball with $E = E_0$. Conservation of momentum slices that ball by the hyperplane $\vec{p}_1 + \vec{p}_2+ \vec{p}_3 = \vec{P}_0$, forming a 6-dimensional ball whose boundary, a sphere, is precisely 3-particle phase space.\footnote{Recent work~\cite{Larkoski:2020thc} proposes a convenient spinor interpretation of phase space as a double quotient of the unitary group $({\rm U}(n) / {\rm U}(n-2))/{\rm U}(1)^n$. Ref.~\cite{Larkoski:2020thc} further decomposes ${\rm U}(n) / {\rm U}(n-2)$ as a twisted product of $S^{2n-1}$ and $S^{2n - 3}$, and gives a measure which is defined simply in terms of each factor. Note that global topology is still spherical; as in the Hopf fibration, the twisted product of spheres is another sphere.} Note also that this argument generalizes straightforwardly to $n$-particle phase space, which has the topology of $S^{3n-4}$.

Visualizing the geometry and topology of high-dimensional manifolds can be difficult, but the Dalitz plot introduced in Sec.~\ref{sec:bumphunt} provides a convenient starting point. A point within the Dalitz triangle fixes the energies of the final-state particles. Momentum conservation implies the final-state particles are coplanar in the COM frame, and thus their orientations are determined by three Euler angles (\textit{i.e.} an element of SO(3)) which fix the unit normal vector to the event plane and the orientation within the event plane. Locally, then, the geometry of 3-body phase space is $\mathbb{R}^2 \times {\rm SO}(3)$. At the boundaries of the triangle, a pair of particles becomes collinear and define an event vector rather than an event plane, which introduces a redundancy because many elements of SO(3) contain the same $S^2$ which orients the event vector. Furthermore, at the vertices of the triangle, a particle becomes soft (\textit{i.e.} its energy goes to zero). The properties of the boundaries and the corners are particularly important for relating the underlying topology to the extrinsic geometry. At the boundaries, uniform sampling in the Dalitz plane leads to effective oversampling with respect to the round metric on $S^5$ because of the redundancy of SO(3) rotations when two vectors are collinear, much as in Sec.~\ref{sec:phasespace} where the double cone sampled uniformly in height oversampled the tips compared to the uniform sampling of the 2-sphere. Furthermore, the embedding of phase space in $\mathbb{R}^{12}$ is non-differentiable at the corners where $E_i \to 0$, leading to a singularity in the extrinsic curvature, a higher-dimensional analogue of the tips of the double cone.

Based on the results of our low-dimensional examples, these topological features should be apparent when uniformly-sampled phase space is used to train an autoencoder with latent dimension 5. In any realistic physics application, the distribution of events will also be weighted by the matrix element for the relevant process, which could have almost arbitrary dependence on the Dalitz plane variables in a model-independent search of the kind autoencoders are useful for. As we have seen in Secs.~\ref{sec:circle} and~\ref{sec:sphere} with the undersampled $S^1$ and $S^2$, the sampling distribution can interplay with the data topology in interesting ways. For the example which follows, we will take a constant matrix element, leaving an exploration of the effects of some common forms of matrix elements for future work.

\begin{figure}[t!]
\begin{centering}
	\includegraphics[width = 0.42\textwidth]{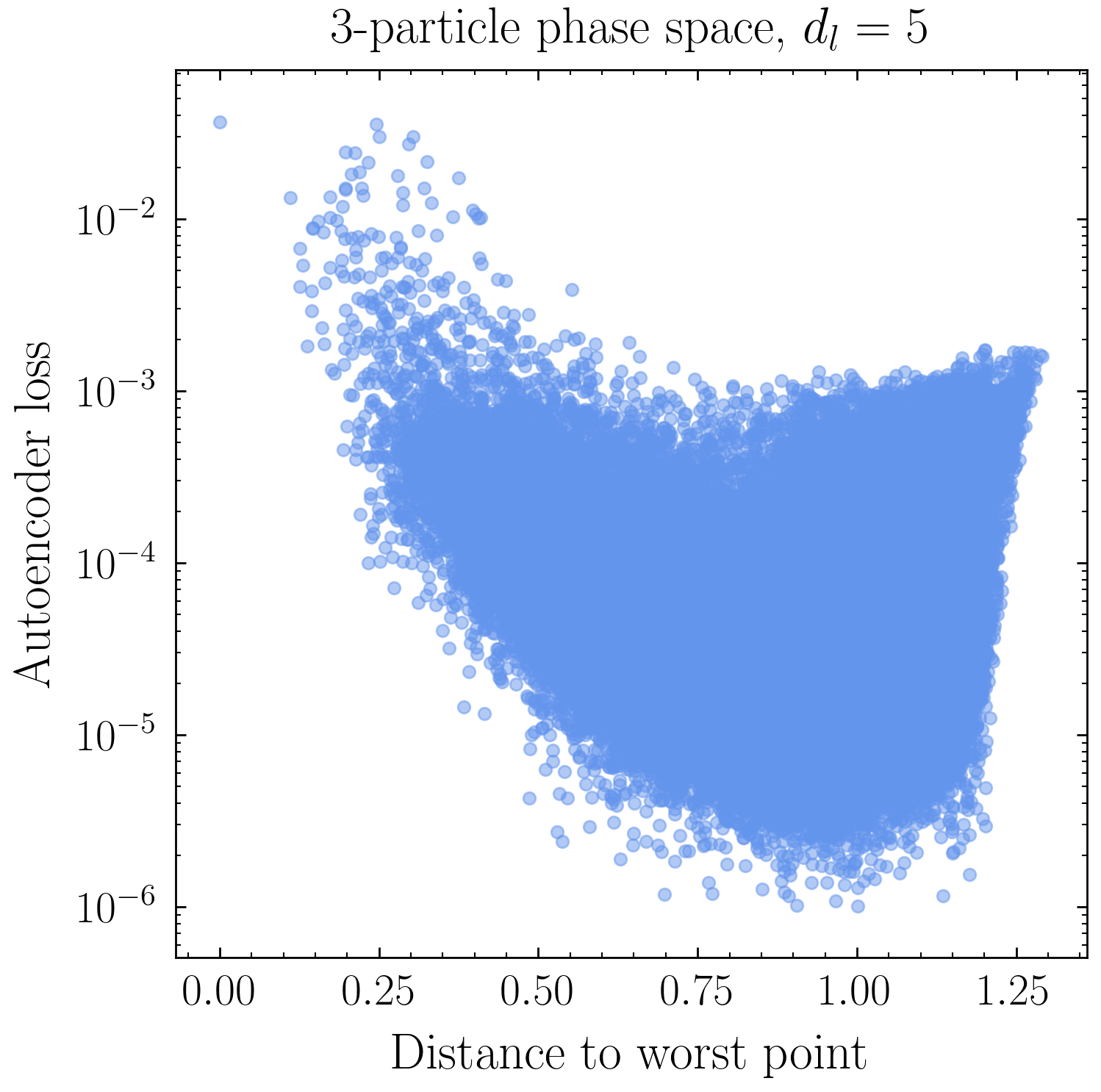}\qquad
	\includegraphics[width = 0.49\textwidth]{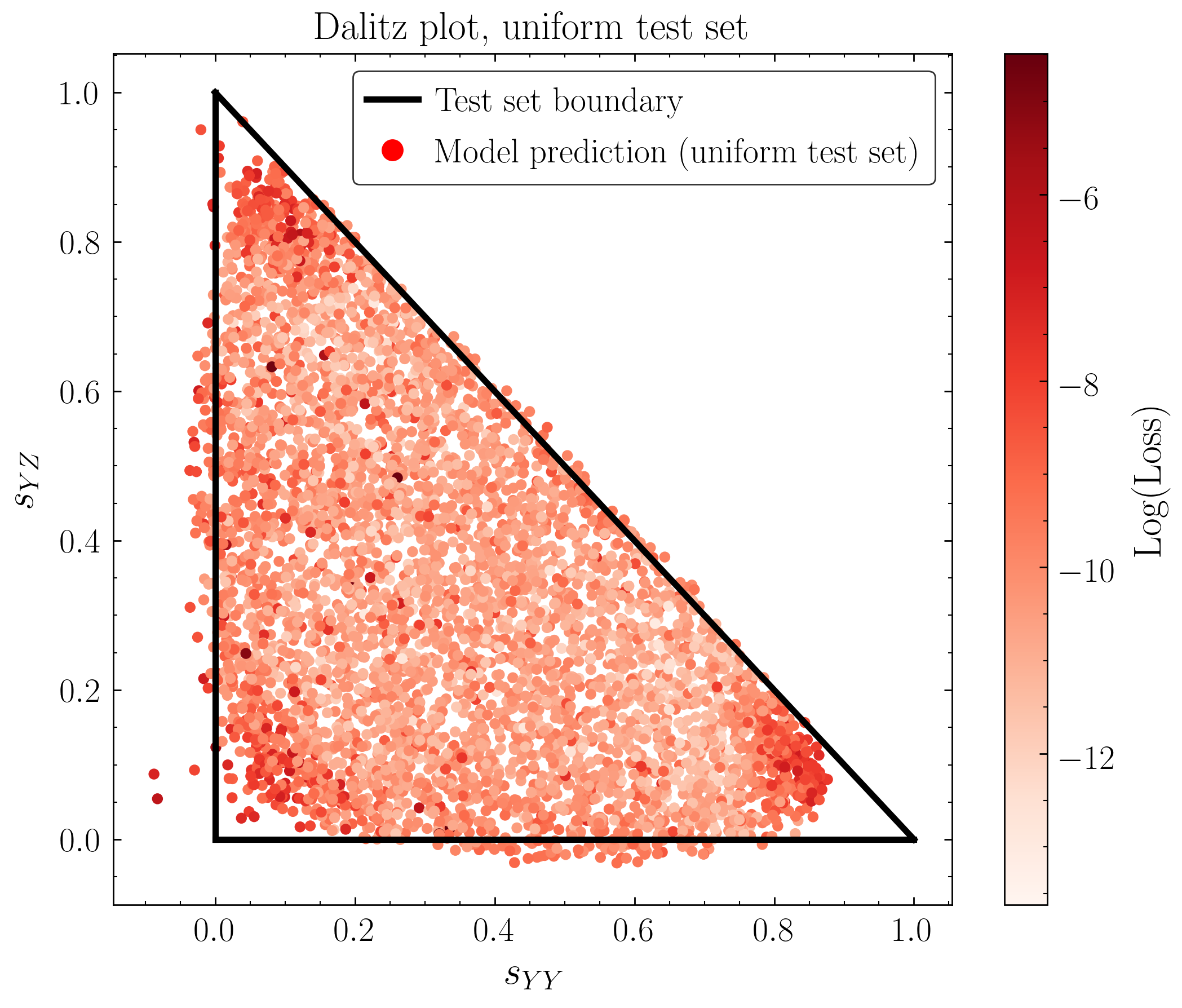} \\
	\includegraphics[width=0.4\textwidth]{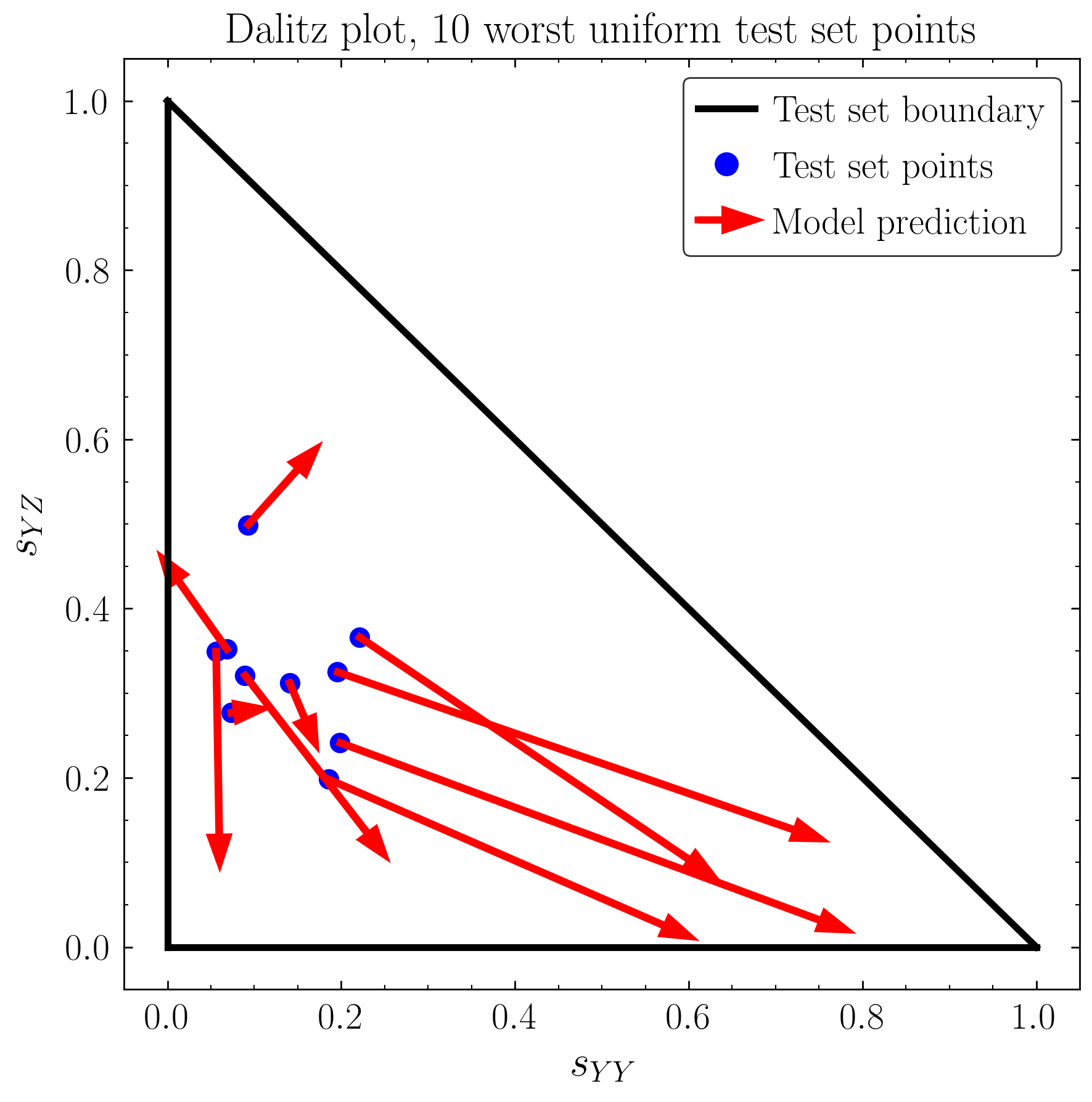}
	\caption{\textbf{Left:} Loss-versus-distance plot for massless 3-particle phase space with $d_l = 5$. The test set is uniformly sampled using the Lorentz-invariant measure (\ref{eq:dPhi3}) from an initial state with unit energy. The resulting loss shows the single break point characteristic of $S^5$ (Fig.~\ref{fig:Sn}, center). \textbf{Right:} Dalitz plot showing the distribution of the model prediction (red) on a uniformly-sampled test set (sampled from the interior of the outlined black triangle). \textbf{Bottom}: The 10 worst points from the uniform test set and their model predictions. The largest-loss points are localized near a generic point in the interior of the Dalitz triangle, while at the corners the loss is lower even as the reconstruction is poor, in close analogy to the double-cone example of Sec.~\ref{sec:cone}.}
	\label{fig:PhaseSpace}
\end{centering}
\end{figure}

Here, we sample events \emph{uniformly} from massless 3-particle phase space~(\ref{eq:dPhi3}) -- as opposed to our sideband distribution in Sec.~\ref{sec:bumphunt} -- and train with a 7-layer autoencoder with $d_l = 5$. The loss-versus-distance plot is shown in Fig.~\ref{fig:PhaseSpace} (left); as expected, the largest loss is localized near a point, reflecting the topology of $S^5$ (compare with Fig.~\ref{fig:Sn}, center). The embedding in $\mathbb{R}^{12}$ does not change the topology, so just as ${\rm SU}(2) \subset \mathbb{R}^{8}$ had the same loss-versus-distance plot as the standard embedding of the $n$-sphere in $\mathbb{R}^{n+1}$ (Fig.~\ref{fig:Sn}, right), $\mathcal{M}_{n = 3} \subset \mathbb{R}^{12}$ exhibits the same topological features as $S^5 \subset \mathbb{R}^{6}$.

To visualize the autoencoder reconstruction, we plot the output of the model on the Dalitz plane in Fig.~\ref{fig:PhaseSpace} (right), exactly analogous to our bump hunt example in Fig.~\ref{fig:SidebandDalitz} of Sec.~\ref{sec:bumphunt}. The corners of the triangle, where the extrinsic curvature is singular, are not reproduced well, and there is a local maximum of the loss at each corner. The behavior is a straightforward higher-dimensional analogue of the double cone of Sec.~\ref{sec:cone}. However, the 10 worst points (and hence the global maximum of the loss) are located near a generic point inside the Dalitz triangle, as shown in Fig.~\ref{fig:PhaseSpace} (bottom).\footnote{Note that since the Dalitz plot is a 2-dimensional projection of 5-dimensional phase space, points that are somewhat distant in the Dalitz plot can still be ``close'' in the SO(3) coordinates over each point; the loss-versus-distance plot of Fig.~\ref{fig:PhaseSpace} (left) makes clear that the 10 worst points are indeed close in $\mathcal{M}_{n=3}$.} In contrast to the sideband test set of Fig.~\ref{fig:SidebandDalitz}, or the double cone with the equator excised, there is no undersampled region where the break point is preferred. The points near the break point are mapped far away in the Dalitz plane, which is consistent with their large loss under the Euclidean metric. We emphasize once again that none of these features have anything to do with anomalies, because we have uniformly sampled phase space according to the Lorentz-invariant measure, so any point is as ``typical'' as any other. While it is true that the autoencoder task is only to minimize the Euclidean distance between the model and the data point-by-point in phase space, the spurious features in the predicted distribution point to correlations which will be imprinted on the loss distribution, which is the desired diagnostic for anomaly detection.

\begin{figure}[t!]
\begin{centering}
	\includegraphics[width = 0.5\textwidth]{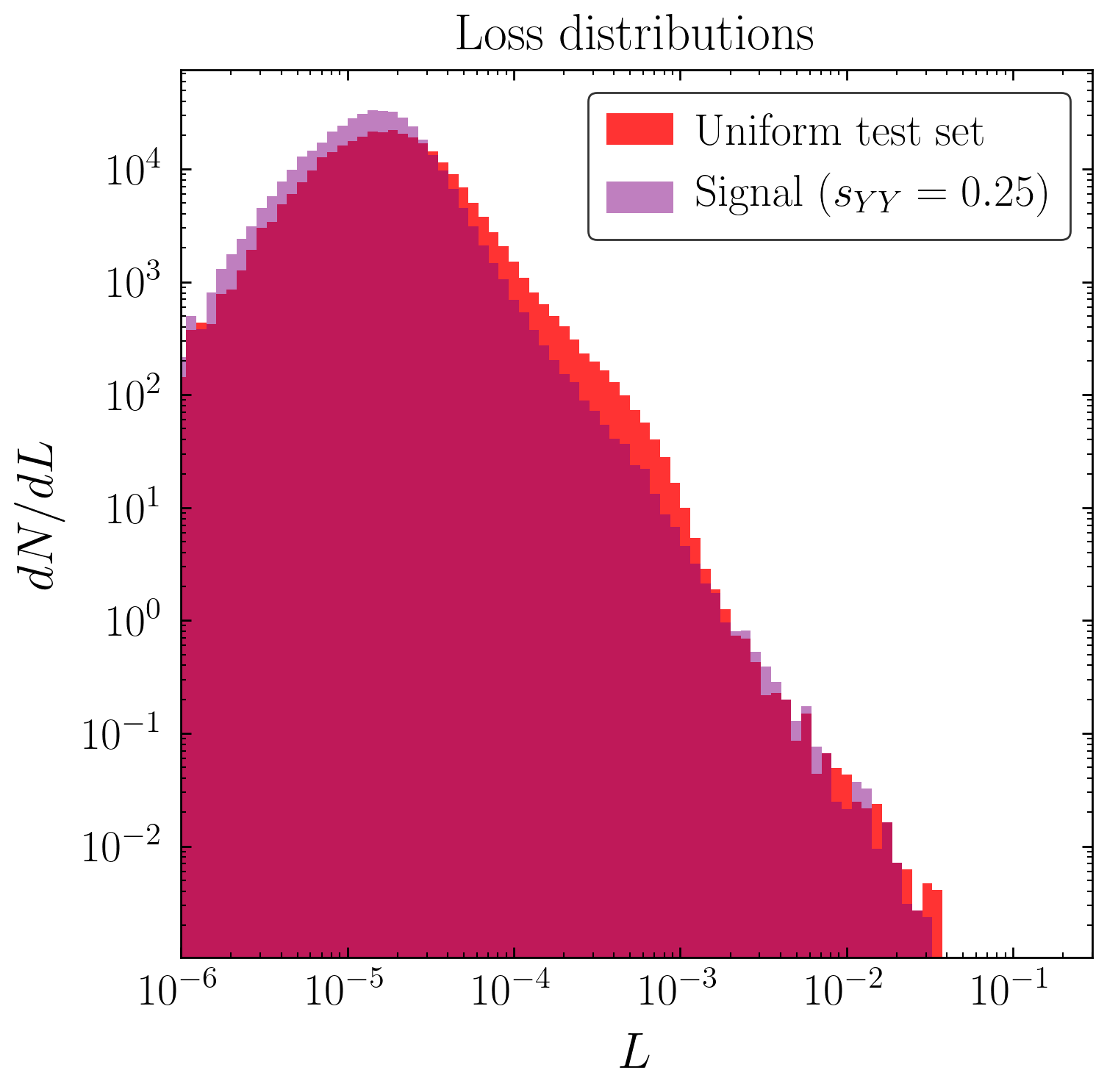}
	\caption{Normalized loss distributions for the uniform test set (Fig.~\ref{fig:PhaseSpace}) and signal test set with $s_{YY} = 0.25$ from Fig.~\ref{fig:SignalDalitz}. The loss tail for the signal events is smaller than for the background events, the opposite of the desired behavior.}
	\label{fig:SignalLossUniform}
\end{centering}
\end{figure}

From a topological perspective, the failure of the bump hunt described in Sec.~\ref{sec:bumphunt} is now straightforward to understand. With 3-body phase space having the local geometry of $\mathbb{R}^2 \times {\rm SO}(3)$, the submanifold with $s_{YY}$ equal to a certain value in the interior of the Dalitz triangle -- \emph{i.e.} the \emph{signal} -- is much like the equator of the sphere in Sec.~\ref{sec:sphere} or the equator of the double cone in Sec.~\ref{sec:cone}, and interpolating through this region is topologically trivial. The $S^5$ topology means that a break point \emph{must} exist, where the latent representation rips the data manifold and test set points near the break point are mapped far away. If the autoencoder is trained on a distribution with an undersampled region, the break point will typically be placed nearby (Fig.~\ref{fig:SidebandDalitz}), but the rest of the submanifold of fixed $s_{YY}$ will be reconstructed with low loss like any other generic point in phase space, as we saw in Fig.~\ref{fig:SignalDalitz}. Said another way, the autoencoder will detect a \emph{point} in phase space as anomalous (representing a particular orientation of the final-state 3-vectors), but this is only a set of measure zero on the \emph{submanifold} of desired anomalous events. The situation is even worse if the training set is uniform in phase space, because there is no guarantee that the break point will even lie on the signal submanifold; as seen in Fig.~\ref{fig:SignalLossUniform}, the loss tail from pure signal events is smaller than the loss tail from the background events because the network can achieve near-perfect reconstruction on the 4-dimensional signal submanifold with $d_l = 5$. It is clear that if trained on a smooth background distribution (no sidebands), the autoencoder cannot detect anomalies in a test set with both signal and background, since even the 100\% signal sample is indistinguishable from the background. Finally, the effective oversampling of the boundaries of the triangle with respect to the standard round metric on $S^5$ is analogous to the double-cone example of Sec.~\ref{sec:cone}: even though the average loss is minimized by placing the break point in the interior of the Dalitz triangle, the model will struggle to reconstruct the corners where the extrinsic curvature is large, introducing additional distortions in the loss distribution. This strongly suggests that caution is warranted when using an autoencoder as an anomaly detector for real physics events, where the nontrivial matrix element will induce a non-uniform distribution in the Dalitz plane.

\section{Changing the latent dimension}
\label{sec:ChangeD}

As most of our examples have focused on the case $d_l = d$, it is worth asking to what extent the topological obstructions to autoencoder reconstruction that we have identified are robust to changes in the latent dimension. Here we briefly summarize a series of examples illustrating that increasing $d_l$ beyond the intrinsic dimension of the data manifold is not guaranteed to cure topological issues; further details are provided in App.~\ref{app:MoreExamples} for the interested reader. Of course, if $d_l \geq \din$, the network will learn the identity map, which will not detect any anomalies, so we focus on the case $d < d_l < \din$.
\begin{itemize}
\item A circle may be embedded in $\mathbb{R}^3$ as a knot, with nontrivial extrinsic topology; even for $d_l = 2$, where perfect reconstruction is theoretically possible (as the circle does embed in the plane), the training process gets stuck at a local minimum with self-intersections in the latent space. However, the performance can be substantially improved by \emph{modifying the loss function} to force the network to learn a latent representation without self-intersections. 
\item The torus $T^2$ may be embedded in $\mathbb{R}^3$ as the standard ``donut'' embedding, or in $\mathbb{R}^4$ as a direct product of two circles, the Clifford torus. For $\din = 4$, the global loss minimum for $d_l = 3$ is the donut embedding. After training an ensemble of networks, the latent representations fall into two qualitative categories: infrequently, the network finds the global loss minimum, but more often, the latent map pinches one of the circles in two locations along the torus, yielding poor reconstruction. This demonstrates that even though an embedding may be topologically possible, a randomly-initialized autoencoder is not guaranteed to find it, raising concerns about the robustness of autoencoder performance on data manifolds with nontrivial topology.
\item The group manifold SO(3) is locally isomorphic to SU(2), but has the topology of the real projective space $\mathbb{RP}^3$, consisting of identifying antipodal points on ${\rm SU}(2) \cong S^3$. With $\din = 9$ (\textit{i.e.} flattening the $3 \times 3$ SO(3) matrix into a 9-component real vector), this global topological obstruction prevents good reconstruction even up to $d_l = 5$, which is the dimension in which SO(3) may be embedded in $\mathbb{R}^n$.
\end{itemize}

While some of these examples may represent more complicated topology than a generic data set in the wild (or even a practically-relevant data set in high-energy physics), they are important illustrations of the fact that simply increasing the size of the latent space does not guarantee that an autoencoder trained on data sampled from a topologically-nontrivial manifold can achieve low uniform reconstruction error.

\section{Conclusions}
\label{sec:conclusion}

The importance of understanding the topology of input data has been recognized since the 1960s and was a pressing issue for Rosenblatt, the inventor of the perceptron (see \cite{rosenblatt1961principles}). At the time, the question was not about a latent representation, but about the extrinsic topology of the input, \textit{e.g.}\ whether a circle was inside or outside a square in an input image or whether two components of an image are connected or not. Early neural network models struggled to identify these very global features of input images, as was famously elucidated by Minsky and Papert in their famous critique of the perceptron model \cite{miskypapertperceptron}. 

In this paper we have attempted to further understand this intertwining of data topology and neural network performance via an extensive study of a rich variety of low-dimensional input data sets exhibiting both nontrivial intrinsic and extrinsic topology. In particular, we have identified several situations where the \emph{global} topological features of the data set pose an obstruction to faithfully compressing the data even when the latent space dimension is equal to the intrinsic dimension of the data, which is a \emph{local} feature. As an application, we have shown that in the canonical example of anomaly-finding in high-energy physics, a ``bump hunt,'' an neural network autoencoder trained on data drawn from $n$-particle phase space (with $n$ fixed) inevitably results in order-1 reconstruction error for generic points in the training set, and moreover fails to flag as anomalies events with invariant mass values that are entirely absent from the training set. 

Since the issues of large reconstruction error are entirely due to the topologically-impossible task of trying to cover a whole manifold with a single chart, they could in principle be ameliorated by training multiple networks -- the latent representations of which would represent independent charts -- and using the regions of faithful reconstruction in pairwise overlaps of charts to construct transition functions. Indeed, an ensemble of networks has already been used in the context of weakly-supervised learning for collider physics to mitigate the trials factor or ``look-elsewhere'' effect \cite{Aad:2020cws}. If the large-loss points are uniformly distributed across the data manifold, a simple (but computationally-expensive) way to do this would be to independently train a large number of randomly-initialized networks and take the median of the outputs; specifically, for each test set point, sort the list of autoencoder losses from each network and define the output of the ensemble to be the autoencoder output corresponding to the median loss in this list. A more sophisticated solution would correlate the network parameters to construct transition functions directly, along the lines of \cite{korman2018autoencoding}. Such a strategy of multiple network realizations would also be helpful in cases like the 2-sphere with the excised equator, where the trained ensemble doesn't concentrate on a single minimum. Alternatively, one could let a single network find the transition functions itself using an architecture consisting of parallel sub-networks for each chart, perhaps with suitable encouragement by modifying the loss function.\footnote{We thank Ben Lehmann for this observation.}  However, for submanifold-type anomalies -- such as the bump hunt in phase space -- each of the networks or sub-networks will be able to smoothly interpolate through the anomalous submanifold, and no anomalies would be detected.

As our examples with the undersampled circle and phase space have shown, some knowledge of the data topology can be very useful in interpreting the output of an autoencoder. Specifically, knowing that there must be points with large loss could motivate a network architecture which correlates that loss with the data distribution. It would be particularly interesting to investigate how the topology of phase space is imprinted on jet substructure observables where the number of final-state particles is not fixed, and additional soft and collinear radiation ``dresses'' the parton-level phase space with events near the boundaries of the higher-dimensional simplex defining the analogue of the Dalitz plot for hadron-level phase space. Conversely, the autoencoder itself can be a useful diagnostic of the data topology, by examining whether the points with large loss are correlated in distance. 

Far from being some esoteric feature, we might expect that some nontrivial topology is generic for data sets containing features with any degree of rotational symmetry, which applies to a number of examples outside of physics such as 3-dimensional objects viewed from different perspectives. Indeed, Refs.~\cite{DBLP:journals/corr/abs-1801-10130,kondor2018clebschgordan} have used the nontrivial topology of the sphere (in particular, distortions resulting from planar projections) to motivate SO(3)-equivariant networks to perform machine learning tasks on datasets which live on spheres, which may have applications for learning observables which are functions on phase space. There has been some very interesting recent work on estimating the intrinsic dimensionality of generic data sets \cite{camastra2016intrinsic,sharma2020neural}, but these techniques rely on various proxies for the data dimension after the data has already passed through layers of the neural network, which as we have argued is a good probe of local dimension but necessarily misses the global topological features. Depending on how that local dimension is being used in the downstream machine learning task, it might be necessary to adapt such methods further to account for cases of nontrivial topology. Given the considerable recent work which has focused on incorporating nontrivial priors about the data set, including symmetry properties, into the network architecture, we hope this work has motivated including data topology into that set of priors.

\acknowledgments{We thank Marat Freytsis for collaboration in the early stages of this work, and Tim Cohen, Jeff Dror, Christina Gao, Jim Halverson, Andrew Larkoski, Ben Lehmann, Tom Melia, Mark Neubauer, Jesse Thaler, Sho Yaida, and Dave Zhao for helpful conversations. This work is supported by the National Science Foundation under Cooperative Agreement PHY-2019786 (The NSF AI Institute for Artificial Intelligence and Fundamental Interactions, \url{http://iaifi.org/}).}

\begin{appendix}

\section{Hyperparameters}
\label{app:hyperparams}

Our autoencoder neural networks were fully-connected nets constructed with \texttt{Pytorch}, using default initializations and trained with stochastic gradient descent (SGD) for 20,000 epochs. In the examples described in the main text, we used tanh activations for all layers except the output of the encoder and the output of the decoder; other activation functions and training algorithms for the $S^1$ autoencoder are discussed in App.~\ref{app:MoreCircle} below. The hyperparameters for each of the examples are shown in Tab.~\ref{tab:hyperparameters}.

\begin{table}[htp]
\caption{Autoencoder hyperparameters}
\begin{center}
\begin{tabular}{|c|c|c|c|c|}
\hline
\textbf{Example} & \textbf{Section} & \textbf{Sample Size} & \textbf{Learning Rate} & \textbf{Batch Size}  \\
\hline
\hline
Unit $S^1$ & \ref{sec:circle}& $N_{\rm train} = N_{\rm test} = 1000$ & $10^{-2}$ & 64 \\
\hline
Sparse $S^1$ & \ref{sec:circle}, App.~\ref{app:sparse} & $N_{\rm train} = 12, 20$ & $10^{-2}$ & $N_{\rm train}$ \\
\hline
$S^2$ & \ref{sec:sphere} & $N_{\rm train} = N_{\rm test} = 3000$ & $10^{-2}$ & 64 \\
\hline
Paraboloid & \ref{sec:sphere} & $N_{\rm train} = N_{\rm test} = 3000$ & $10^{-2}$ & 64 \\
\hline
Double cone & \ref{sec:cone} & $N_{\rm train} = N_{\rm test} = 3000$ & $10^{-2}$ & 64 \\
\hline
$S^4$ & \ref{sec:dimn}  & $N_{\rm train} = N_{\rm test} = 10^{4}$ & $10^{-2}$ & 64 \\
\hline
$S^5$ & \ref{sec:dimn} & $N_{\rm train} = N_{\rm test} = 10^{5}$ & $0.1$ & 512\\
\hline
SU(2)& \ref{sec:dimn}, App.~\ref{sec:su2so3}  & $N_{\rm train} = N_{\rm test} = 10^{4}$ & $10^{-2}$ & 64 \\
\hline
Phase space & \ref{sec:bumphunt}, \ref{sec:phasespace} & $N_{\rm train} = N_{\rm test} = 10^{5}$ & $0.1$ & 512\\
\hline
Trefoil & App.~\ref{app:knot} & $N_{\rm train} = N_{\rm test} = 1000$ &  $10^{-2}$ & 64 \\
\hline
Torus & App.~\ref{sec:torus} & $N_{\rm train} = N_{\rm test} = 3000$ & $10^{-2}$ & 64 \\
\hline
SO(3) & App.~\ref{sec:su2so3}  & $N_{\rm train} = N_{\rm test} = 10^{4}$ & $10^{-2}$ & 64 \\
\hline
\end{tabular}
\end{center}
\label{tab:hyperparameters}
\end{table}%

We increased the number of sample points as the dimension increased, and for the highest-dimensional examples we also increased the batch size by about an order of magnitude and as such increased the learning rate accordingly \cite{l.2018dont}.

\section{Further investigation of the $S^1$ autoencoder}
\label{app:MoreCircle}

In Sec.~\ref{sec:circle}, we argued that the appearance of a break point in an autoencoder with latent dimension 1 is an unavoidable feature of a data set with the topology of $S^1$. In principle, one could imagine that with sufficient training, the break point would be placed in between finitely-spaced training points, such that the network could achieve perfect reconstruction on the training set. In practice, we find that this is not true, and the appearance of a finite-sized break region encompassing multiple training points seems generic and robust with respect to changes in the network hyperparamters and architecture. In this Appendix we will justify these statemements and perform a simple analytical analysis of the network dynamics to relate the topological requirement of a break region to the network parameters which determine it. While nothing in this Appendix has anything to do with physics per se, we find the richness of this simple example worthy of serious investigation in future study as it touches on notions of spontaneous symmetry breaking, topology of finite data sets, and the neural network loss landscape.

\subsection{Changing hyperparameters}
\label{app:S1checks}

\begin{figure}[t!]
	\includegraphics[width = 0.3\textwidth]{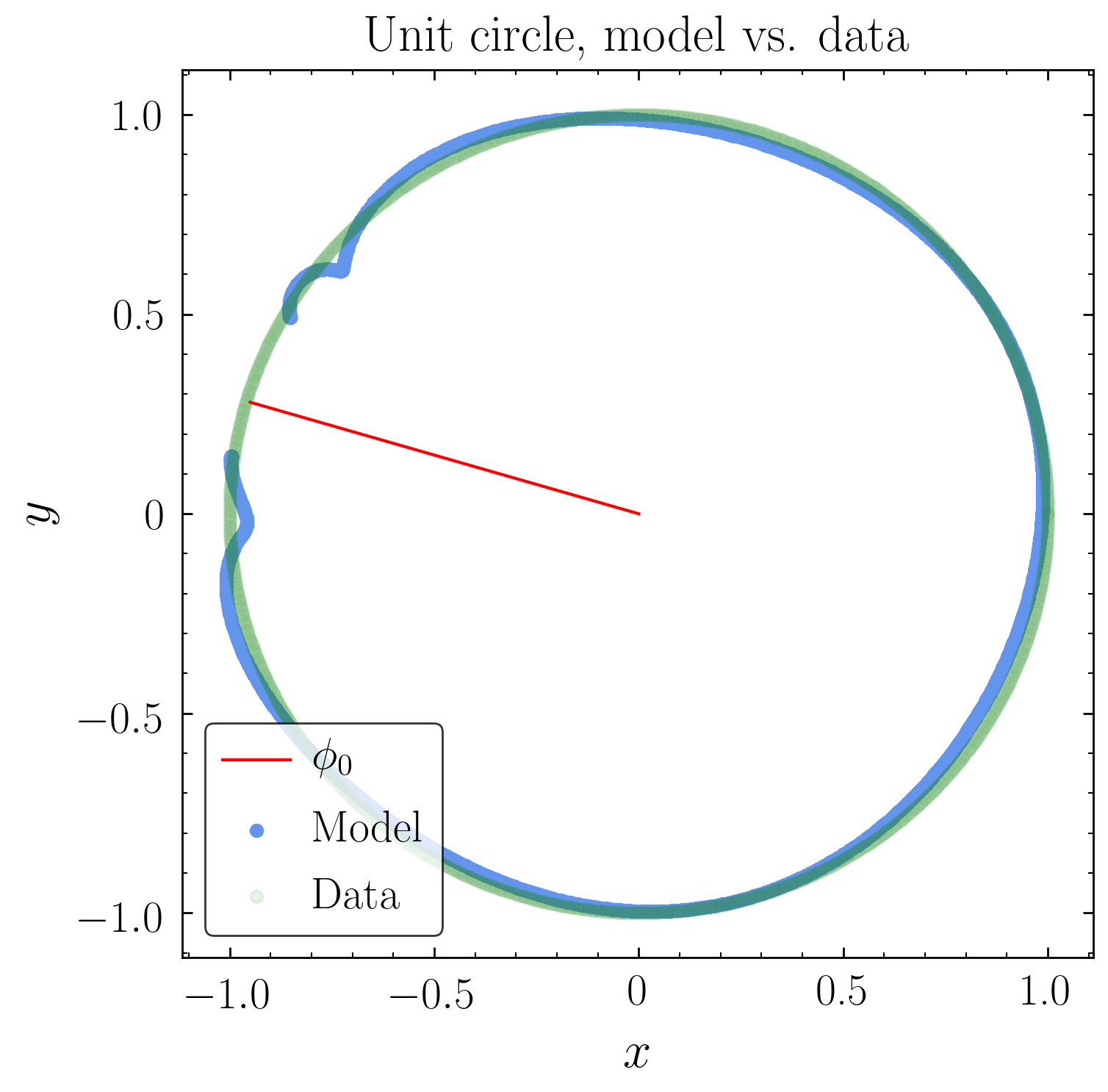} \qquad
	\includegraphics[width = 0.3\textwidth]{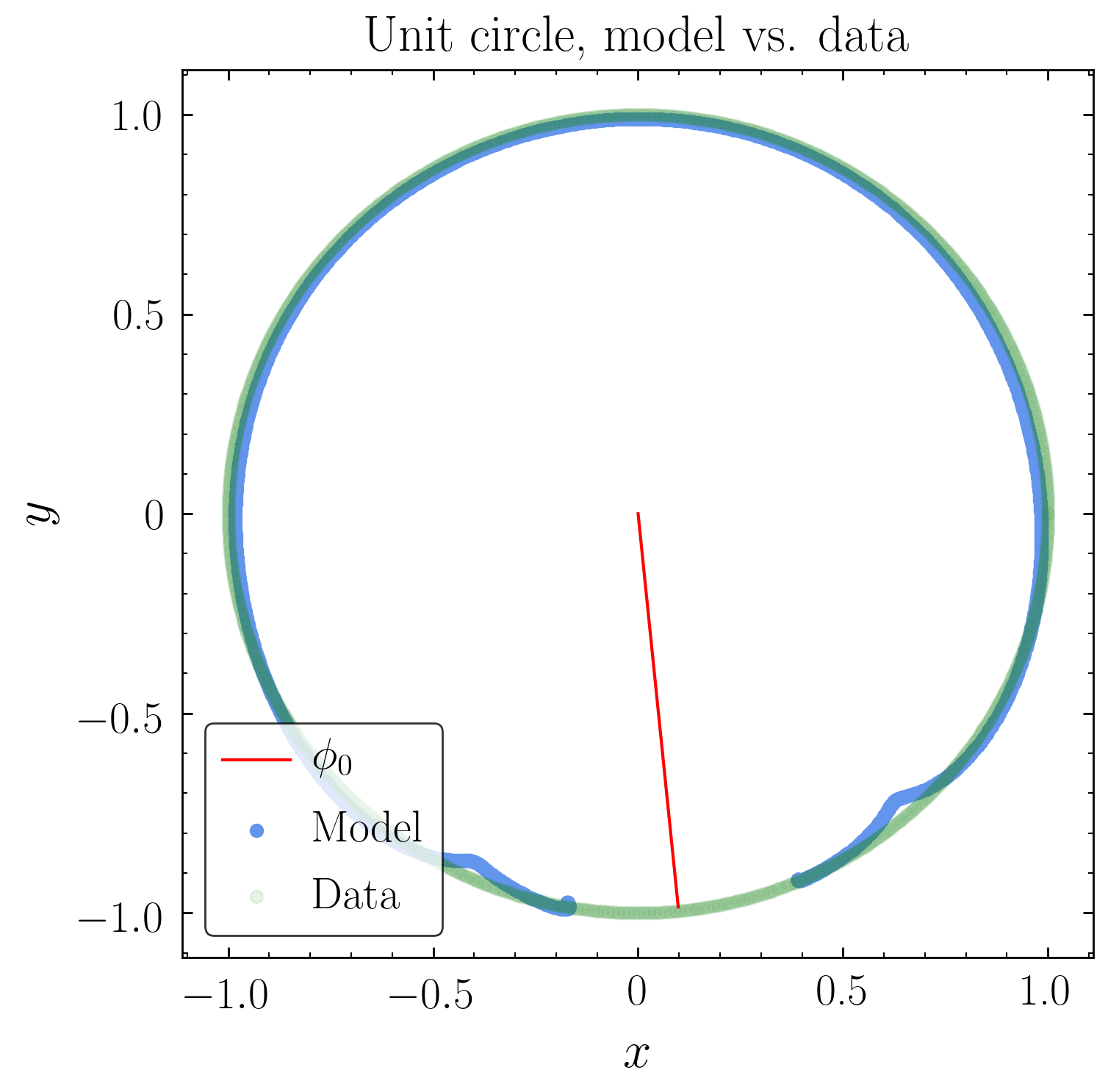} \qquad
	\includegraphics[width = 0.3\textwidth]{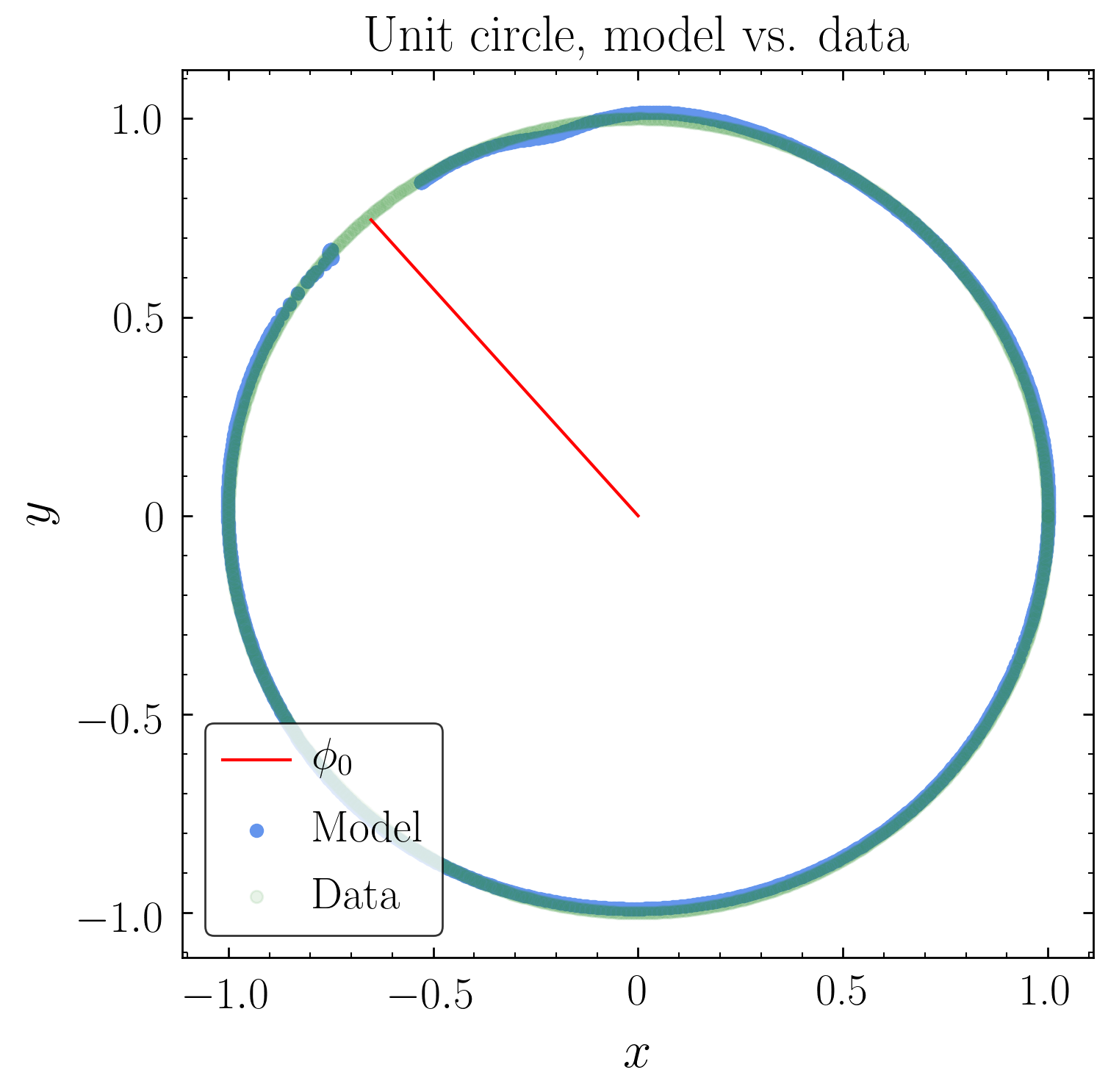}
	\caption{Output of the $S^1$ autoencoder with different network architectures but all other hyperparameters from Tab.~\ref{tab:hyperparameters} held fixed: 5-layer networks with $d_w = 32$ \textbf{(left)} and $d_w = 128$ \textbf{(center)}, and a 7-layer network with $d_w = 64$ \textbf{(right)}. The finite-sized break region persists in all cases, though it is reduced somewhat in the 7-layer network.}
	\label{fig:S1arch}
\end{figure}

\begin{figure}[t!]
	\includegraphics[width = 0.3\textwidth]{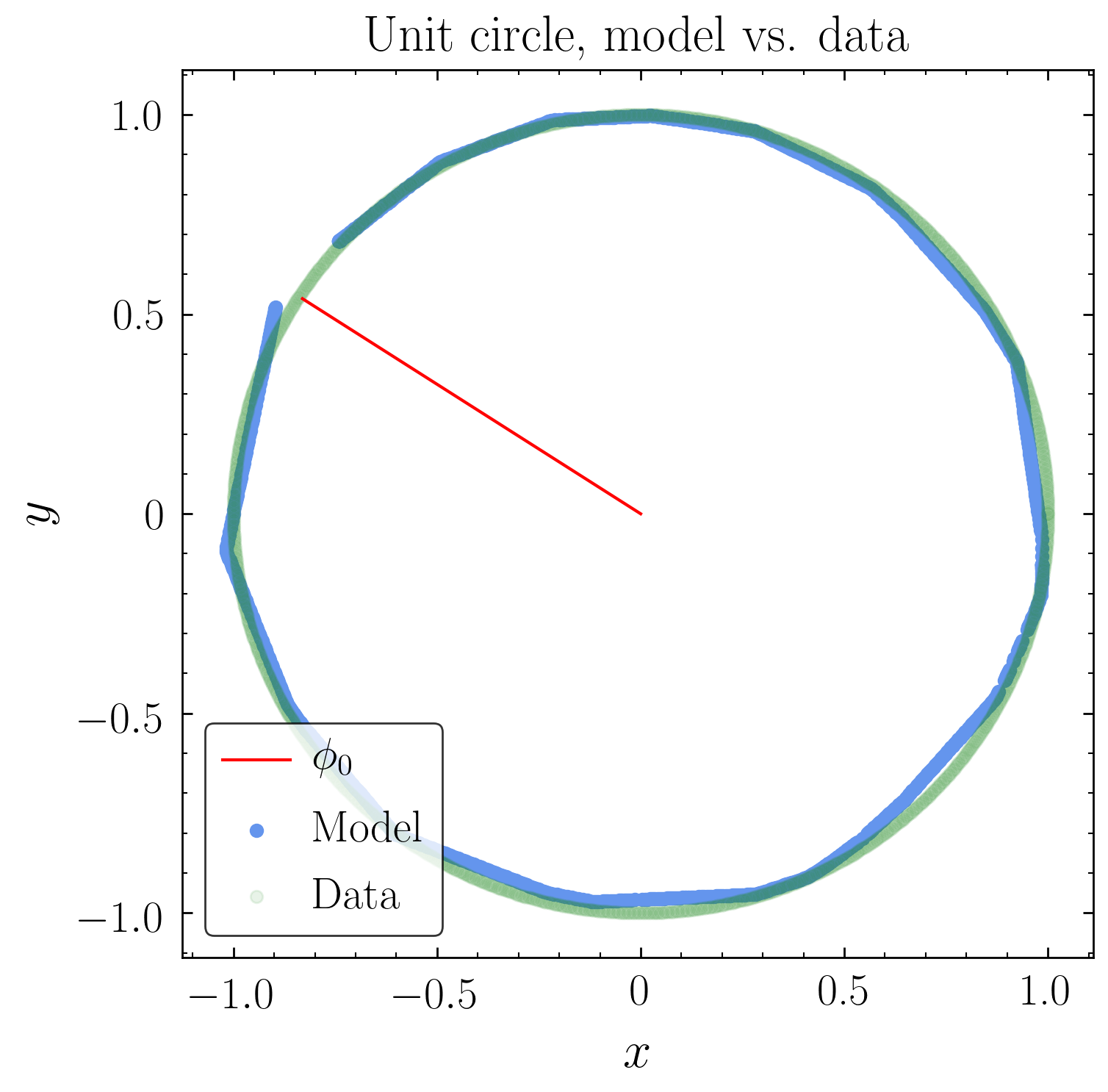} \qquad
	\includegraphics[width = 0.3\textwidth]{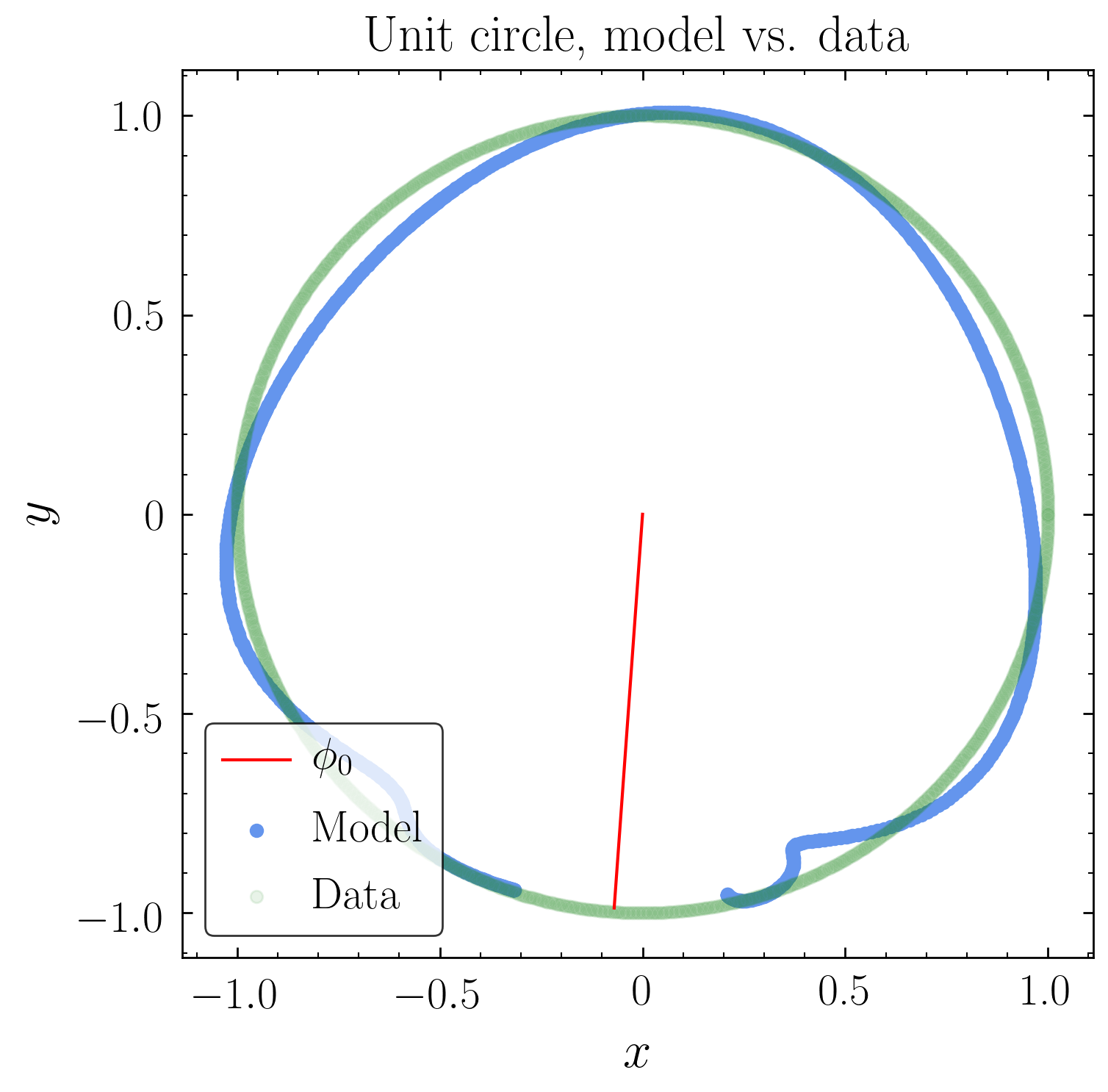} \qquad
	\includegraphics[width = 0.3\textwidth]{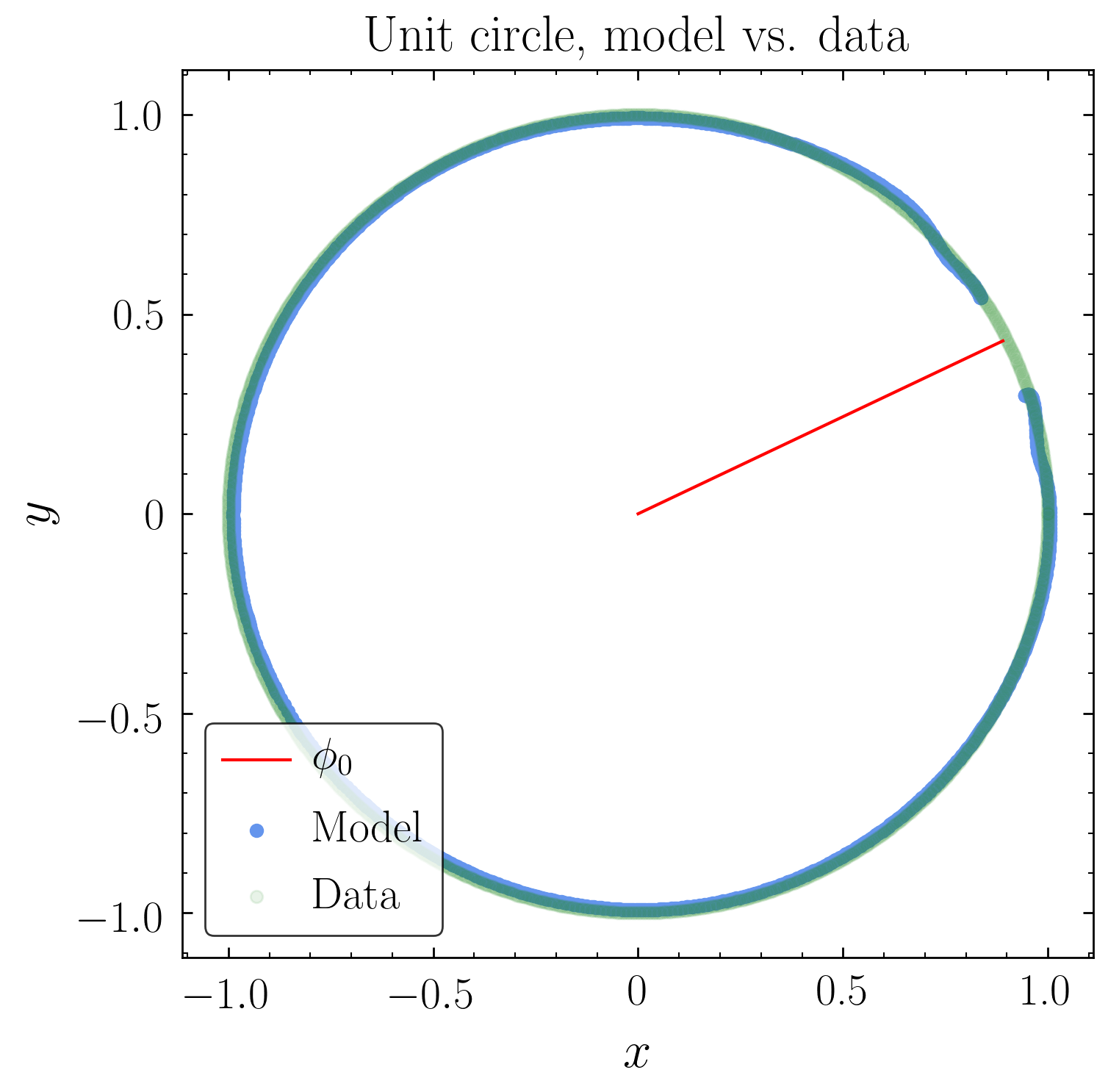} 
	\caption{Output of the $S^1$ autoencoder with different activation functions: ReLU \textbf{(left)}, and $\frac{1}{\beta} \tanh(\beta x)$ with $\beta =0.5$ \textbf{(center)} and $\beta = 5$ \textbf{(right)}.}
	\label{fig:S1act}
\end{figure}

\begin{figure}[t!]
	\includegraphics[width = 0.45\textwidth]{figures/S12Doutput.png} \qquad
	\includegraphics[width = 0.475\textwidth]{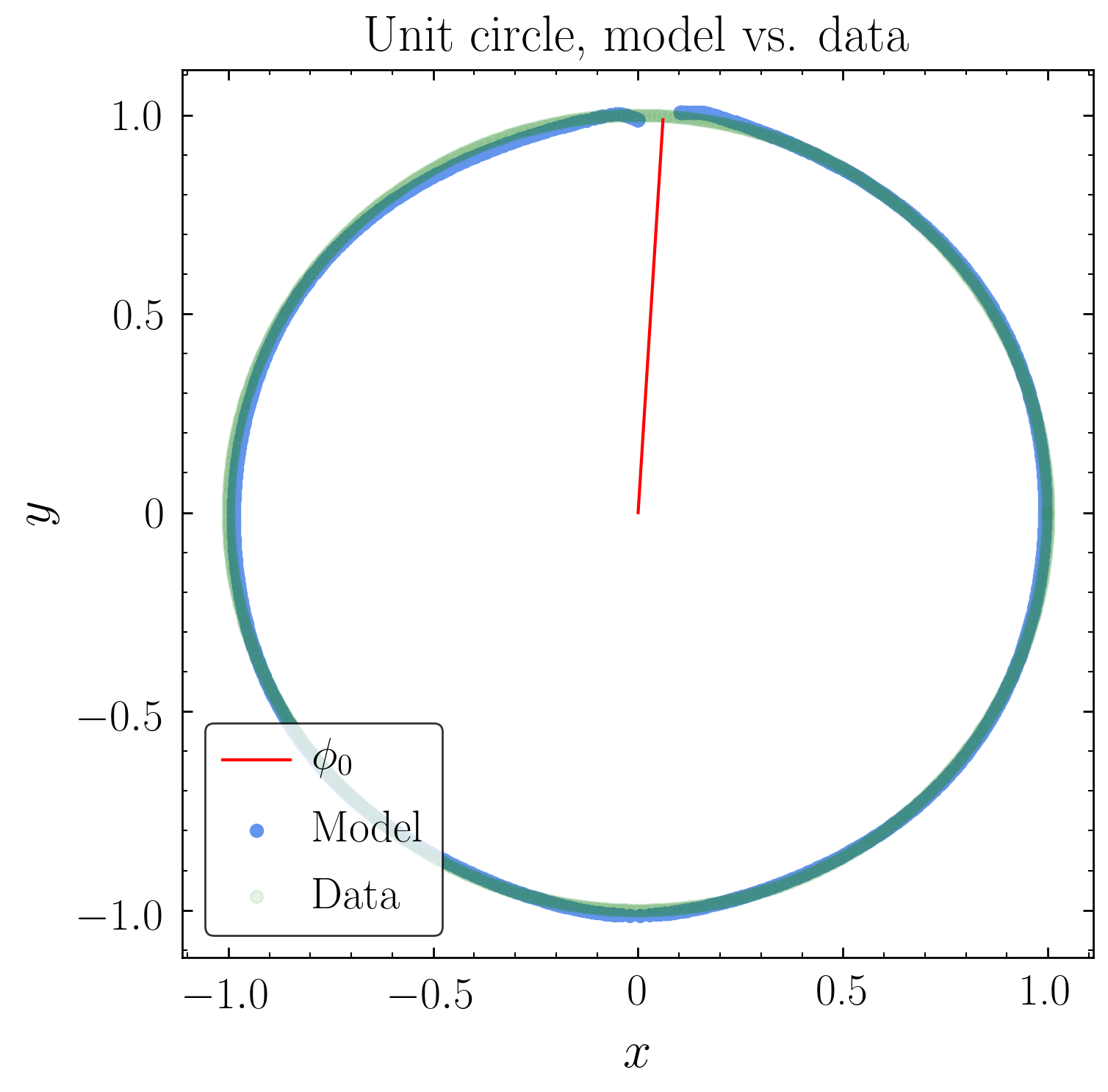} 
	\caption{Output of the $S^1$ autoencoder with identical hyperparameters but different training algorithms: SGD \textbf{(left)} (same as Fig.~\ref{fig:S1Uniform}), and Adam \cite{kingma2019method} \textbf{(right)}.}
	\label{fig:S1traindiff}
\end{figure}

\begin{figure}[t!]
\begin{center}
	\includegraphics[width = 0.5\textwidth]{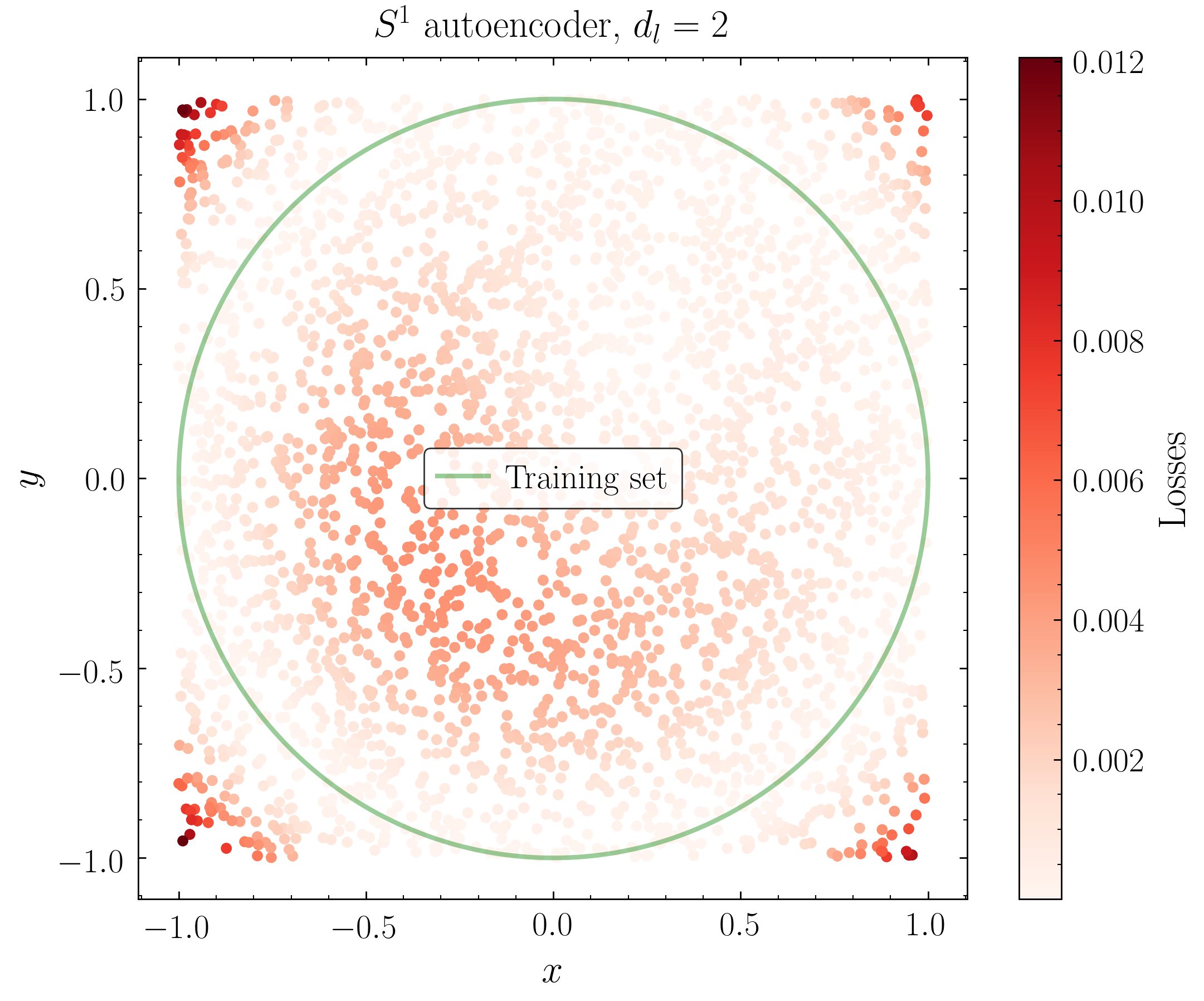}
	\caption{The $S^1$ autoencoder with $d = 2$ learns the identity map on all of $\mathbb{R}^2$. Losses are concentrated outside the region $x^2 + y^2 = 1$ which defined the training set, because extrapolation is required in that region.}
	\label{fig:S1D2}
\end{center}
\end{figure}

We first note that the persistence of the break region is insensitive to changes in the width or depth of the network. Fig~\ref{fig:S1arch} shows the $S^1$ autoencoder with three different architectures: a 5-layer network with $d_w = 32$ (left) and $d_w = 128$ (center), and a 7-layer network with $d_w = 64$ (right). In addition, we experimented with changing the activation function: Fig.~\ref{fig:S1act} shows results for our default architecture with a ReLU activation function (left), as well as modified tanh activation functions $\frac{1}{\beta} \tanh(\beta x)$ with varying ``temperature'' $\beta$ (center and right). The ReLU and $\beta > 1$ activation functions seem to result in a somewhat smaller gap, but one which is still easily visible and encompasses multiple data points from the training set of 1000 equally-spaced points. We also tried a different training algorithm: Fig.~\ref{fig:S1traindiff} shows the results for the Adam \cite{kingma2019method} optimizer compared to SGD. As expected, Adam converges to the gradient descent minimum faster than SGD. This results in a smaller gap for the same amount of training, though once again the finite gap remains even after 20,000 epochs. Finally, we verified that the topological obstruction was indeed arising from a latent dimension $d_l = 1$ rather than any issues with inadequate network capacity. After training the default $S^1$ autoencoder with $d_l = 2$ on 1000 equally-spaced points on the unit circle, Fig.~\ref{fig:S1D2} shows the output of the autoencoder on a test set of 3000 points uniform on the whole square $-1 \leq x, y \leq 1$. The loss is smallest on the training set (green), as expected, but the fact that the loss is of the same order everywhere else in the square except at the corners strongly suggests that the network is learning the trivial map on $\mathbb{R}^2$ for $x^2 + y^2 \leq 1$.

\subsection{Sparse circle}
\label{app:sparse}

\begin{figure}[t!]
	\includegraphics[width = 0.3\textwidth]{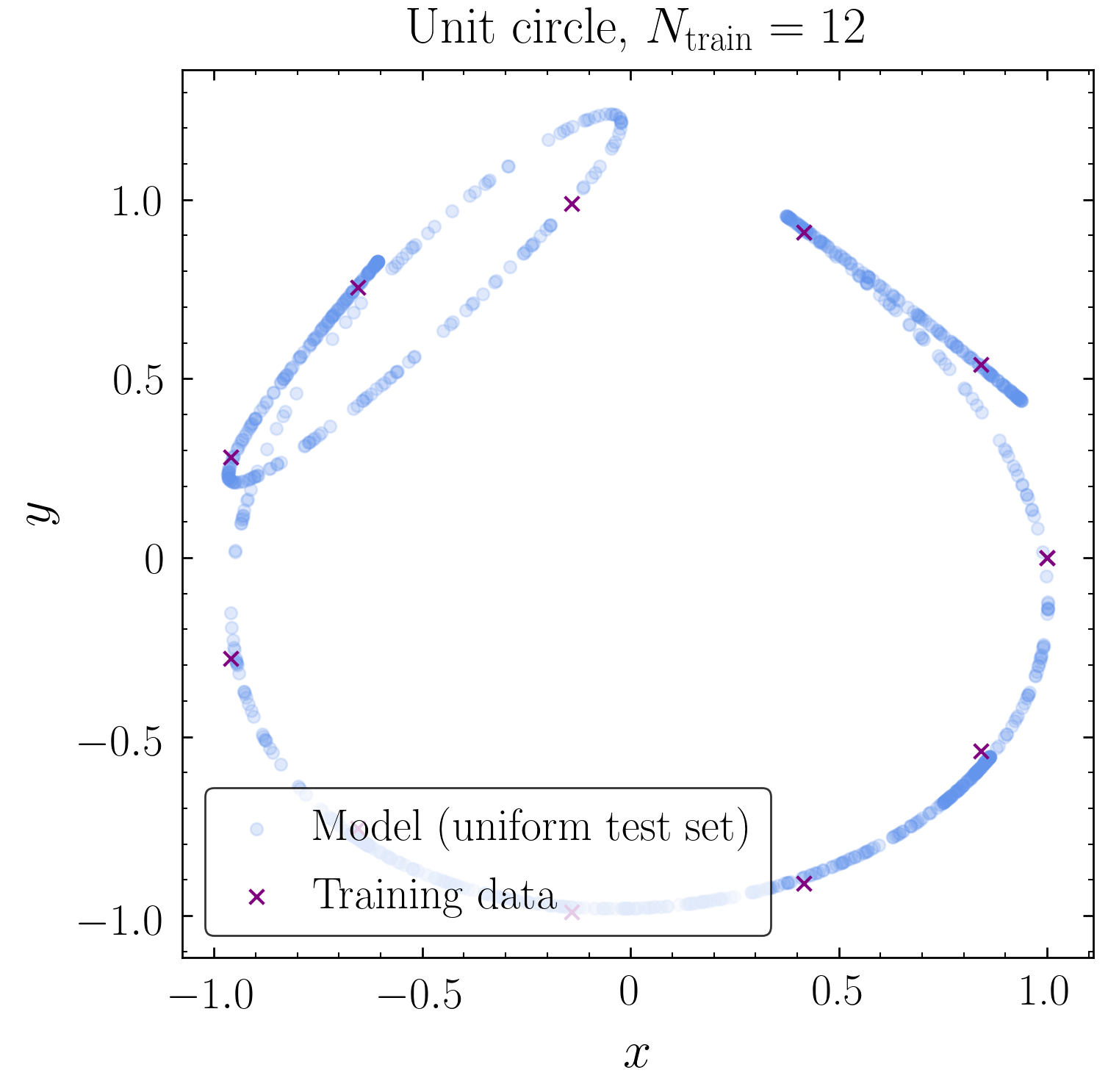} \qquad
	\includegraphics[width = 0.3\textwidth]{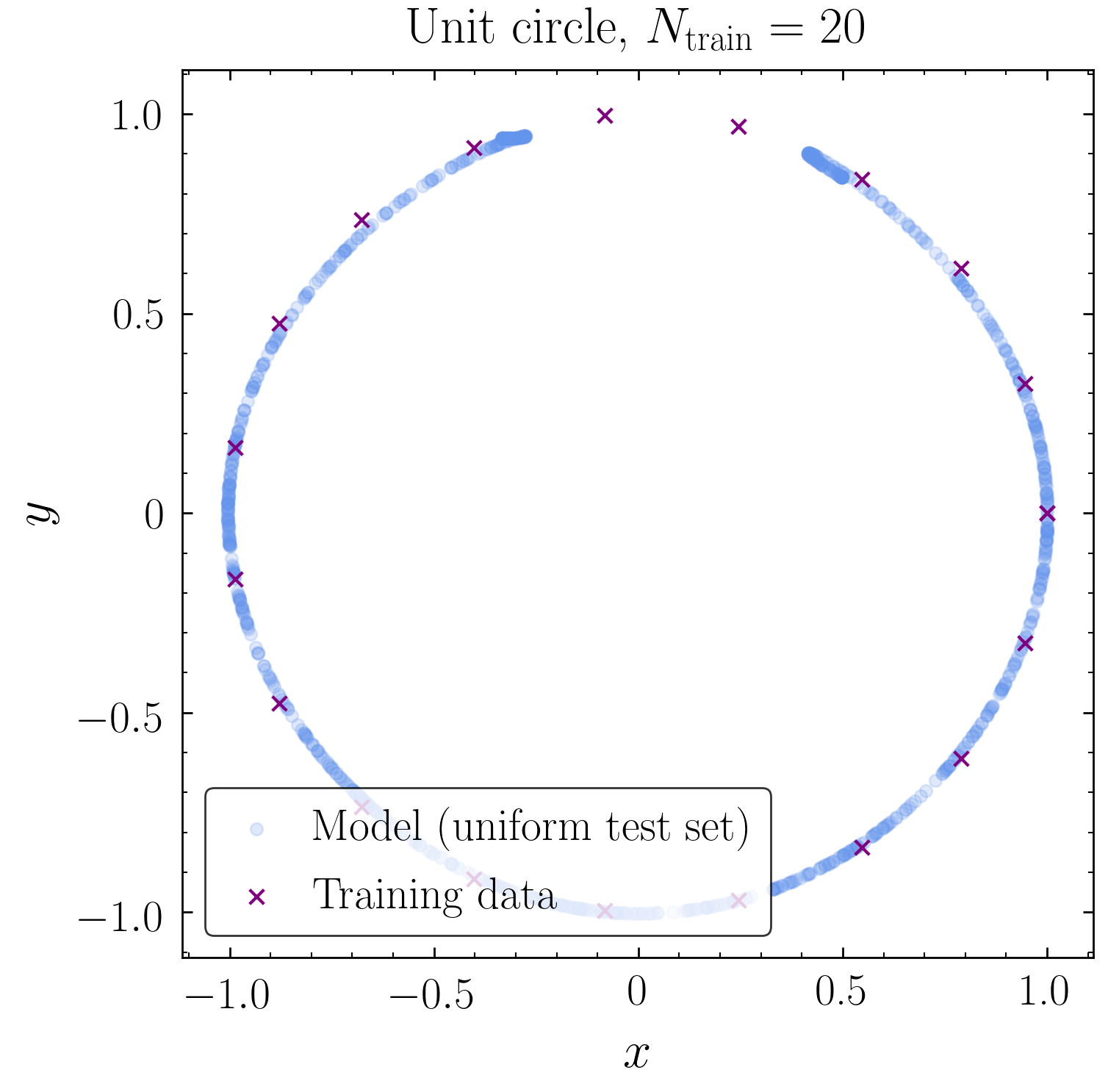} \qquad
	\includegraphics[width = 0.3\textwidth]{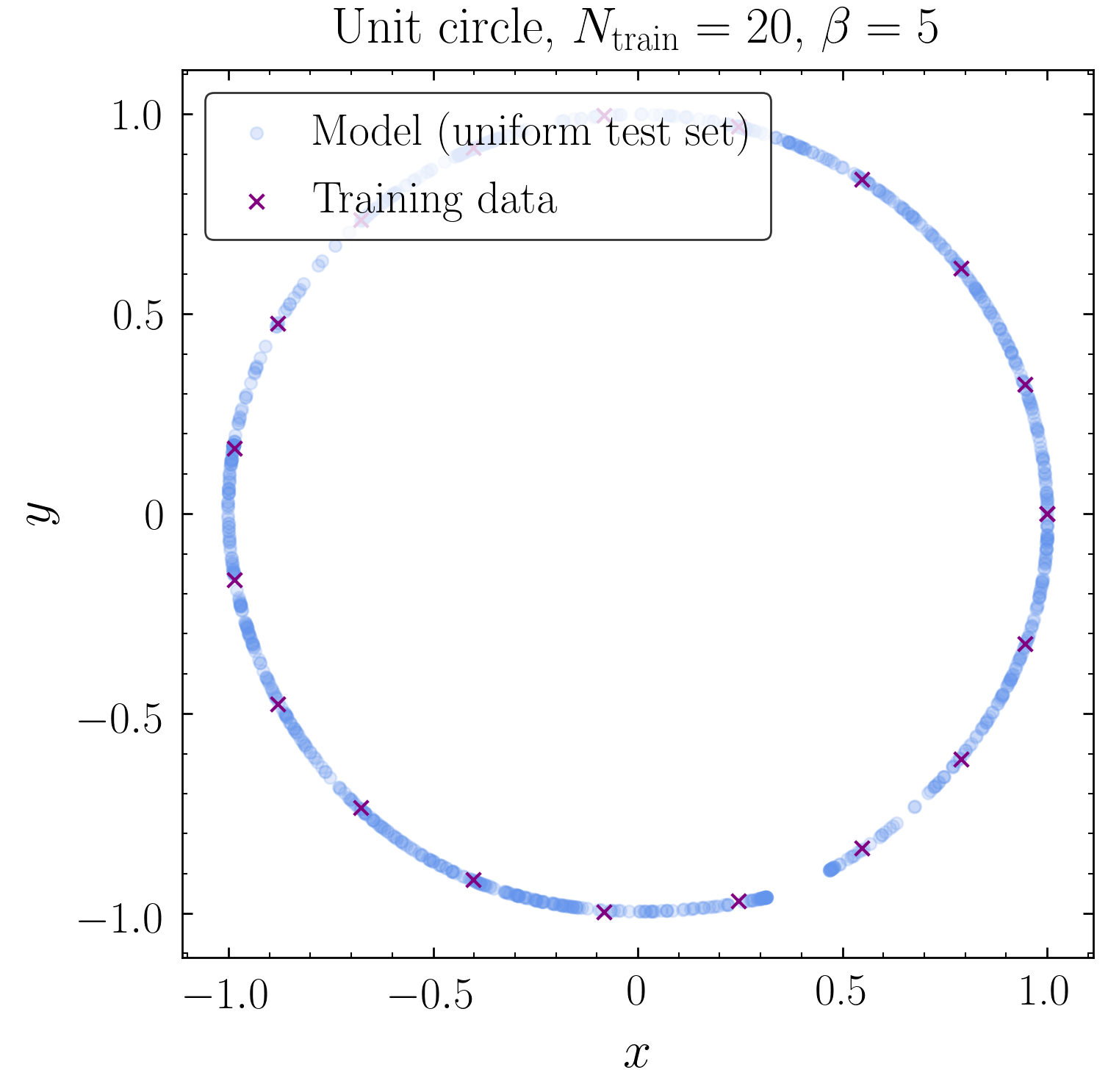} 
	\caption{Output of the $S^1$ autoencoder with sparse training sets of size 12 \textbf{(left)} and 20 \textbf{(center and right)} after 100,000 epochs of GD training. The right plot shows the results for a modified activation function $\frac{1}{\beta} \tanh(\beta x)$ with $\beta =5$.}
	\label{fig:S1Sparse}
\end{figure}

The $S^1$ autoencoder exhibits interesting behavior when the size of the training set is very small, as shown in Fig.~\ref{fig:S1Sparse}. For the same hyperparameters as given in Tab.~\ref{tab:hyperparameters} but with 100,000 epochs of training, a training size of $N_{\rm train} = 12$ allows the network to memorize the training set but at the cost of a rather poor reconstruction on a larger test set of size $N_{\rm test} = 1000$ uniformly sampled from the circle. This is obviously an avatar of overfitting. Next, increasing to $N_{\rm train} = 20$ gives the familiar behavior with a break region as described above, with the network unable to memorize the full training set because the spacing between training points is smaller than the break region. As anticipated above, by changing the activation to $\frac{1}{\beta} \tanh(\beta x)$ with $\beta = 5$, the size of the break region can be reduced, allowing the network to perfectly reconstruct the training set while still maintaining accurate reconstruction of the larger test set by placing the break region between training points. However, for $N_{\rm train} \gtrsim 100$, even the $\beta = 5$ activation function cannot memorize the training set after 100,000 epochs.

\subsection{Dynamics of training in the $S^1$ encoder}
\label{sec:SolveCircle}

To gain some analytic understanding of the behavior of the circle autoencoder with $d_l = 1$, we examine the structure of the autoencoder network explicitly. The encoder map $\fenc(\x)$ with $\x = (x,y)$ is a map $\mathbb{R}^2 \to \mathbb{R}^1$, while the decoder map $\fdec(q)$ is a map $\mathbb{R}^1 \to \mathbb{R}^2$. Restricting the input data to the unit circle, the encoder can be thought of as a map $\fenc(\phi)$ from $S^1$ to $\mathbb{R}^1$, with $x = \cos \phi$ and $y = \sin \phi$. The model map is $f(\x) = \fdec(\fenc(\x))$, and the loss function is the mean squared error,
\be
L = \frac{1}{N}\sum_{j = 1}^N ||f(\x_j) - \x_j||^2 ,
\ee
where $\x_j$ and $f(\x_j)$ are points in $\mathbb{R}^2$ and the norm is the usual Euclidean norm.

Explicitly, the encoder map with a single hidden layer of width $d_w$ is
\be
\label{eq:fenc}
\fenc(\phi; \theta_\alpha) = b_2 + \sum_{i = 1}^{d_w} W_i^{(2)}\sigma(W_{ix}\cos \phi + W_{iy} \sin \phi + b_i) ,
\ee
where $\sigma$ is the activation function on the hidden layer output, and $\theta_\alpha = \{W_{ix}, W_{iy}, W_{i}^{(2)}, b_i, b_2 \}$ are the encoder parameters: $W_{ix}$ and $W_{iy}$ are the weights going to the hidden layer, $W_{i}^{(2)}$ are the weights going to the output of the encoder, and $b_i$ and $b_2$ are the biases. It will be convenient to define
\be
z_i(\phi) = W_{ix}\cos \phi + W_{iy} \sin \phi + b_i
\ee
as the pre-activation at neuron $i$ of the first hidden layer. A necessary condition for the network to be at a loss minimum after training is that the gradient of the loss with respect to the encoder parameters $\theta_\alpha$ vanishes:
\be
\nabla_\alpha L = \frac{2}{N} \sum_{j = 1}^N \left [(f(\x_j) - \x_j) \cdot \left ( \frac{ df^{\rm dec}}{d \fenc}  \right )\right]\nabla_\alpha \fenc(\phi_j) = 0 .
\ee
The point of this expression is to note that at the loss minimum, one of three things must be true (absent accidental orthogonalities), data point by data point: either the reconstruction is perfect ($f(\x_j) = \x_j$), or the derivative of the decoder vanishes, or the gradient of the encoder vanishes.

Because of the topological issues previously noted in Sec.~\ref{sec:circle}, it is impossible for the network to satisfy the first condition near the break point. The second condition, the vanishing of the decoder derivative, would imply that the decoder is independent of the latent representation to first order, which would mean the network is not actually learning anything from the latent representation and nearby points in the latent space get mapped to the same point in the model.\footnote{Of course, a good decoder will map nearby latent points to \emph{nearby} points in the model, but this implies a nonzero (if small) derivative.} We therefore expect that near the break point $\phi_0$, the third condition holds, $ \nabla_\alpha \fenc(\phi_0) = 0$. To the extent that this is true, the appearance and position of the break point is entirely driven by the encoder, which greatly simplifies the analysis since there is only a single hidden layer. We will use the explicit expression (\ref{eq:fenc}) to relate $\partial \fenc/\partial \phi$ to $ \nabla_\alpha \fenc(\phi)$.

Treating the encoder as a function of the input variable $\phi$, we have 
\be
\label{eq:dfdtheta}
\frac{\partial \fenc}{\partial \phi} = \sum_{i = 1}^{d_w} W_i^{(2)}(-W_{ix} \sin \phi + W_{iy} \cos \phi)\sigma'(z_i) ,
\ee
where $\sigma'$ is the first derivative of the activation function. Similarly, treating the encoder as a function of $\theta_\alpha$, the derivatives of $\fenc$ with respect to the network parameters are
\begin{align}
\frac{\partial \fenc}{\partial W_i^{(2)}} & = \sigma(z_i), \\
\frac{\partial \fenc}{\partial W_{ix}} & = W_i^{(2)} \cos \phi \,\sigma'(z_i) , \\
\frac{\partial \fenc}{\partial W_{iy}} & = W_i^{(2)} \sin \phi \, \sigma'(z_i) , \\
\frac{\partial \fenc}{\partial b_i} & = W_i^{(2)}\sigma'(z_i) , \\
\frac{\partial \fenc}{\partial b_2} & = 1.
\end{align}
Note that because $\partial \fenc/\partial b_2$ never vanishes, there are no true critical points for $\fenc$.
However, since there are $4d_w$ additional parameters, if $d_w \gg 1$ then the gradient will be dominated by the other parameters, so we can attempt to minimize those derivatives to find a quasi-minimum.

To make further progress, let's suppose the activation function $\sigma(z)$ vanishes only at $z = 0$ and furthermore that $\sigma'(0) \neq 0$, which is true for example for the tanh, and popular smooth approximation of the ReLU (though not ReLU itself) such as the GELU \cite{hendrycks2016gaussian} and SWISH \cite{ramachandran2017searching} activations. The derivatives with respect to the second-layer weights, $\partial \fenc/\partial W_i^{(2)}$, can only vanish if $z_i = 0$, but since $\sigma'(0) \neq 0$ by assumption, the remaining derivatives with respect to the first-layer weights and biases can only vanish if $W_i^{(2)} = 0$. Therefore, the global quasi-minimum is for all of the second-layer weights to vanish, which a trivial model with poor reconstruction error, since from Eq.~(\ref{eq:dfdtheta}), $\fenc(\phi)$ is then independent of $\phi$. Empirically, what the network tries to do instead is minimize all but a few of the $W_i^{(2)}$; the ones that remain nonzero stop evolving when their corresponding pre-activations vanish. Indeed, let $i^* = {\rm argmax} |W_i^{(2)}|$, and $\phi_0$ be a solution to $z_{i^*}(\phi) = 0$:
\be
W_{{i^*}x} \cos \phi_0 + W_{{i^*} y} \sin \phi_0 + b_{i^*} = 0
\label{eq:preactvanish}
\ee
Then by Eq.~(\ref{eq:dfdtheta}), $|\partial \fenc/\partial \phi|$ is large at $\phi = \phi_0$ since it is dominated by $|W_{i^*}^{(2)}|$ and $\sigma'(z_{i^*}) \neq 0$.  At most input values $\phi$, the derivative of the encoder is small and nearly constant, allowing it to approximate a linear map where the encoder learns the $\phi$ parametrization. Since $\fenc$ is continuous, however, there is a short interval in $\phi$ where $\fenc$ must retrace the path traversed by the rest of the input domain, incurring a large derivative; the ``break point'' $\phi_0$ is near the center of that interval (see Fig.~\ref{fig:S1Uniform}).  Note further that this quasi-minimum has a flat direction at the break point $\phi_0$:
\be
\left . \left (W_{i^*x} \frac{\partial \fenc}{\partial W_{i^*x}} + W_{i^*y} \frac{\partial \fenc}{\partial W_{i^*y}} +  b_{i^*} \frac{\partial \fenc}{\partial b_{i^*}}\right) \right |_{\phi = \phi_0} = 0,
\ee
which follows from Eq.~(\ref{eq:preactvanish}) and thus implies that the first-layer weights and biases can continue to evolve even when the behavior of $\fenc$ near $\phi_0$ doesn't change.

The analysis is somewhat different for a ReLU activation; in that case, a quasi-minimum can be found when all $z_i < 0$ since $\sigma(z_i) = \sigma'(z_i) = 0$ for $z_i < 0$. However, $\partial \fenc/\partial \phi$ is proportional to $\sigma'(z_i)$, and thus would vanish everywhere which would map all of the input data to a single point in the latent space. Thus, the competing requirements of simultaneously needing $\sigma'(z_i) \neq 0$ and  $0 < \sigma(z_i) \ll 1$ push $z_{i^*}$ towards zero as in the case of a tanh activation, leading to qualitatively similar behavior. The discontinuous derivative makes this case more difficult to analyze analytically, though, so we focus our discussion on smooth activation functions from here on but show an example below of the network dynamics with ReLU activation.

To summarize, at the end of training, the break point $\phi_0$ typically corresponds to a solution to $z_{i^*} = 0$ where $i^* = {\rm argmax} |W_i^{(2)}|$.  We have also qualified this statement with ``typically'' because it may happen that two of the weights have similar magnitudes, and it might be the case that $\phi_0$ is determined by the second-largest one, as we discuss below.

\begin{figure}[t!]
	\includegraphics[width = 0.32\textwidth]{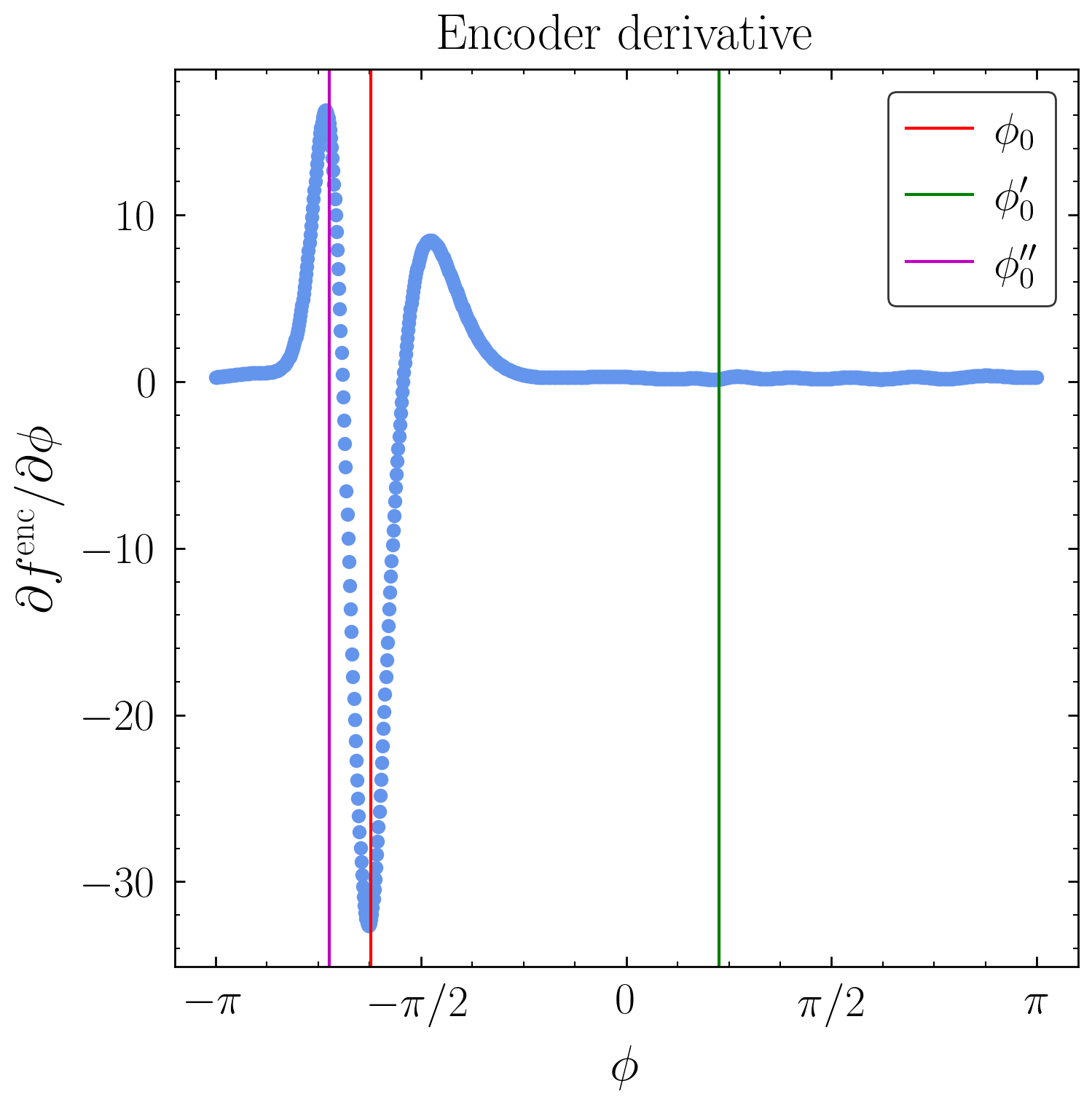}
	\includegraphics[width = 0.315\textwidth]{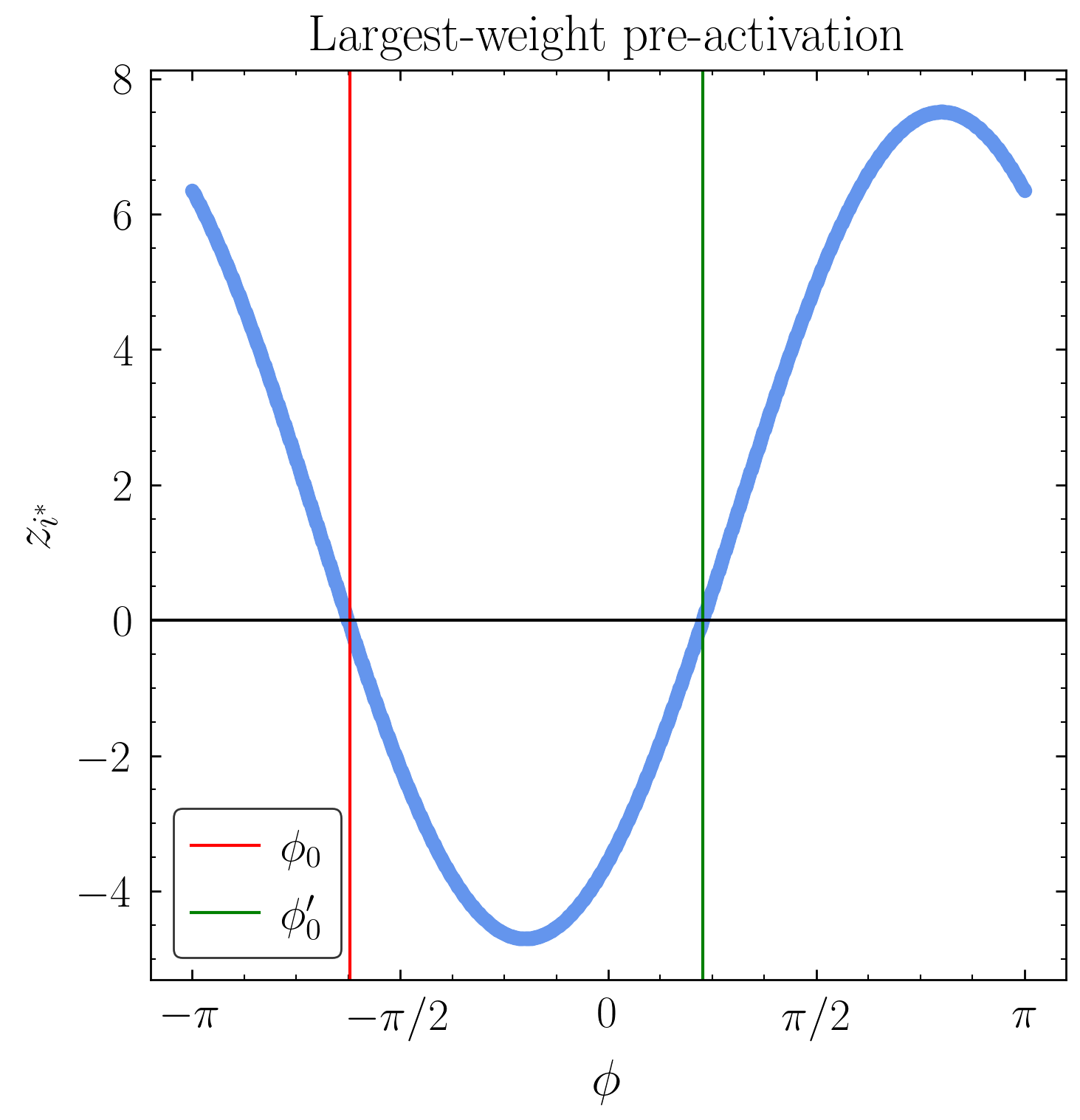}
	\includegraphics[width = 0.315\textwidth]{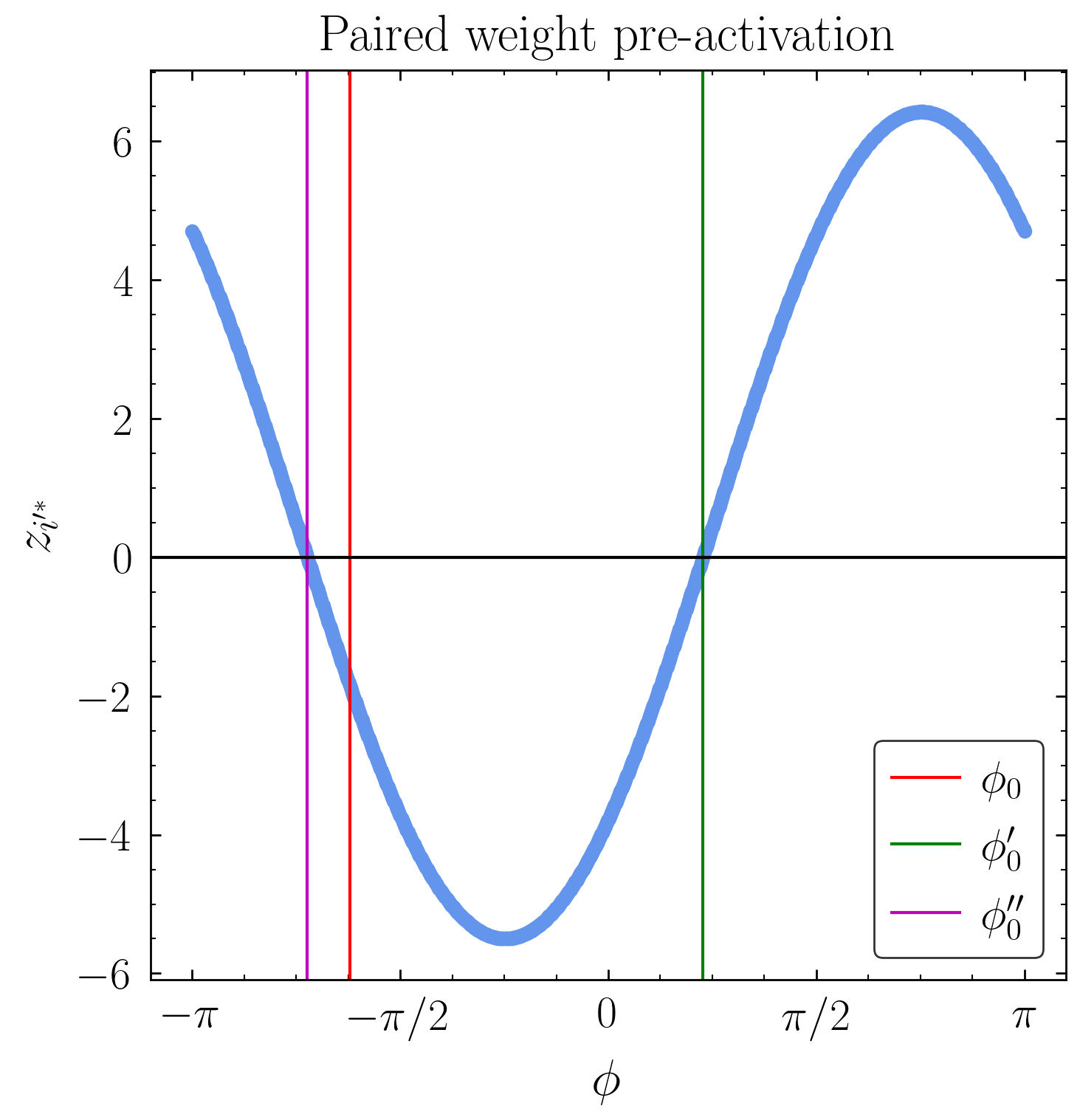}
	\caption{\textbf{Left:} encoder derivative $\partial \fenc/\partial \phi$ as a function of input data coordinate $\phi$. The break point $\phi_0$ is shown in red. \textbf{Center and right:} Pre-activations $z_{i}(\phi)$ of the hidden layer node $i^*$ with the largest outgoing weight \textbf{(center}) and of the paired node $i'^*$ \textbf{(right)} as a function of input $\phi$ for the trained circle network. Note that the point of largest derivative of the encoder, which corresponds to the break point $\phi_0$, is one of the zeros of the pre-activation corresponding to the largest magnitude weight $|W^{(2)}_{i*}|$. The second zero of the largest weight, $\phi_0'$, is also a zero of the paired pre-activation $z_{i^\prime*}$; the second zero of the paired weight, $\phi_0''$, is also a point of large encoder derivative.
	}
	\label{fig:circletrained}
\end{figure}

\begin{figure}[t!]
	\includegraphics[width = 0.32\textwidth]{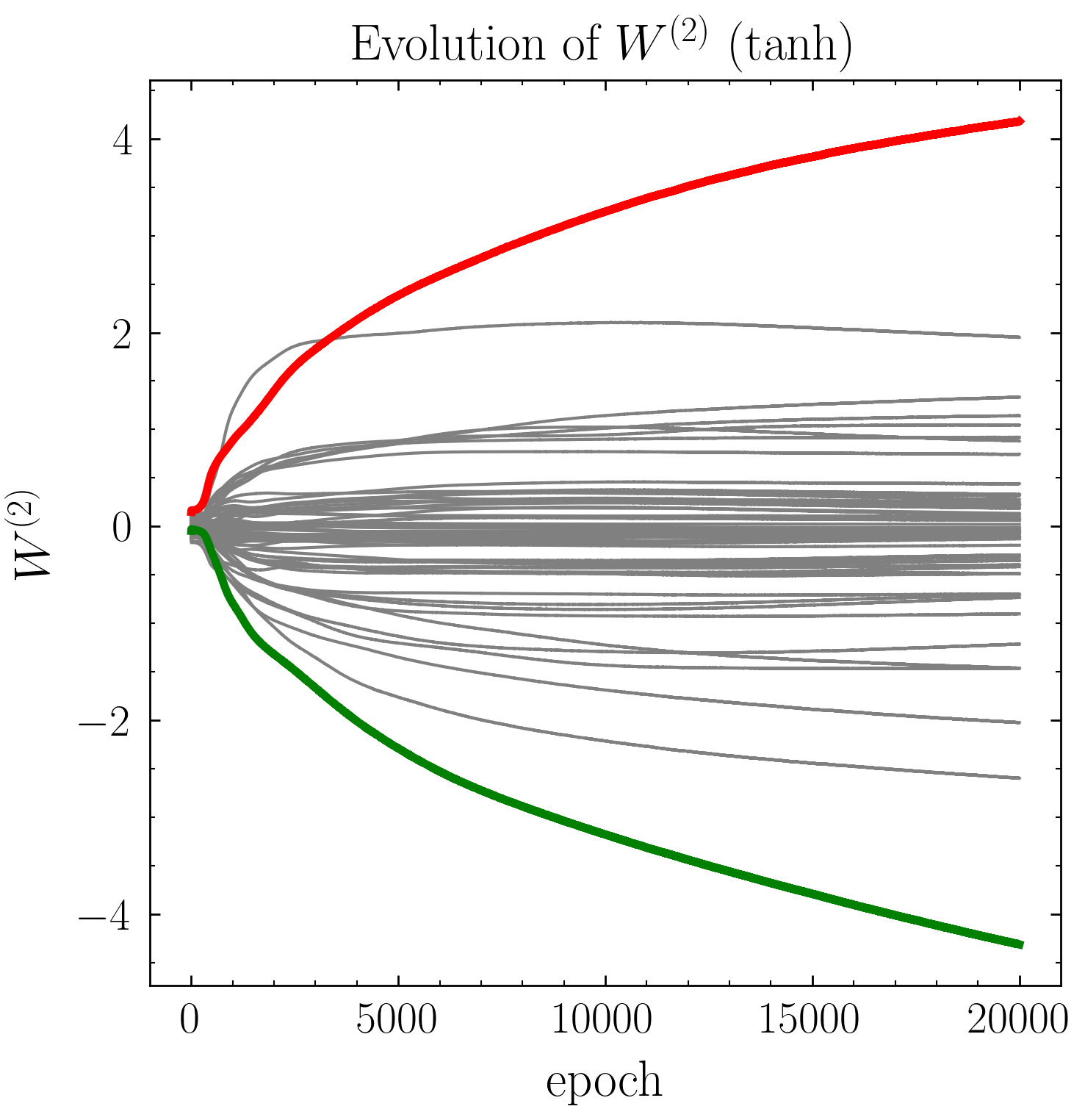}
	\includegraphics[width = 0.33\textwidth]{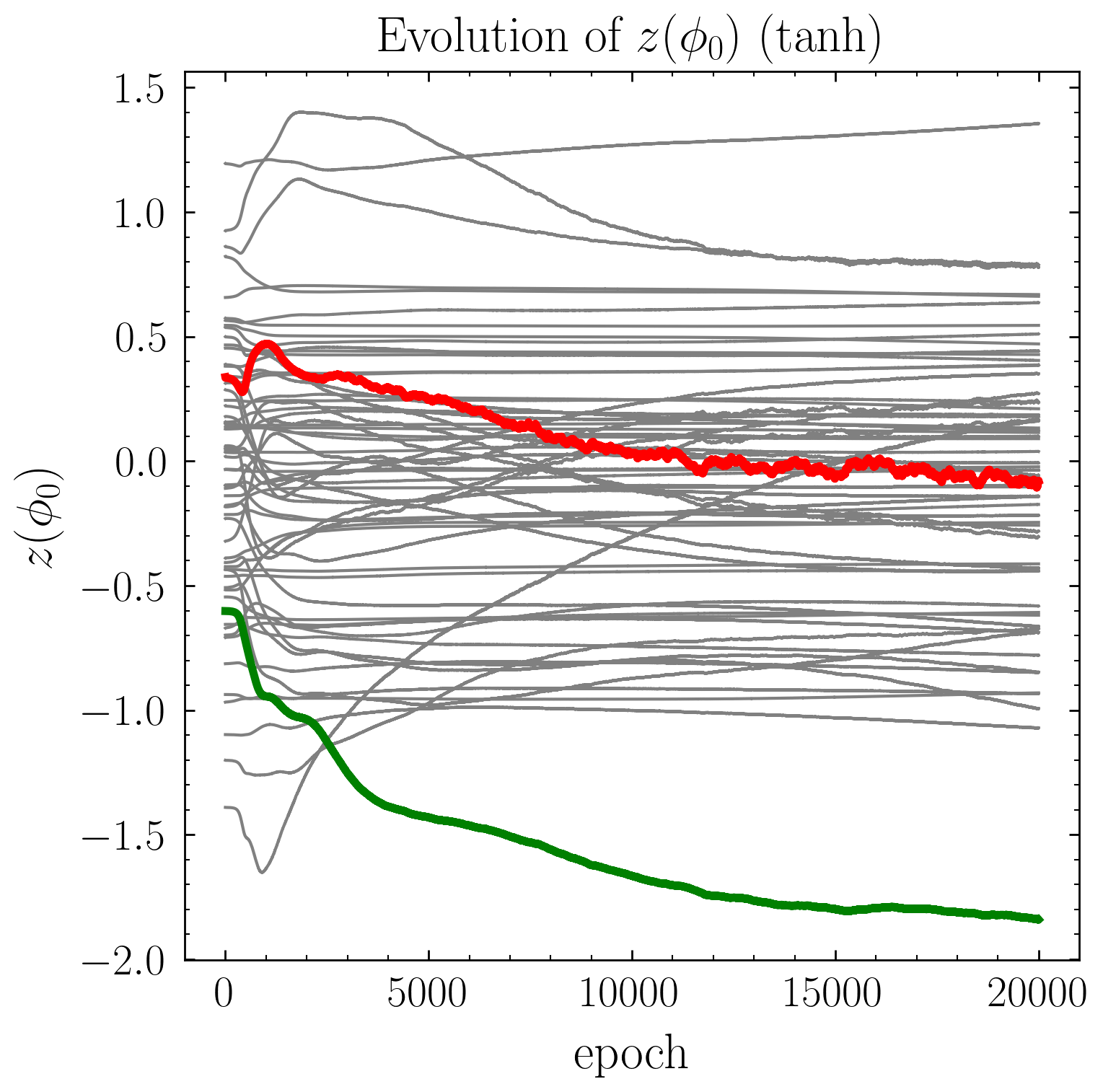}
	\includegraphics[width = 0.32\textwidth]{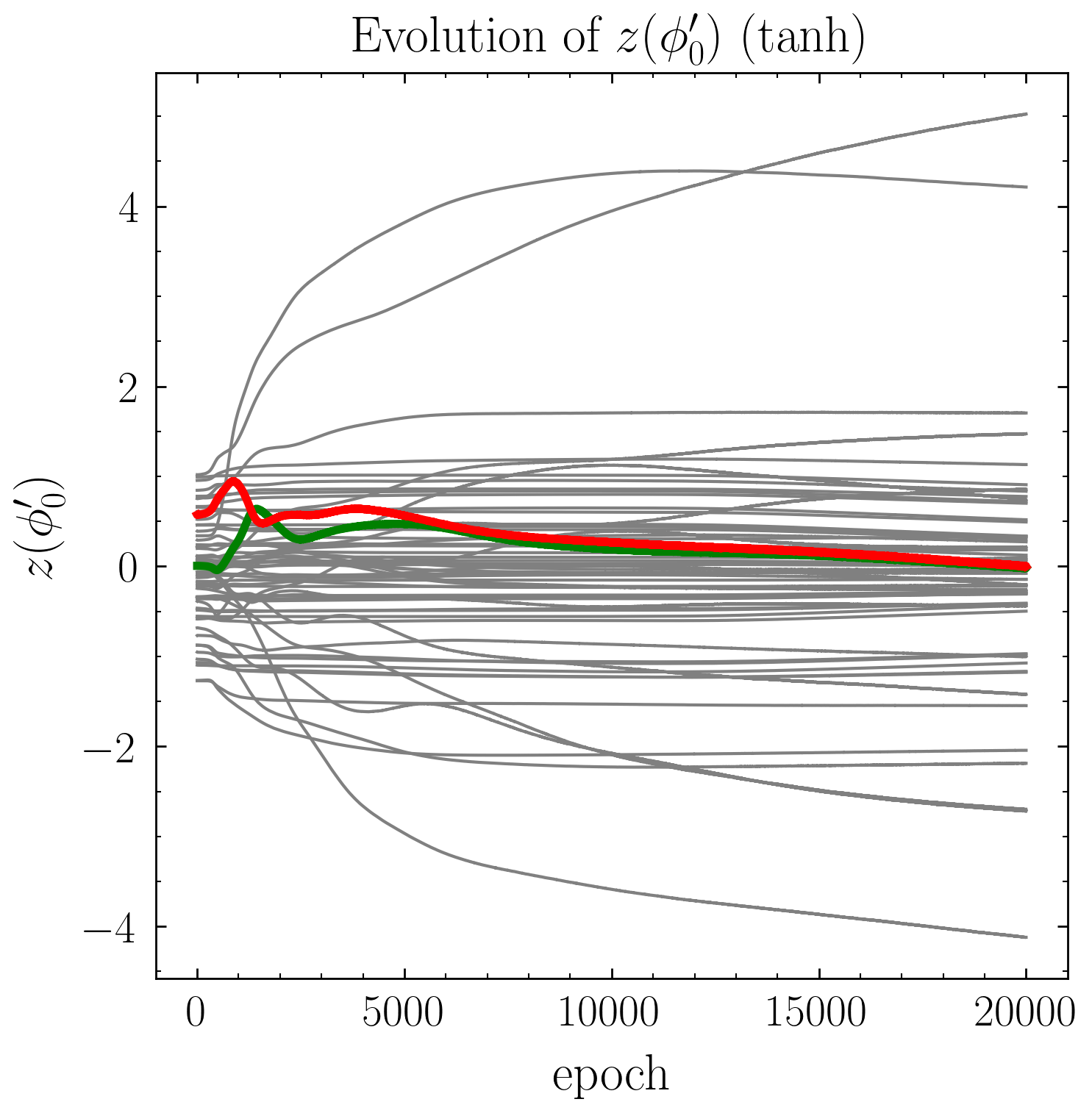}
	\caption{Evolution of the encoder parameters during training, with the weights and pre-activations at nodes $i^*$ and $i'^*$ shown in red and green, respectively. Weights evolve in pairs to large positive and negative values \textbf{(left)}. The pre-activation $z_{i^*}(\phi_0)$ is driven to zero while $z_{i'^*}(\phi_0) \neq 0$ \textbf{(center)}, resulting in a break point where $\fenc$ has large derivative, while $z_{i^*}(\phi_0')$ and $z_{i^*}(\phi_0')$ are both driven to zero \textbf{(right)}.}
	\label{fig:W2evolve}
\end{figure}

Fig.~\ref{fig:circletrained} shows $z_{i^*}(\phi)$ and $\partial \fenc/\partial \phi$ for the trained network shown in Fig.~\ref{fig:S1Uniform} with $\sigma = {\rm tanh}$, which was initialized with random weights and biases drawn from a uniform distribution between $-1/\sqrt{d_w}$ and $1/\sqrt{d_w}$ (the default in \texttt{Pytorch}). As anticipated by the analysis above, the magnitude of the derivative of the encoder is largest at the break point $\phi_0$ (red), which satisfies $z_{i^*}(\phi_0) = 0$, where $i^*$ is determined by the largest-magnitude weight. However, since $z_{i^*}(\phi)$ is a linear combination of sines and cosines plus an offset, it can be written as $A \cos(\phi + \delta) + b$. For sufficiently small $b/A$, this function always has two zeros. The second zero, labeled by $\phi_0^\prime$ (green), does not correspond to a large $\partial \fenc/\partial \phi$ despite the fact that its pre-activation is near zero. Instead, what happens is that there is another weight with large magnitude, $W_{i'^*}^{(2)} \approx -W_{i^*}^{(2)}$, whose pre-activation also contains a zero at $\phi_0^\prime$. We can see this empirically in Fig.~\ref{fig:W2evolve}, which shows the evolution of the weights $W^{(2)}$ and the pre-activations $z_i(\phi_0)$ and $z_i(\phi_0')$; as expected from this analysis, weights evolve in tandem to large positive and negative values, with the corresponding pre-activations driven to zero. This paired weight approximately cancels the large derivative at $\phi_0^\prime$ (some remnants of the imperfect cancellation can be seen in the ``wiggles'' of $\partial \fenc/\partial \phi$ at $\phi_0^\prime$), but absent a fine-tuning of the $W_{ix}$ and $W_{iy}$, the second zero $\phi_0''$ will be different from $\phi_0$, so the large derivative at $\phi_0$ remains.

\begin{figure}[t!]
	\includegraphics[width = 0.32\textwidth]{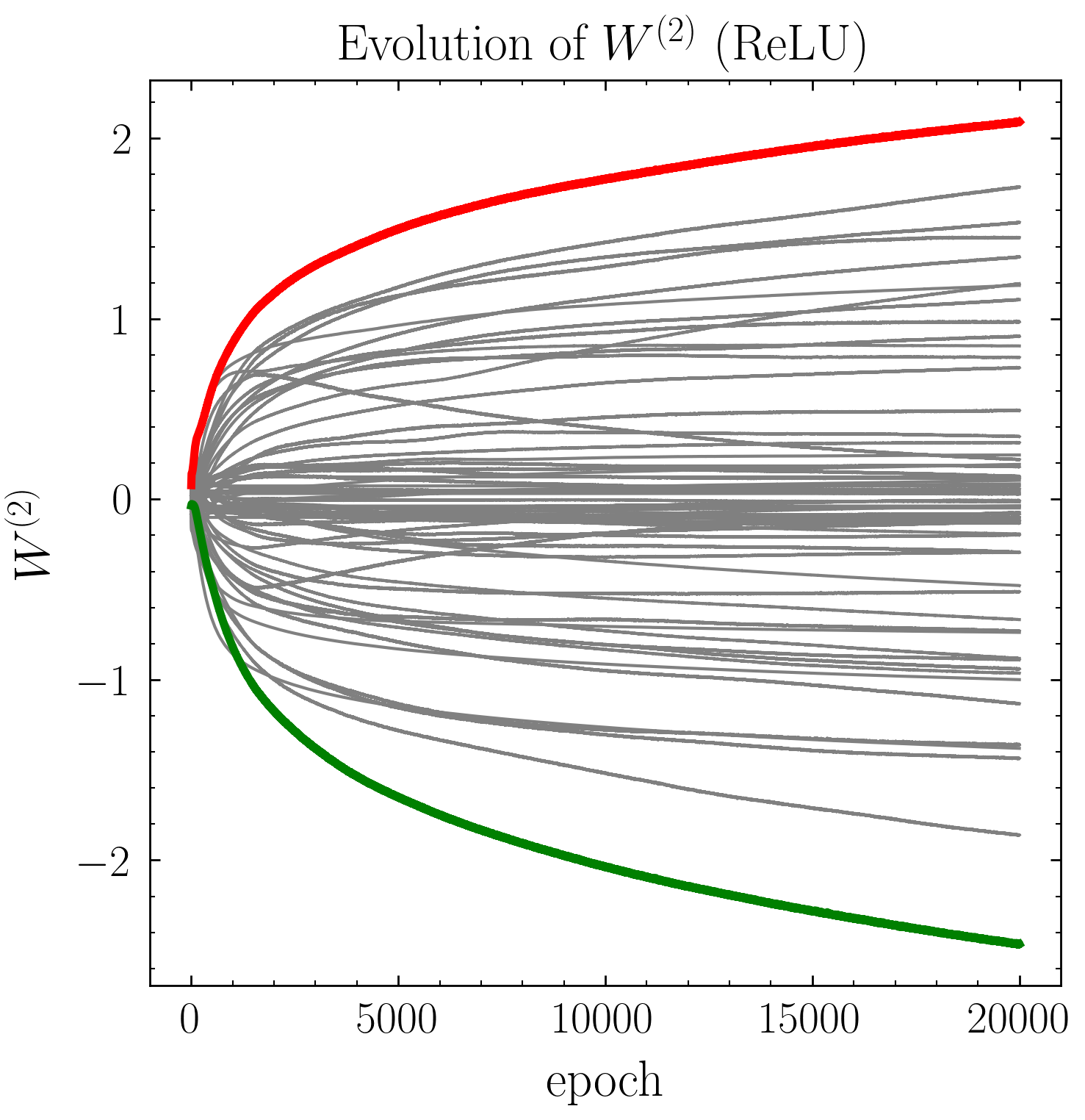}
	\includegraphics[width = 0.33\textwidth]{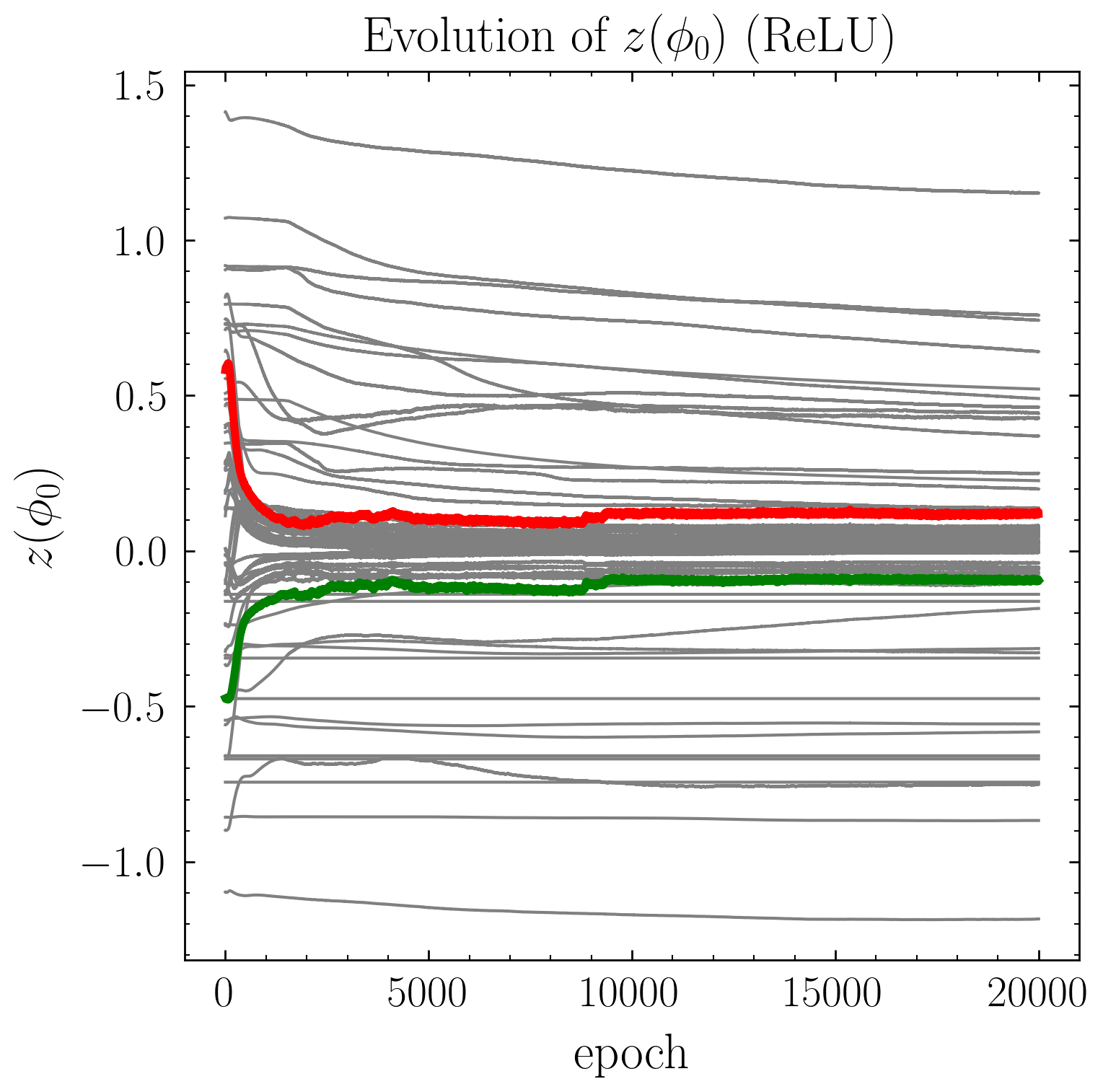}
	\includegraphics[width = 0.32\textwidth]{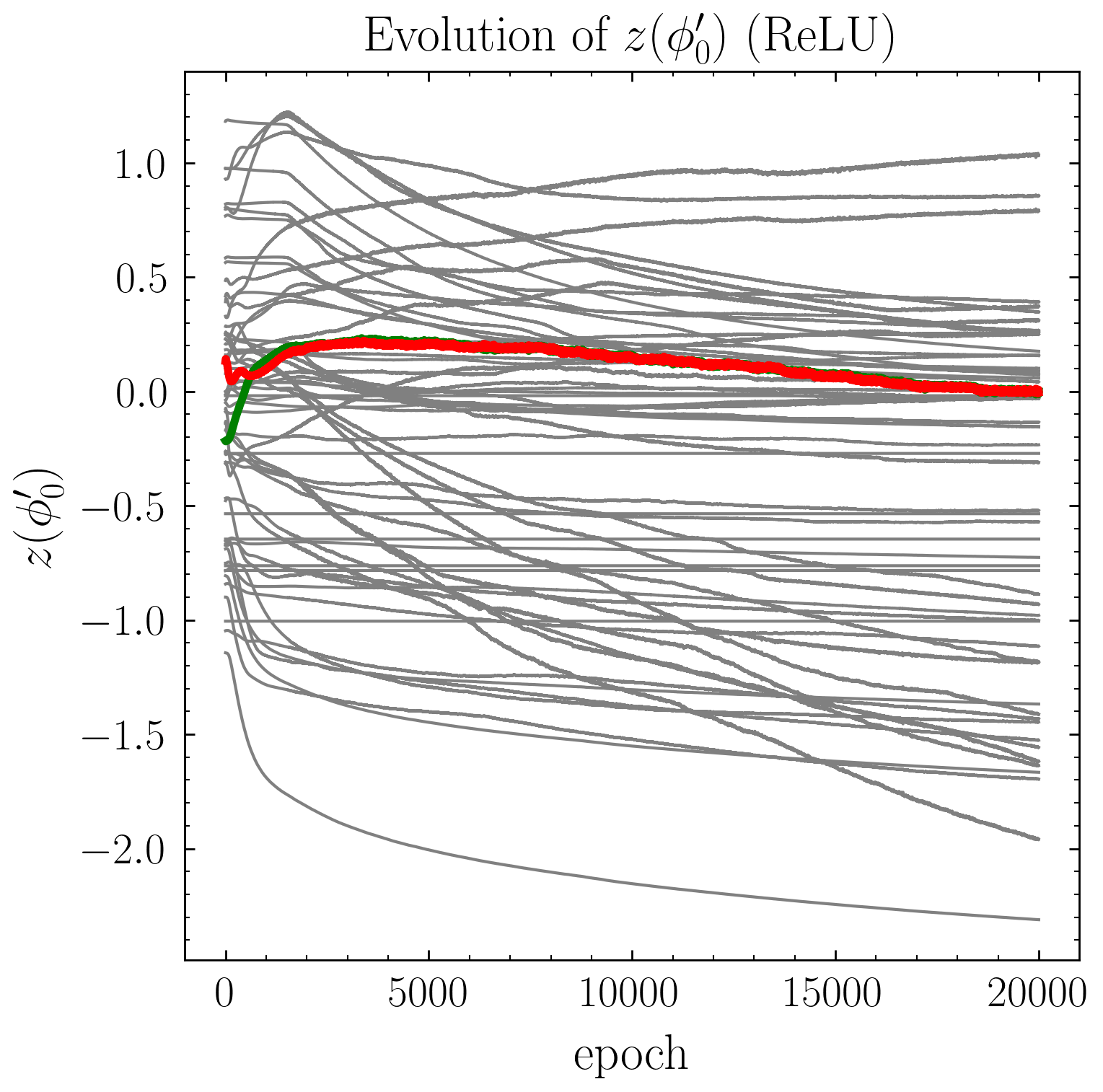}
	\caption{Same as Fig.~\ref{fig:W2evolve} for a ReLU activation function.}
	\label{fig:W2evolveRelu}
\end{figure}

On the other hand, there is now a partially uncancelled zero at $\phi_0''$ (magenta), resulting in a large $\partial \fenc/\partial \phi$ of opposite sign, which can also be seen in Fig.~\ref{fig:circletrained}. The difference $|\phi_0'' - \phi_0|$ is thus responsible for the ``gap'' around the break point, which looks potentially logarithmic as a function of training epoch, since cancelling the zero at $\phi_0''$ requires the network to find its way to a finely-tuned quasi-minimum, or equivalently, for the composition of continuous maps to yield a discontinuous latent representation which has a delta-function derivative. As noted in App.~\ref{app:S1checks}, we have verified that increasing the hidden layer width, or adding another layer to the encoder, does not affect the size of the break region (as measured by the gap in the decoder), which appears to depend mostly on the length of training (for a fixed learning rate) and to some extent on the particular form of the activation function, including the derivative at the origin as measured by $\beta$. Fig.~\ref{fig:W2evolveRelu} shows the evolution of the encoder parameters for a ReLU activation, showing the same qualitative behavior as for tanh. The key differences are that $z_{i^*}(\phi_0)$ is not necessarily driven to zero but can remain positive because the pre-activation derivative $\sigma'(z)$ is identical for any $z > 0$; in addition, the weights $W^{(2)}$ which do not determine the break point are not driven to zero as fast as for the tanh activation. In this example the break point is also determined by the second-largest $|W^{(2)}|$. Nonetheless, the main feature which determines the break point, namely the pair of weights of equal magnitudes evolving in parallel, persists independent of the activation function, since it is required by topology.

\begin{figure}[t!]
\begin{center}
	\includegraphics[width = 0.45\textwidth]{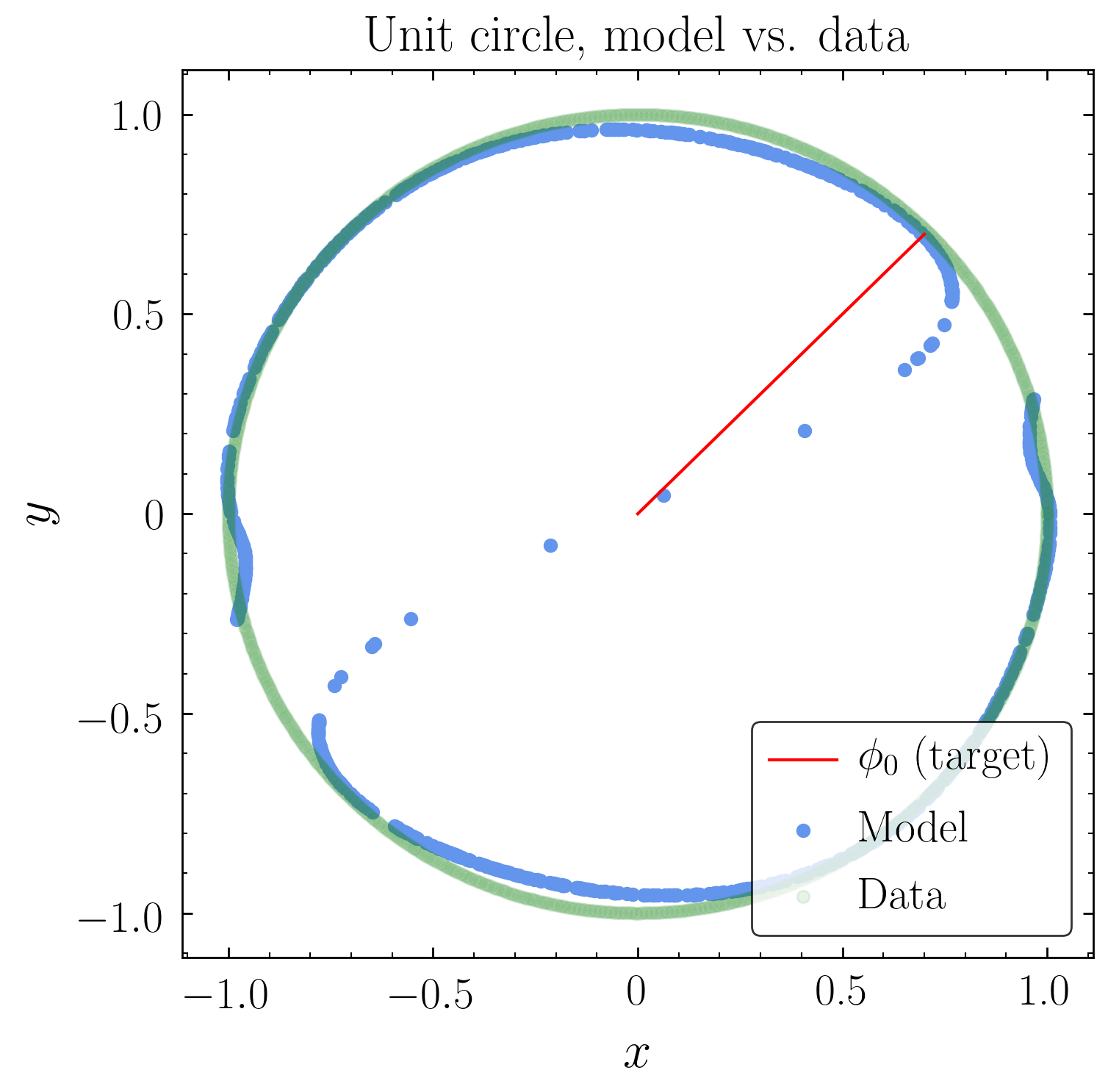} \qquad
	\includegraphics[width = 0.45\textwidth]{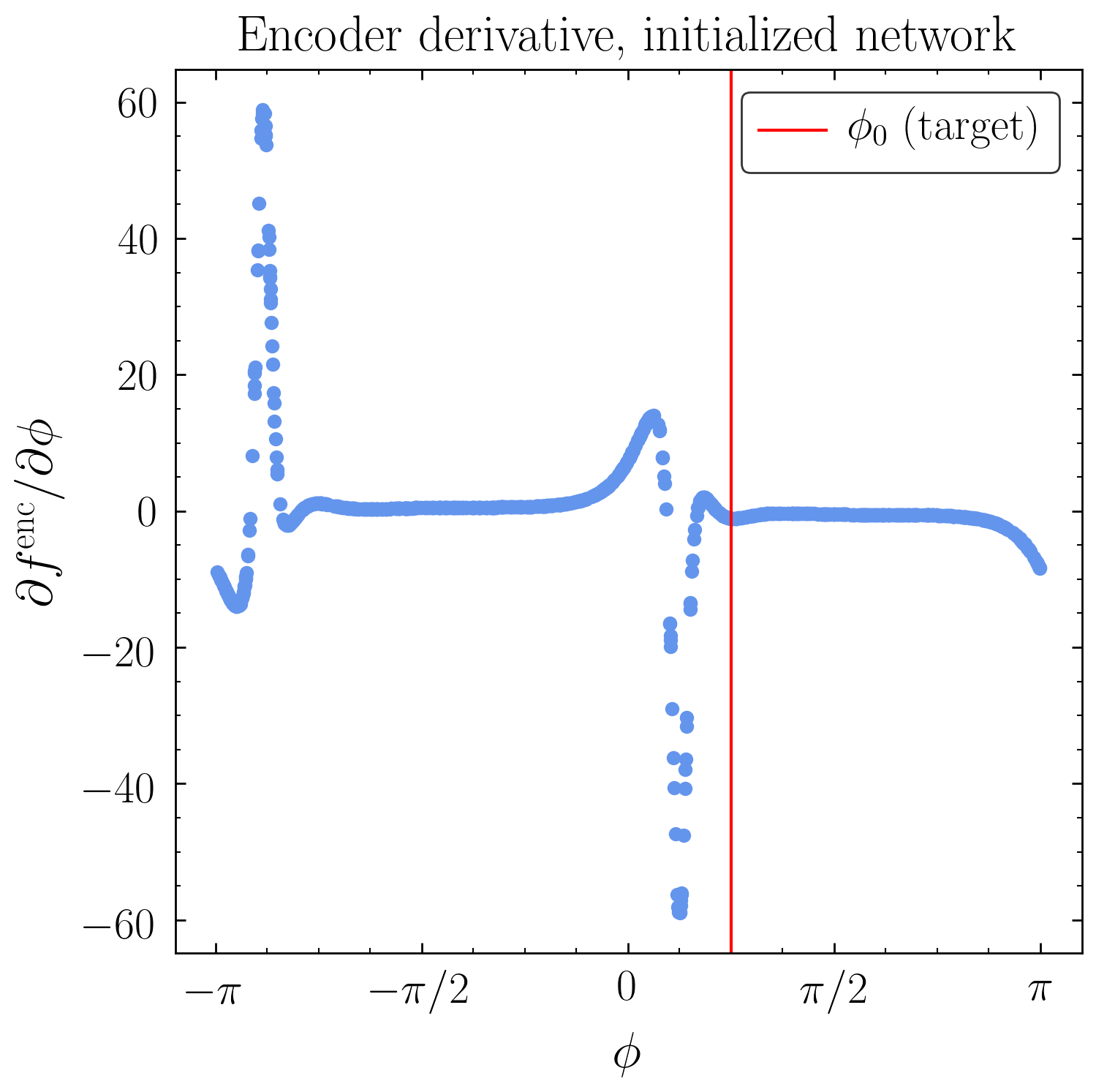}
	\caption{Attempts to determine the break point of the $S^1$ autoencoder by initialization: one weight is initialized to a value 3, larger than the other initialized weights, and the corresponding first-layer parameters are initialized such that $z(\phi_0) = 0$ at $\phi_0 = \pi/4$. \textbf{Left:} Output of the autoencoder. \textbf{Right}: derivative of the encoder $\partial \fenc/\partial \phi$. The network never cancels the second break point, and the trained minimum is poorer quality than the one achieved with random initialization.}
	\label{fig:circleinit}
\end{center}
\end{figure}

Finally, we consider trying to initialize the network to give a break point at a prescribed value of $\phi_0$. Given that $\phi_0$ is determined by the largest second-layer weight $W_i^{(2)}$, we expect that if we initialize one of the weights, say $i = i^*$, to a large value compared to the width of the distribution from which the rest of the weights are drawn ($1/\sqrt{d_w} = 0.125$ for our default network parameters), $\phi_0$ will be determined somehow by the corresponding pre-activation $z_{i^*}$. From the update equations, the network will prefer to move along the flat direction for the first-layer weights and biases, so to choose $\phi_0$ we can also initialize $W_{i{^*}x}$, $W_{i{^*}y}$, and $b_{i^*}$ such that $\phi_0$ is a solution to $z_{i^*} = 0$. Fig.~\ref{fig:circleinit} shows the results of initializing $W^{(2)}_{i^*} = 3$ and $z_{i^*}(\phi_0) = 0$ with $\phi_0 = \pi/4$. The quasi-minimum the initialized network finds is qualitatively different than the randomly-initialized network. One break point ends up close to the target $\phi_0$, but the second zero of the pre-activation $\phi_0^\prime$ remains uncancelled, and the encoder develops two break points, as shown in Fig.~\ref{fig:circleinit}. This behavior is qualitatively similar to the latent representation of the Clifford torus in $\mathbb{R}^4$ in App.~\ref{sec:torus} below. The interplay between initialization and training is fertile ground for future work, especially in this simple example where analytic approaches may be tractable.

\section{Other topologies and geometries}
\label{app:MoreExamples}

\subsection{The trefoil knot: extrinsic topology in dimension 1}
\label{app:knot}

\begin{figure}[t!]
	\includegraphics[width = 0.45\textwidth]{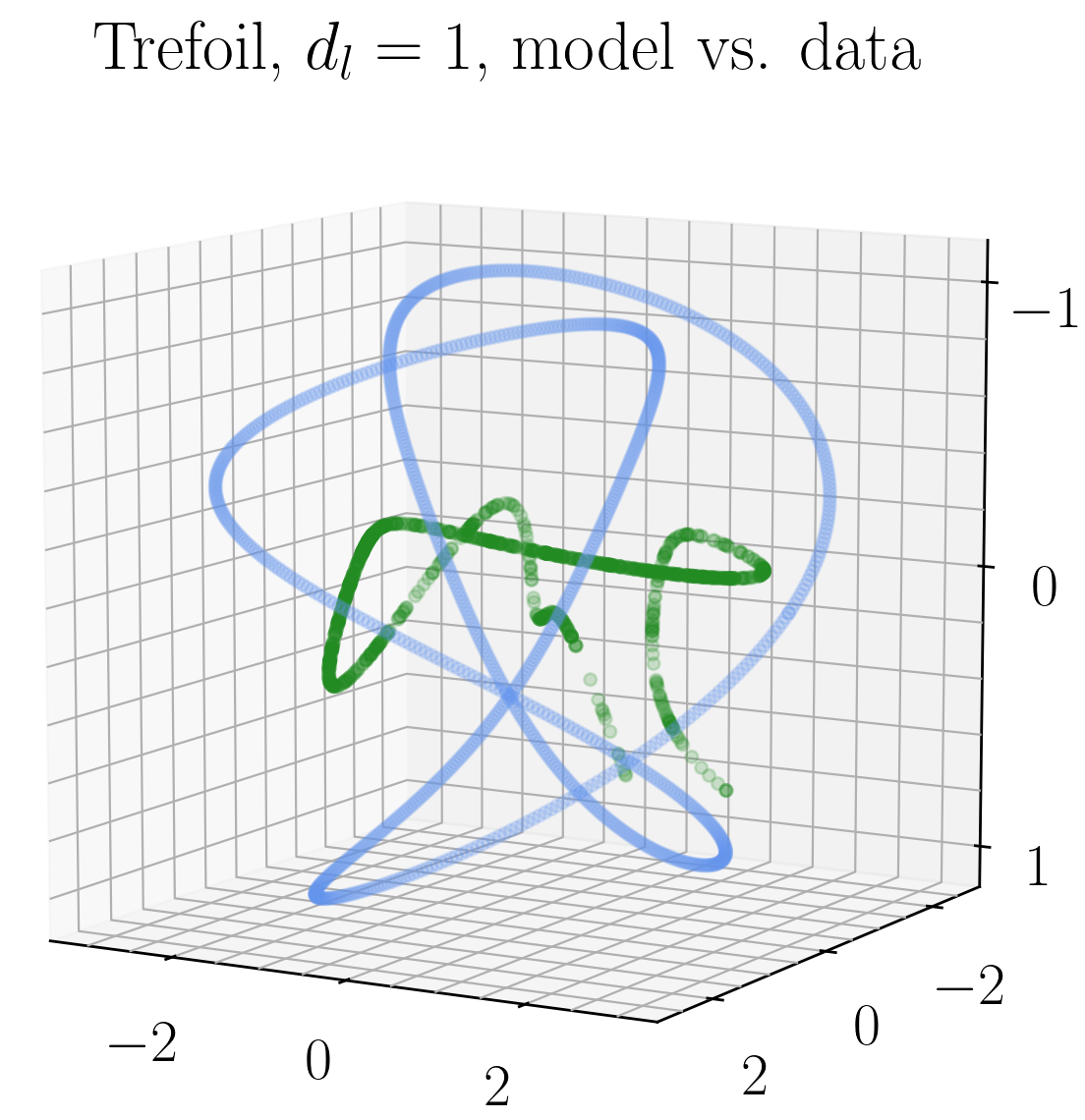} \qquad
	\includegraphics[width = 0.45\textwidth]{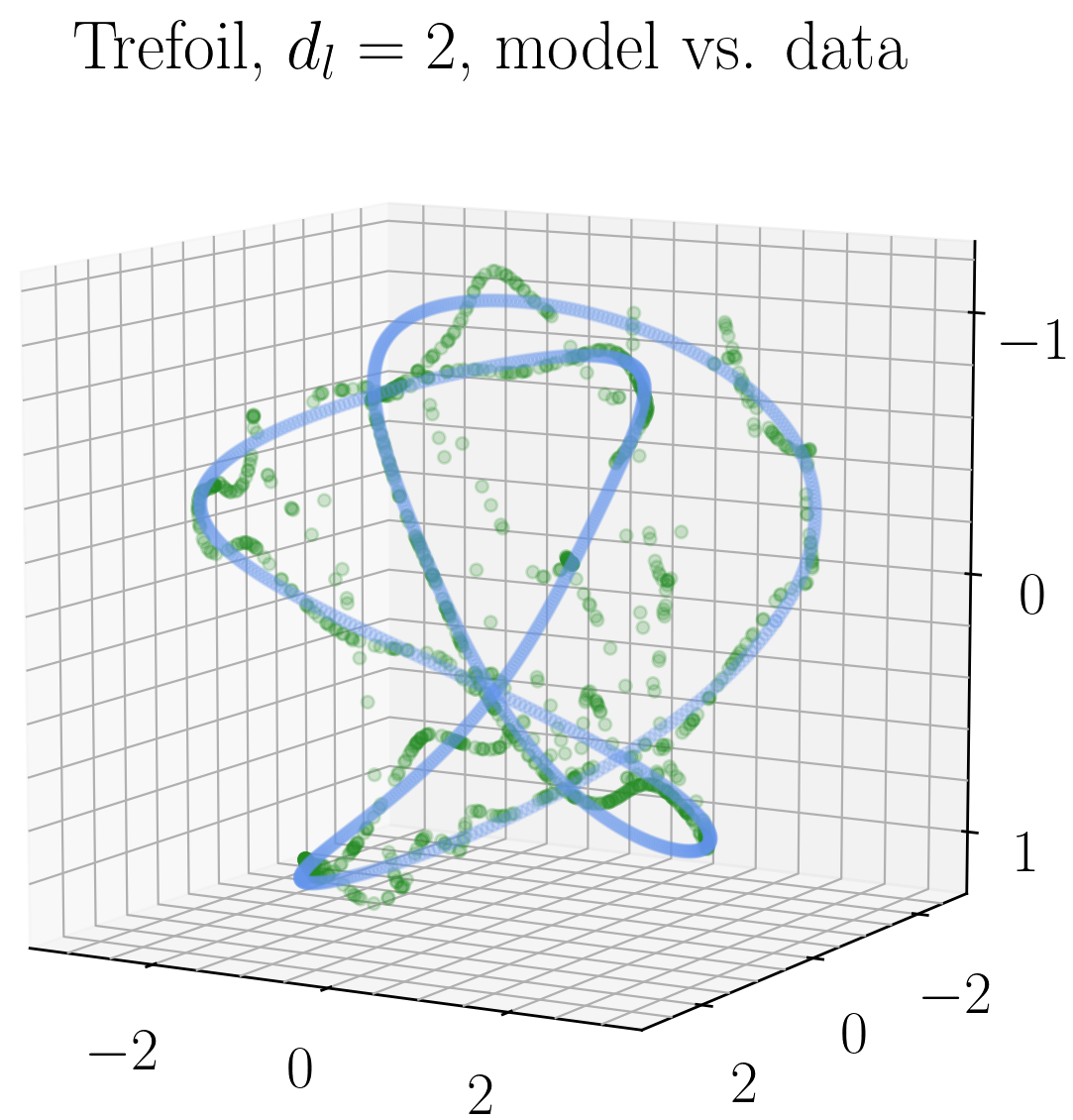} \\
	\includegraphics[width = 0.45\textwidth]{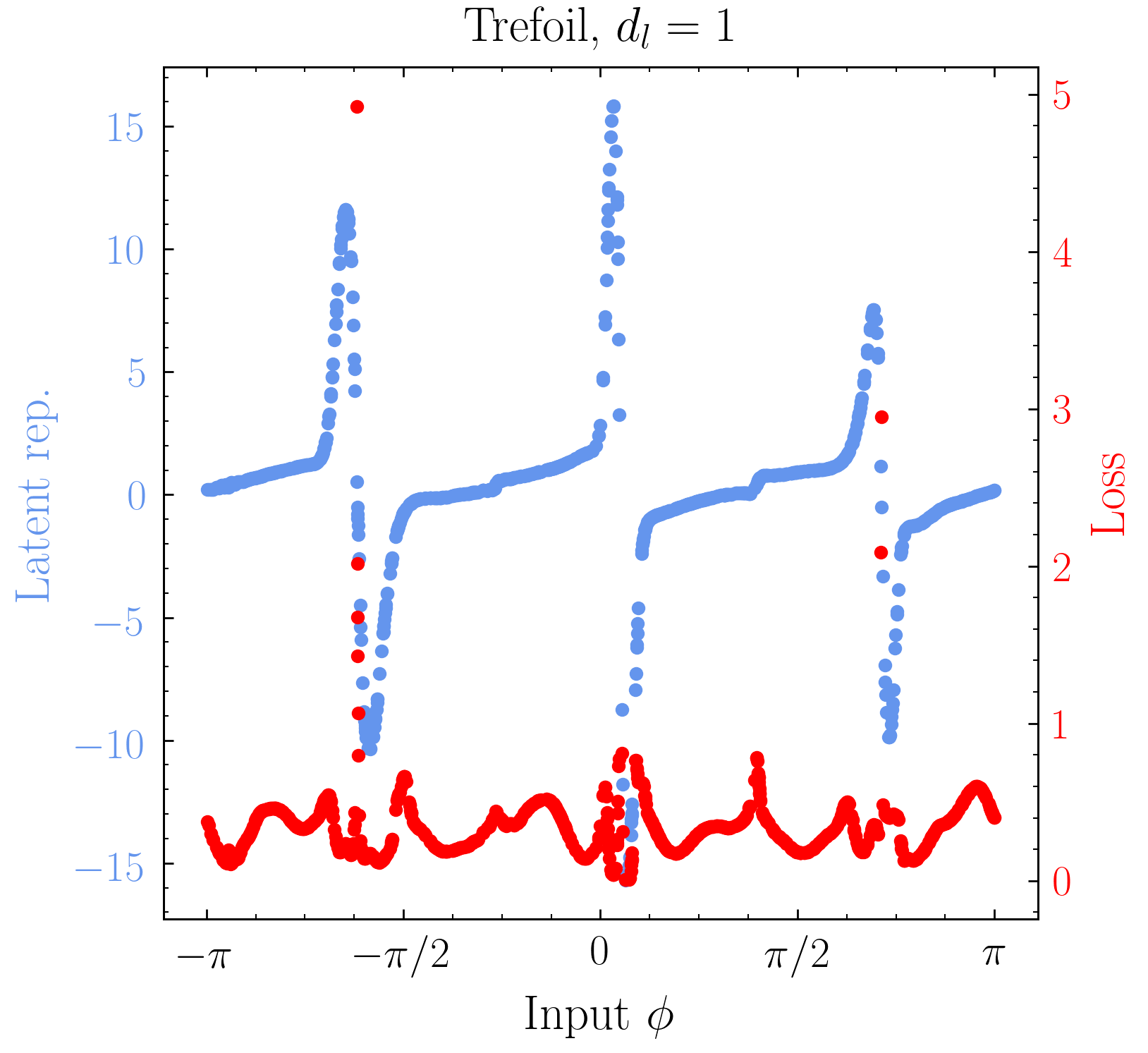} \qquad
	\includegraphics[width = 0.45\textwidth]{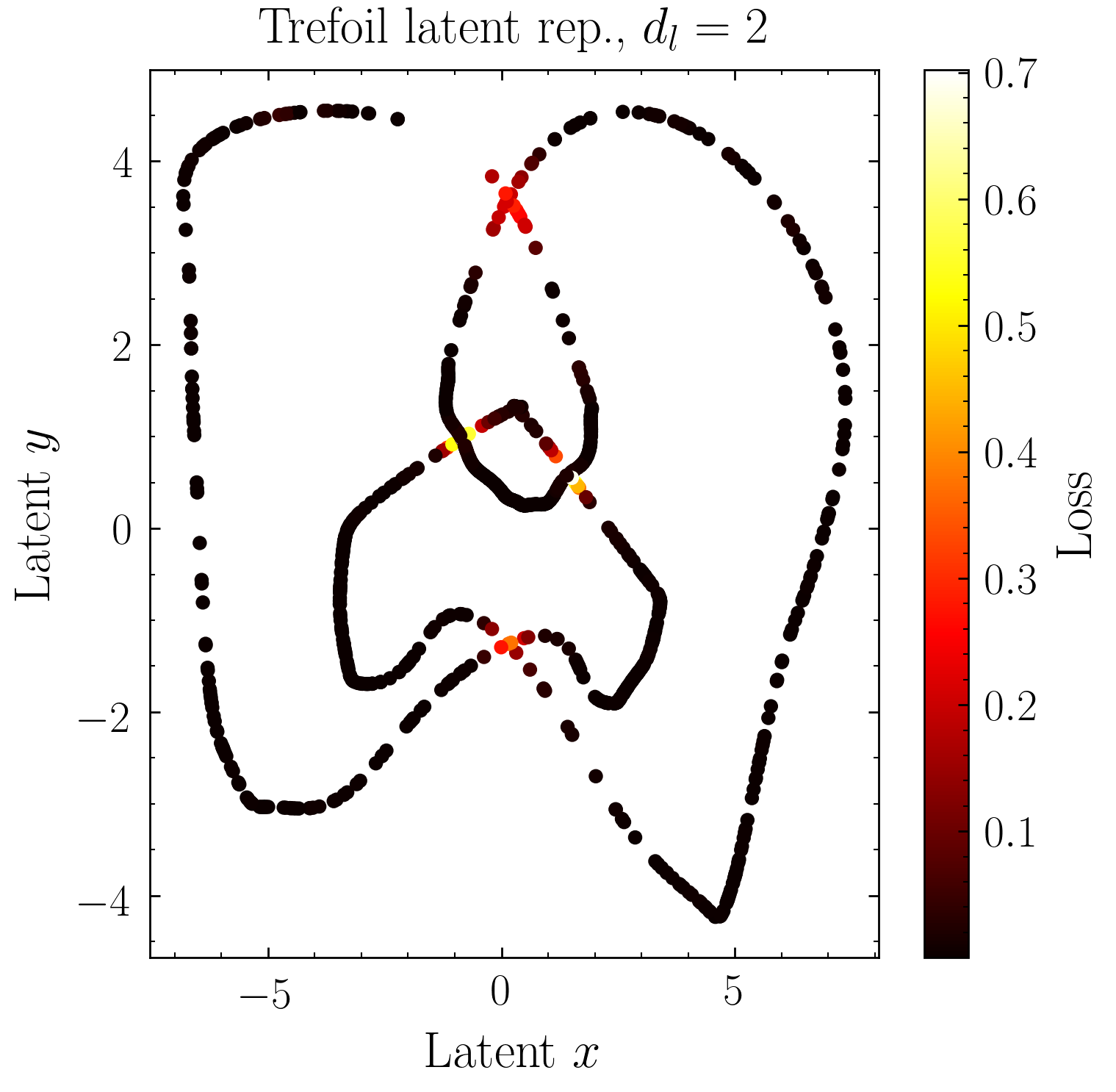}
	\caption{\textbf{Top row:} The trefoil knot (blue) with model output (green) for $d_l = 1$ \textbf{(left)} and $d_l = 2$ \textbf{(right)}. \textbf{Bottom row:} Trefoil latent representations and losses on the test set, for $d_l = 1$ \textbf{(left)} and $d_l = 2$ \textbf{(right)}. For $d_l = 2$, the largest losses are localized to the self-intersections in the latent representation.}
	\label{fig:TrefoilFail}
\end{figure}

Consider a circle embedded with nontrivial extrinsic topology in $\mathbb{R}^3$: namely, the trefoil knot, defined by
\be
x = (R + r + \cos 2\phi)\cos 3\phi, \qquad y = (R + r + \cos 2\phi)\cos 3\phi, \qquad z = r + \sin 2\phi,
\ee
where we take $r = 1$ and $R = 2$ for concreteness. In this context, extrinsic topology refers to the fact that the knot cannot be continuously deformed into the circle in $\mathbb{R}^3$ without tearing, despite these two manifolds having the same intrinsic topology of the circle. As with the $S^1$ autoencoder, we train on an equidistant training set in $\phi$, for both $d_l = 1$ and $d_l = 2$; the results of the output map are shown in the top row of Fig.~\ref{fig:TrefoilFail}. Just as with the circle, the output map contains break points, which in the case of $d_l = 2$ correspond to self-intersections in the latent representation (Fig.~\ref{fig:TrefoilFail}, bottom right). The typical size of the error is much larger for $d_l = 1$, but in both cases the largest errors are confined to neighborhoods of isolated points, as was the case for the circle. Here, though, we are seeing nontrivial intrinsic and extrinsic topology, the latter of which makes it difficult to learn the global geometry of the knot even for $d_l$ larger than the intrinsic dimension of the data, because a generic initialization of the network will lead to a latent representation with self-intersections.

\begin{figure}[t!]
\begin{center}
	\includegraphics[width = 0.45\textwidth]{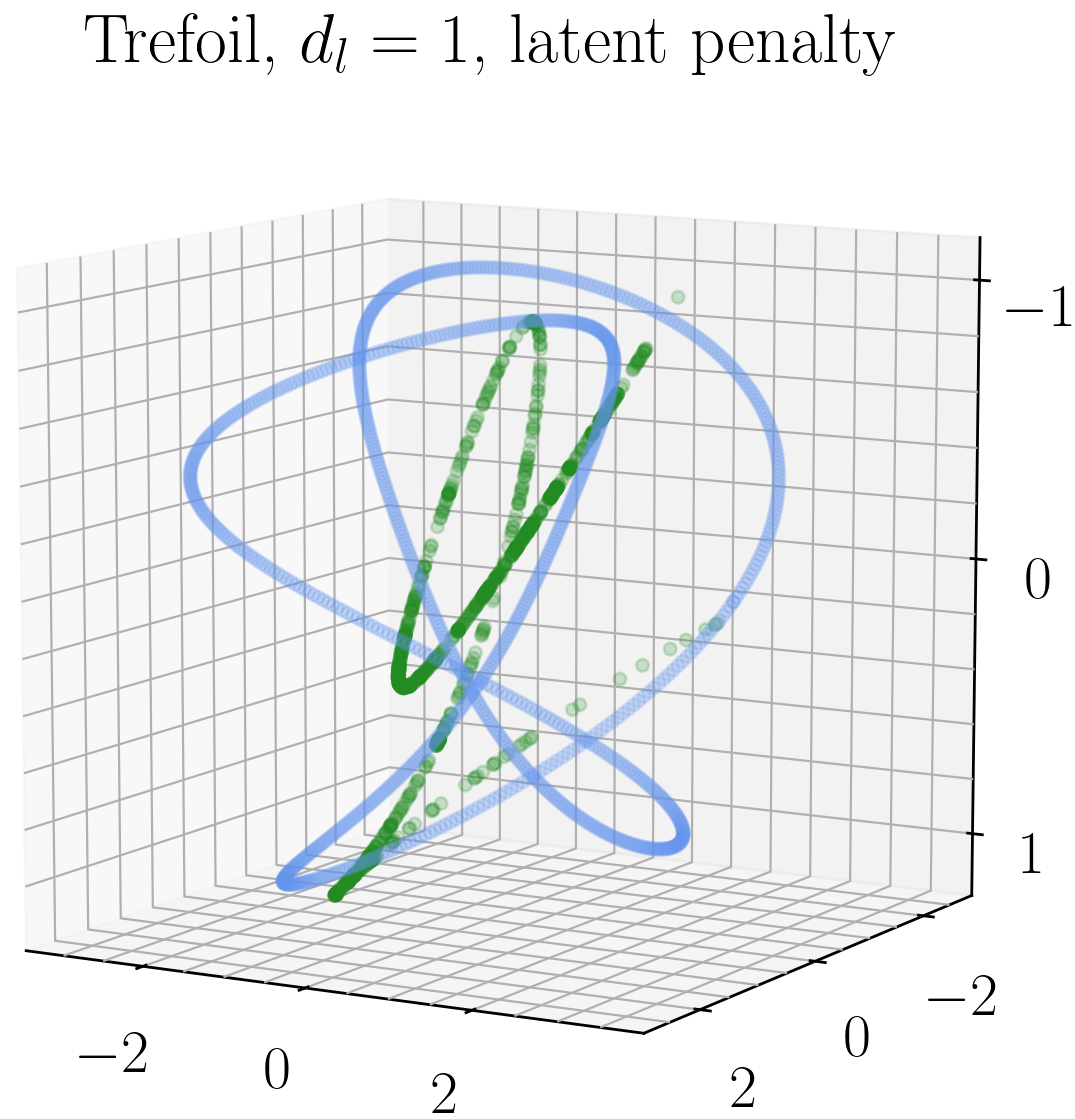} \qquad
	\includegraphics[width = 0.45\textwidth]{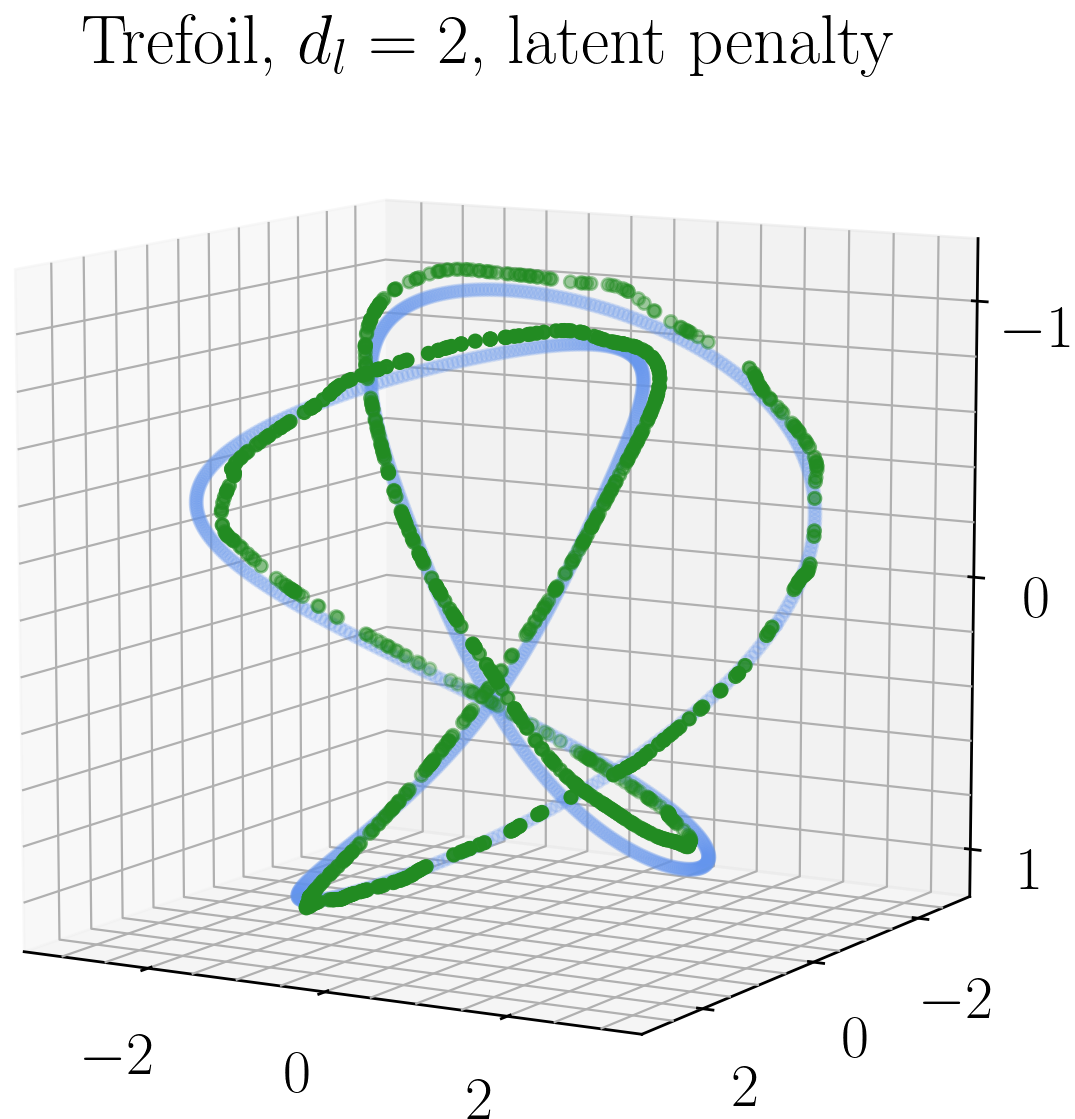} \\
	\includegraphics[width = 0.45\textwidth]{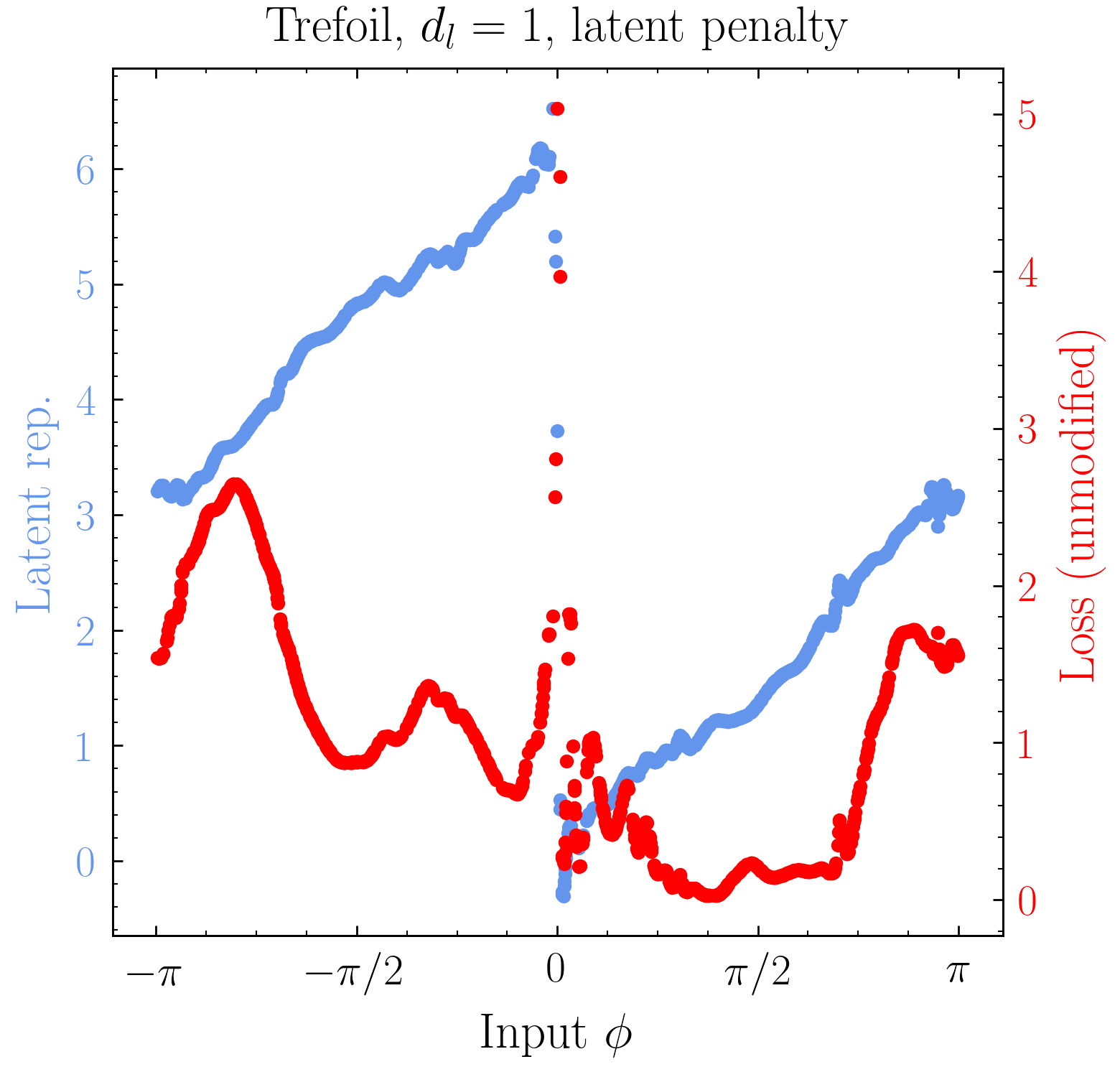} \qquad
	\includegraphics[width = 0.45\textwidth]{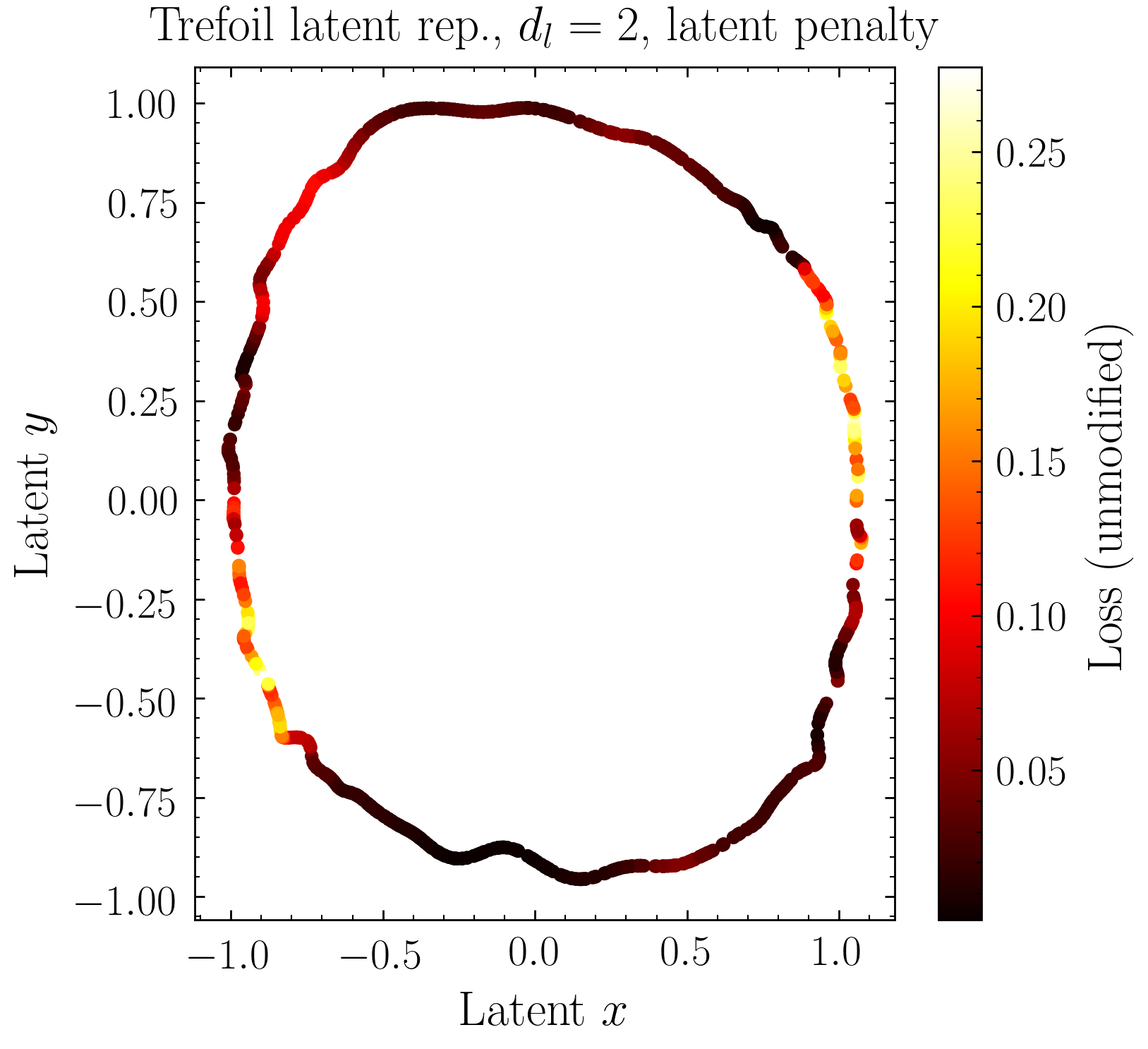}
	\caption{ Model output and latent representation for the trefoil for with the loss function modified to include a penalty when the latent representation deviates from the parametric representation of the data. The \textbf{left} plots show the results for $d_l = 1$ and the \textbf{right} plots show $d_l = 2$. Even with the modified loss, the $d_l = 1$ network still has a break point because of the intrinsic topology of the knot, but forcing the latent representation to approximate a circle for $d_l = 2$ leads to much better reconstruction.
    }
	\label{fig:UntieTrefoil}
\end{center}
\end{figure}

We can cure the topological issues in two ways. First, by taking $d_l = 3$, we can have near-perfect reconstruction of the knot, but at the price of learning the trivial map in the region enclosed by the knot. We can also do something more clever and force the network to learn that the knot is a parametric curve. Consider modifying the loss function to
\be
\widetilde{L} = ||f(\mathbf{x}) - \mathbf{x}||^2 + \lambda ||\fenc(\mathbf{x}) - \mathbf{x}_\phi||^2
\ee
where $f(\mathbf{x})$ is the output of the full network, $\fenc(\mathbf{x})$ is the output of the latent layer (i.e. the encoding of $\mathbf{x}$), $\mathbf{x}_\phi$ is the parametric representation of the knot of the same dimension of the latent layer, and $\lambda$ is a hyperparameter. The new loss $\widetilde{L}$ penalizes the network for learning a latent representation different from the parametric representation by $\phi$; for $d_l = 1$, $\mathbf{x}_\phi = \phi$, and for $d_l = 2$, $\mathbf{x}_\phi = (\cos \phi, \sin \phi)$. Fig.~\ref{fig:UntieTrefoil} shows the results of training with $\widetilde{L}$, setting $\lambda = 10$ with all other hyperparameters the same. For $d_l = 1$, the latent representation cannot overcome the intrinsic $S^1$ topology of the knot, and while the output is clearly better at approximating the shape of the knot than the case for the unmodified loss function, the knot still breaks around a point as did the unit circle. For $d_l = 2$, the latent representation can learn the 2-dimensional representation of the circle, and we get much better reconstruction. We have checked that this network is \emph{not} learning the trivial representation on $\mathbb{R}^3$, since there is still a compression with $d_l < \din$. We conclude that autoencoders can untie knots (i.e. evade obstructions associated to nontrivial extrinsic topology), as long as we tell the network to do so with a suitable modification to the loss. Indeed, this extra term in the loss is a toy example of the incorporation of priors based on topology which can help improve network performance.

\subsection{The torus: quotient spaces}
\label{sec:torus}

\begin{figure}[t!]
\begin{center}
	\includegraphics[width = 0.45\textwidth]{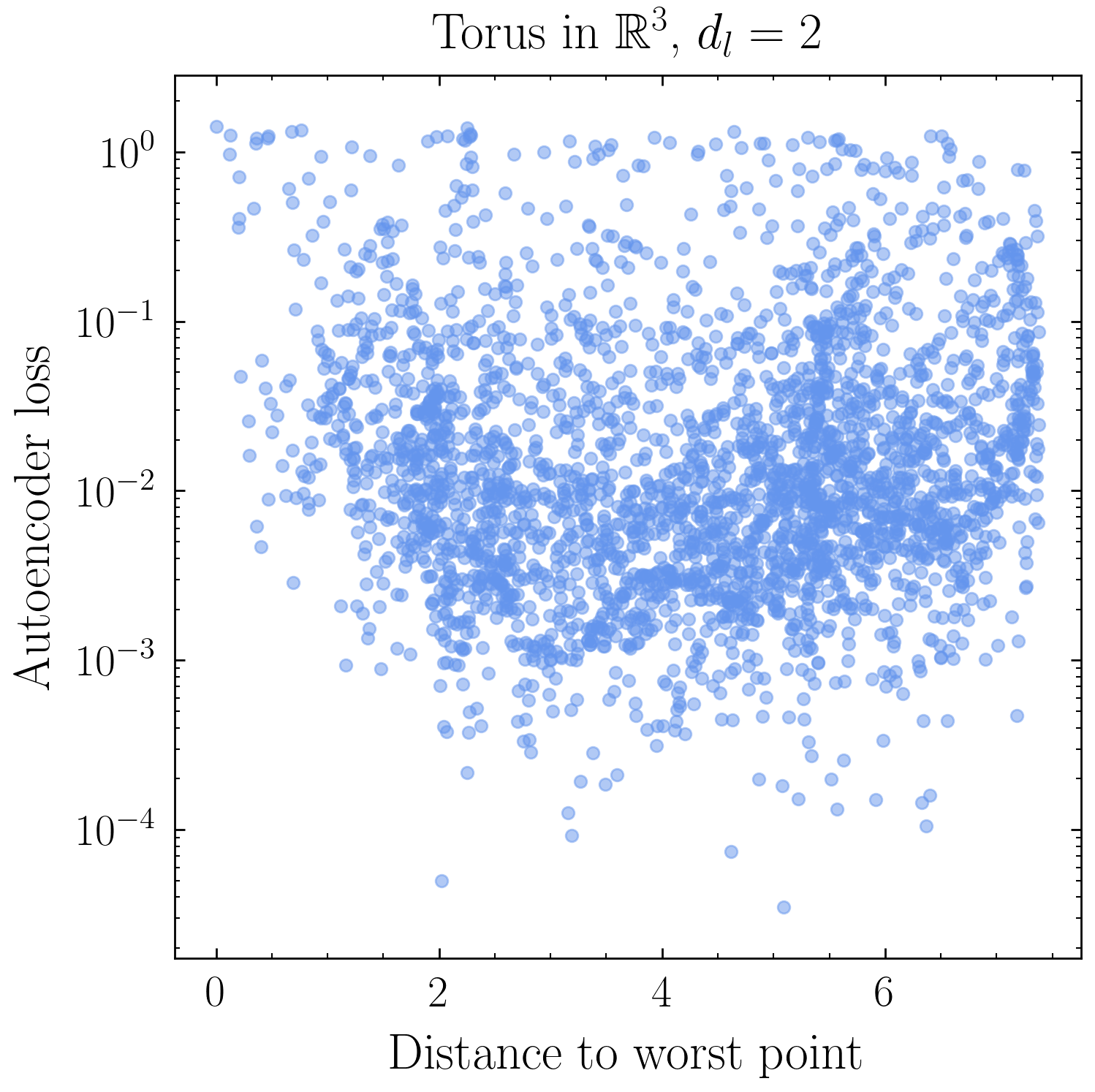}\qquad
	\includegraphics[width = 0.45\textwidth]{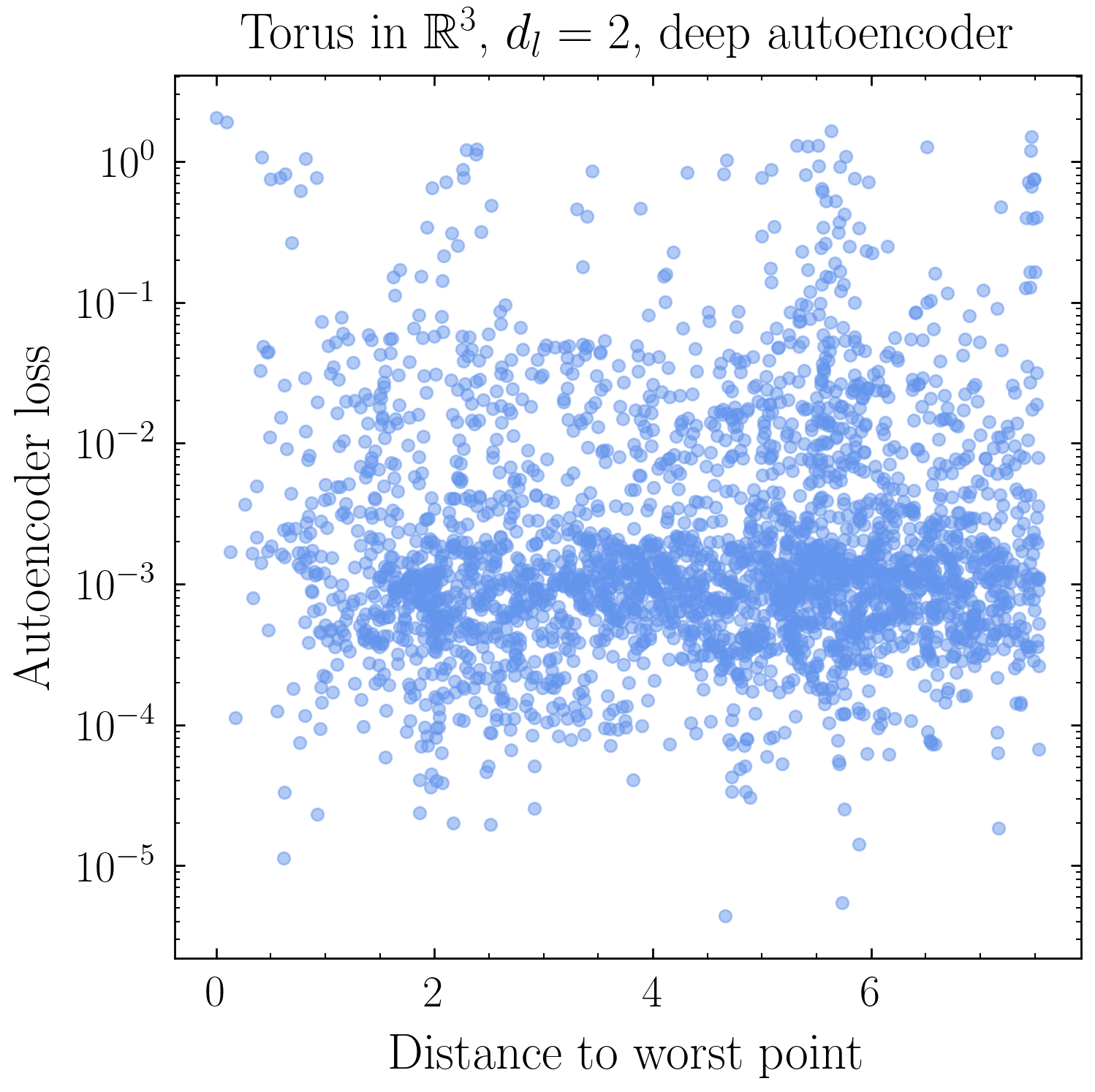}
	\caption{Loss vs. distance plots for the torus $T^2$ in $\mathbb{R}^3$ with $d_l = 2$, for the 5-layer network \textbf{(left)} and the 7-layer network \textbf{(right)}.}
	\label{fig:T2R3}
\end{center}
\end{figure}

Moving to $d = 2$, we consider the torus $T^2$, which can be embedded in $\mathbb{R}^3$ by
\be
\label{eq:TorusR3}
x = (R+r\cos\alpha)\cos\beta, \qquad y = (R+r\cos\alpha)\sin\beta, \qquad z = r \sin \alpha
\ee
We take $r = 1$ and $R = 3$, and generate training and test sets uniformly sampled in $\alpha$ and $\beta$. Since the topology of the torus is that of a quotient space, $S^1 \otimes S^1 = \mathbb{R}^2/\mathbb{Z}^2$, the torus has a nontrivial fundamental group and cannot be covered with a single chart by excising a single point, unlike the sphere. Anticipating that this may make the autoencoder more difficult to train, we use both the 5-layer network and a deeper 7-layer network as defined in Sec.~\ref{sec:Setup}. The results are shown in Fig.~\ref{fig:T2R3}. While the deeper network reduces the loss overall for a generic point on the test set, there are still numerous points with order-1 loss which are far away from the worst point, and numerous points with low loss which are close to the worst point, indicating that the latent representation is non-local. Since at least an $S^1 \wedge S^1$ must be excised from a torus to embed the complement in $\mathbb{R}^2$, this behavior is expected.

\begin{figure}[t!]
\begin{center}
	\includegraphics[width = 0.45\textwidth]{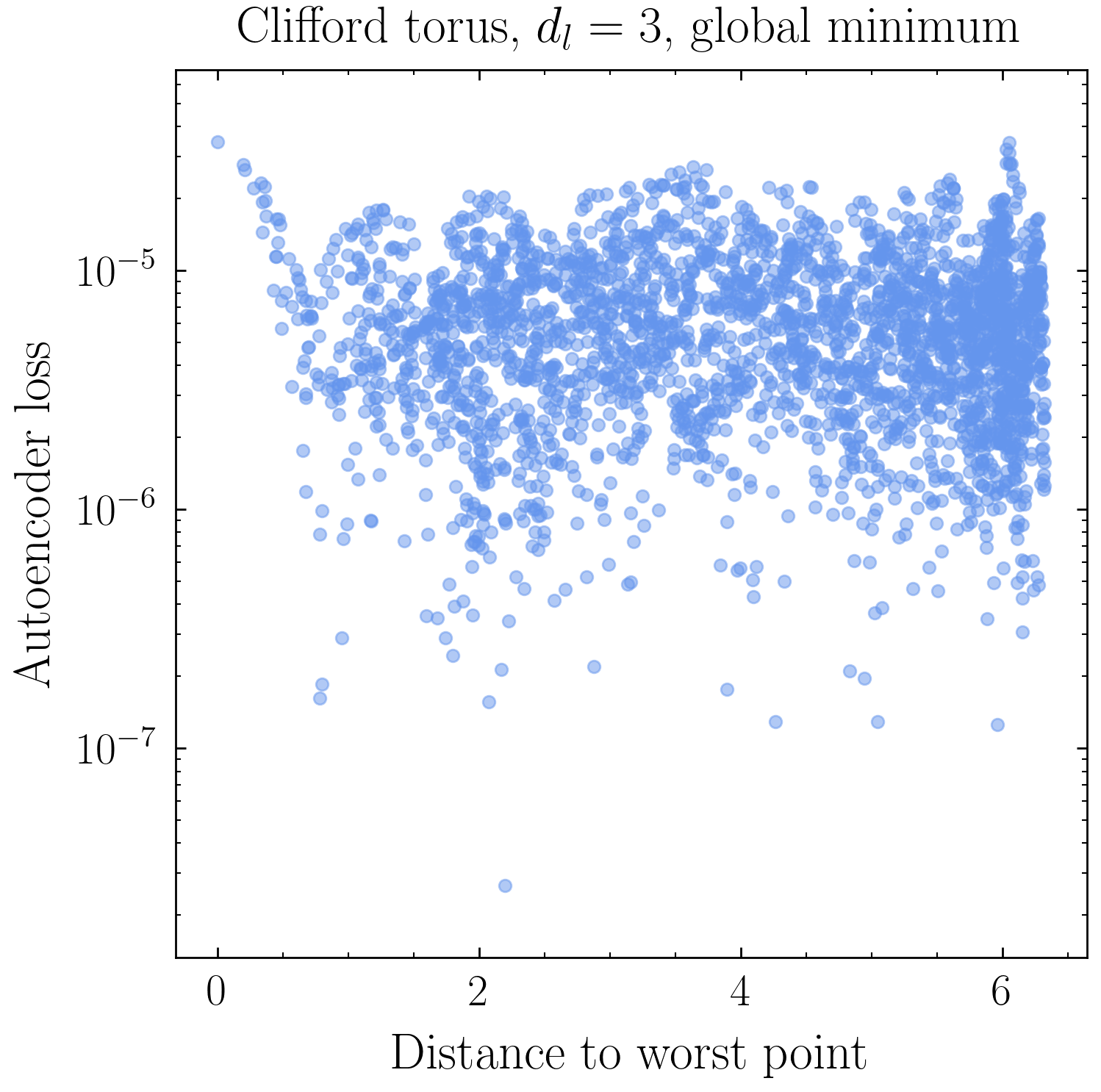} \qquad
	\includegraphics[width = 0.45\textwidth]{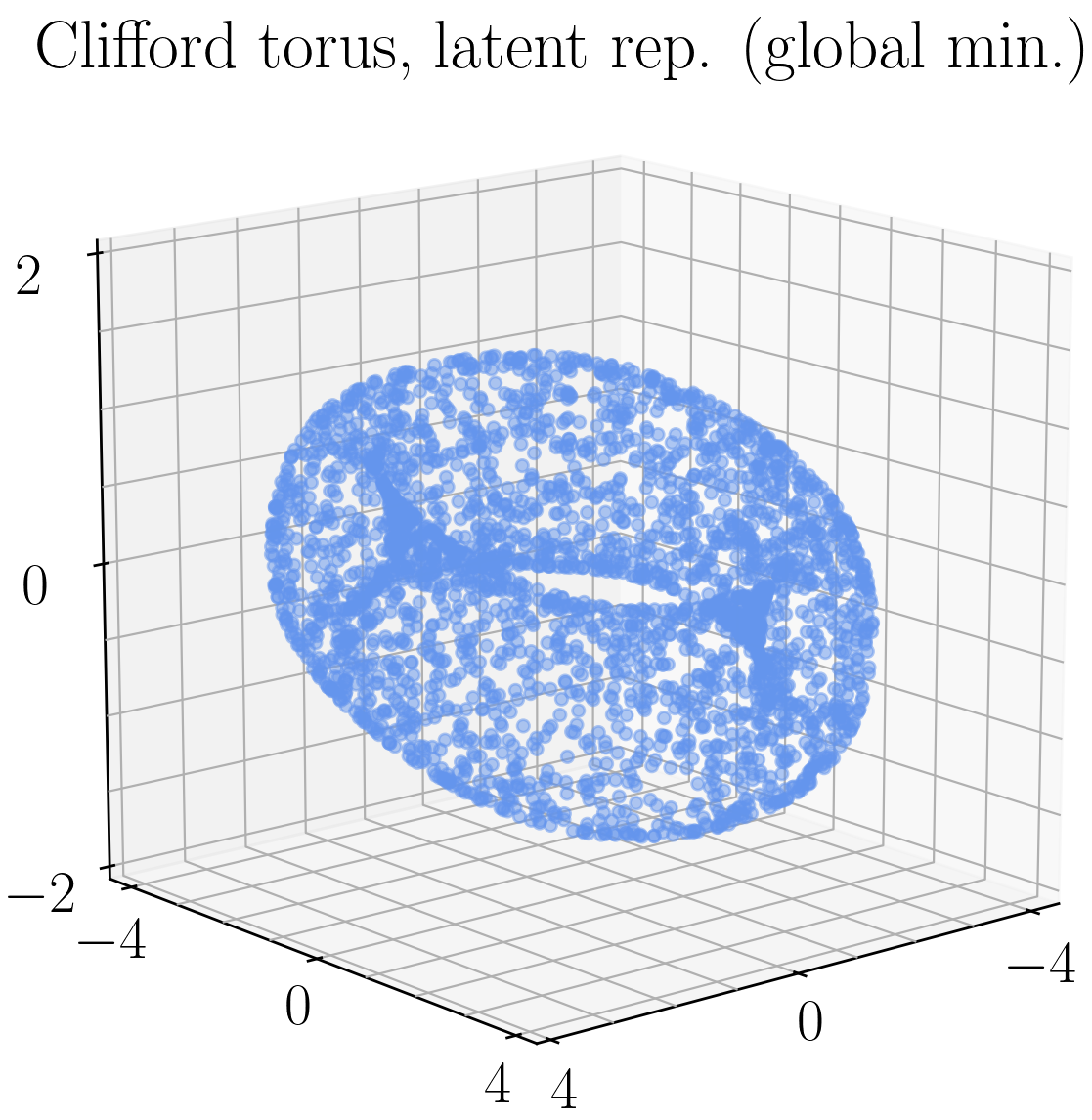}
	\includegraphics[width = 0.45\textwidth]{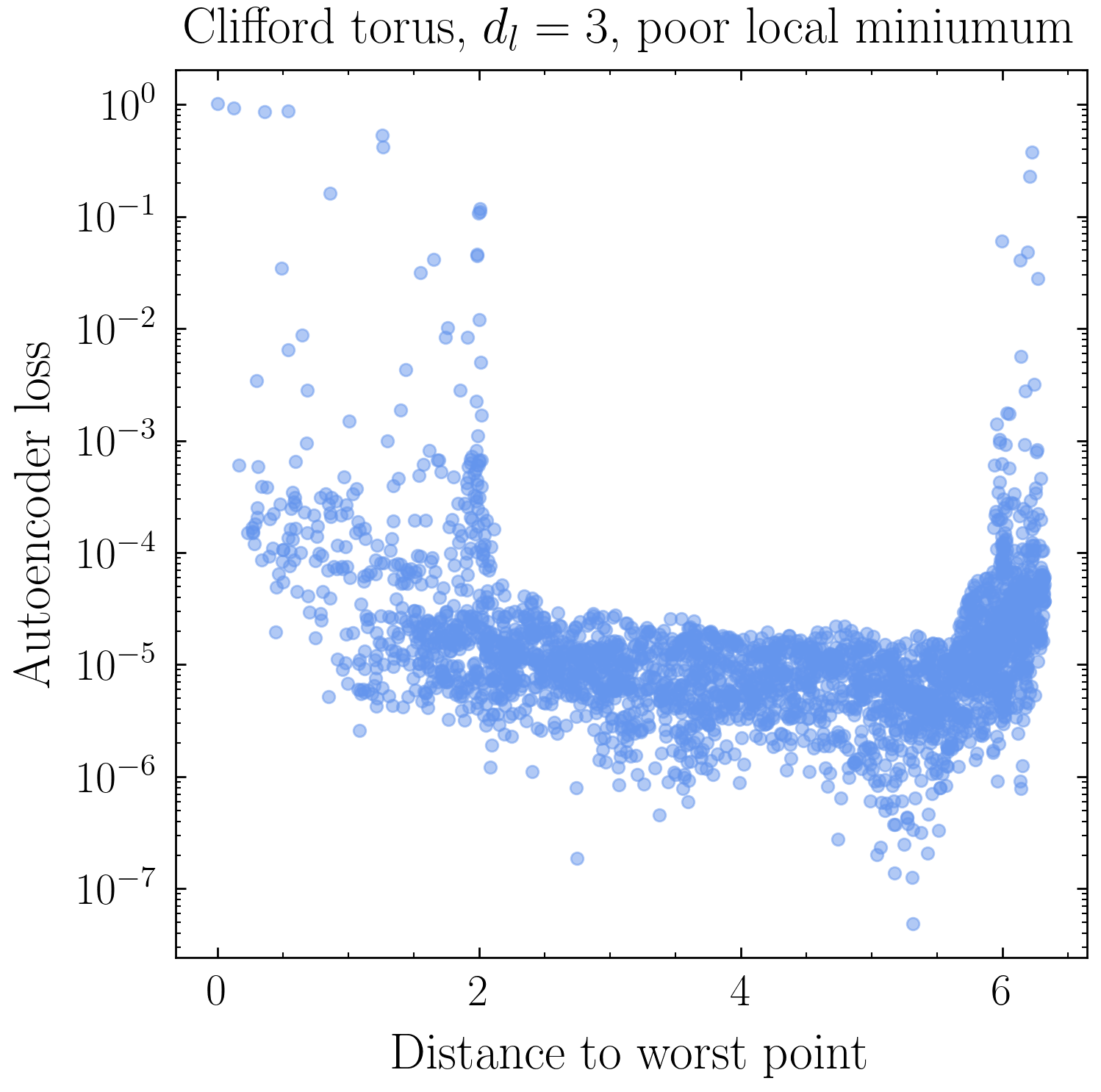} \qquad
	\includegraphics[width = 0.45\textwidth]{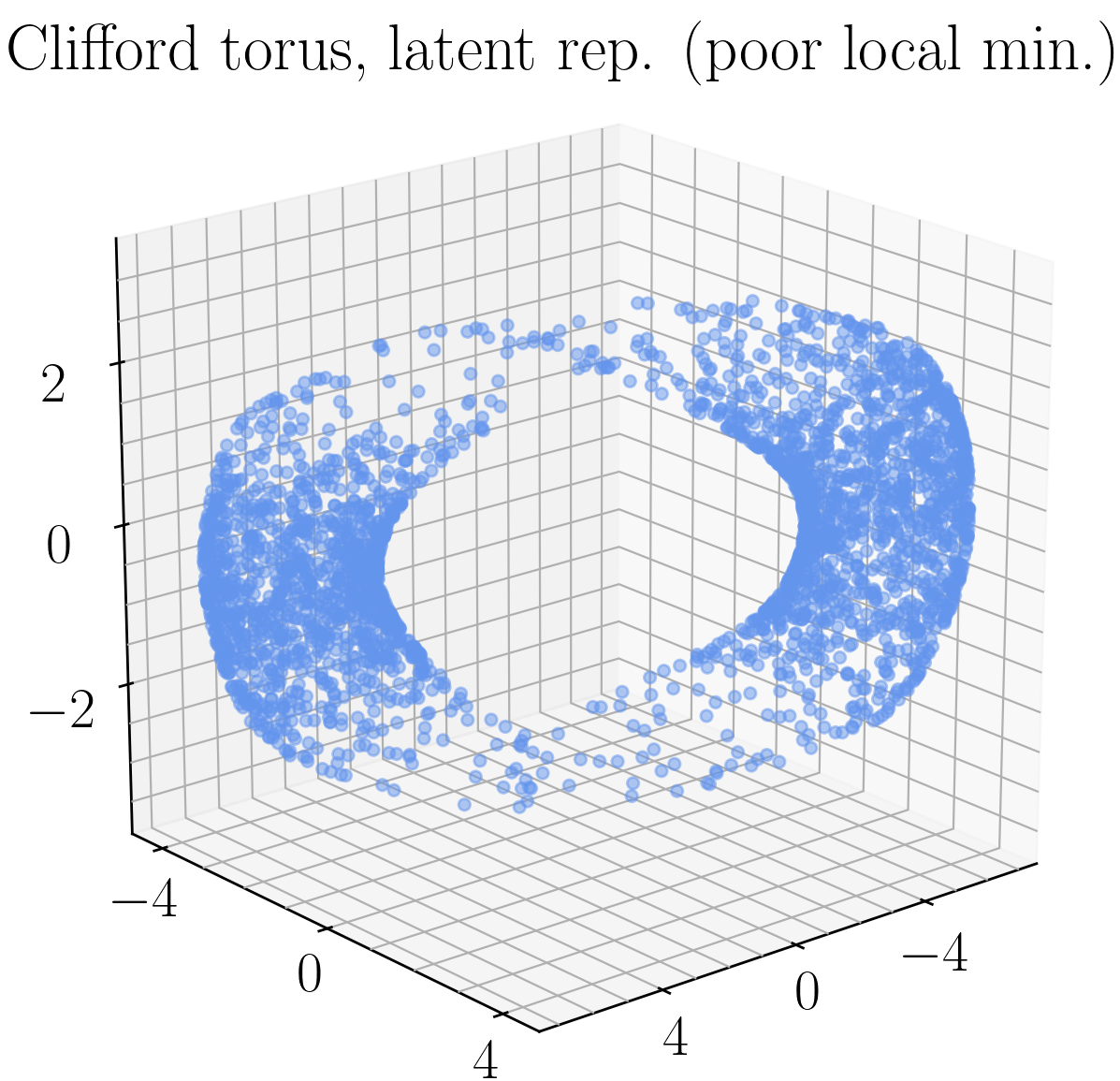} \\
	\caption{Two realizations of an autoencoder trained on the Clifford torus. \textbf{Top:} the network approximates the global minimum of the loss, where the latent representation is the embedding $T^2 \subset \mathbb{R}^3$. \textbf{Bottom:} the network finds a poor local minimum where one circle of the $S^1 \times S^1$ topology is pinched at two points.}
	\label{fig:T2Clifford}
\end{center}
\end{figure}

To explore the role of the extrinsic topology of the data manifold in training an autoencoder, we consider a nontrivial embedding of the torus into a high-dimensional space $\din > 3$ and train an autoencoder with latent dimension $d_l = 3$. Since $\din > d_l$, the autoencoder cannot learn the trivial identity map, but it should be able to learn the standard 3-dimensional embedding given in Eq.~(\ref{eq:TorusR3}) as the latent representation. Indeed, such a high-dimensional embedding exists in $\mathbb{R}^4$, known as the Clifford torus,
\be
(x,y,z,w) = (\cos \alpha, \sin \alpha, \cos \beta \sin \beta).
\ee
Using a training set of uniformly-sampled points on the Clifford torus, and training multiple instantiations of a 7-layer network, we find two qualitatively different results, shown in Fig.~\ref{fig:T2Clifford}. Occasionally, the network will find the global minimum of the loss where the latent representation is homeomorphic to the embedding $T^2 \subset \mathbb{R}^3$. More often, though, the network finds its way to a poor local minimum for the second $S^1$ factor where it ``pinches'' at two points, rather than the optimal global minimum of the embedding in $\mathbb{R}^3$.\footnote{As noted in App.~\ref{app:MoreCircle}, the poor local minimum for $S^1$ which splits at two points rather than one also occurs for some choices of the network initialization for the $S^1$ autoencoder with $d_l = 1$, in particular when one weight is initialized large compared to the others.} Indeed, the Clifford torus parametrization makes the product-space structure of the torus $T^2 = S^1 \times S^1$ explicit, and the latent representation suggests that the autoencoder is learning both circles independently, rather than the global structure required for the embedding in $\mathbb{R}^3$. Thinking about the autoencoder in terms of an ensemble -- defined by the different possible realizations of weights and biases and learning dynamics -- we see that the ensemble doesn't concentrate on one minimum, but rather a discrete set of them. This lack of typicality is problematic from an application standpoint, as the two minima will have wildly different behaviors when employed as anomaly detectors. This example illustrates once again the richness of the autoencoder loss landscape and the dependence of performance on initialization. 

\subsection{SU(2) and SO(3): topology versus geometry, or global versus local}
\label{sec:su2so3}

\begin{figure}[t!]
	\includegraphics[width = 0.45\textwidth]{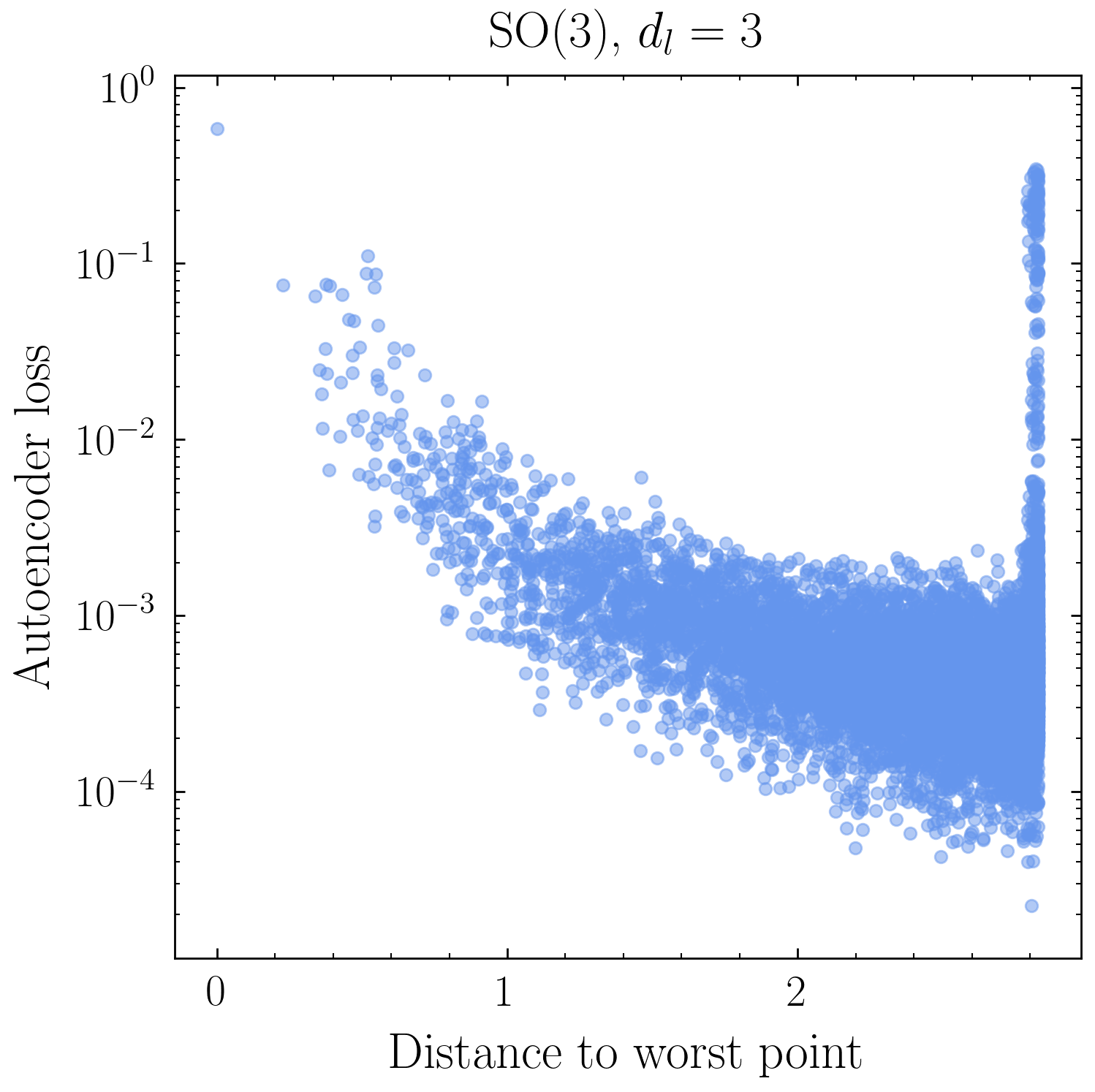} \qquad
	\includegraphics[width = 0.45\textwidth]{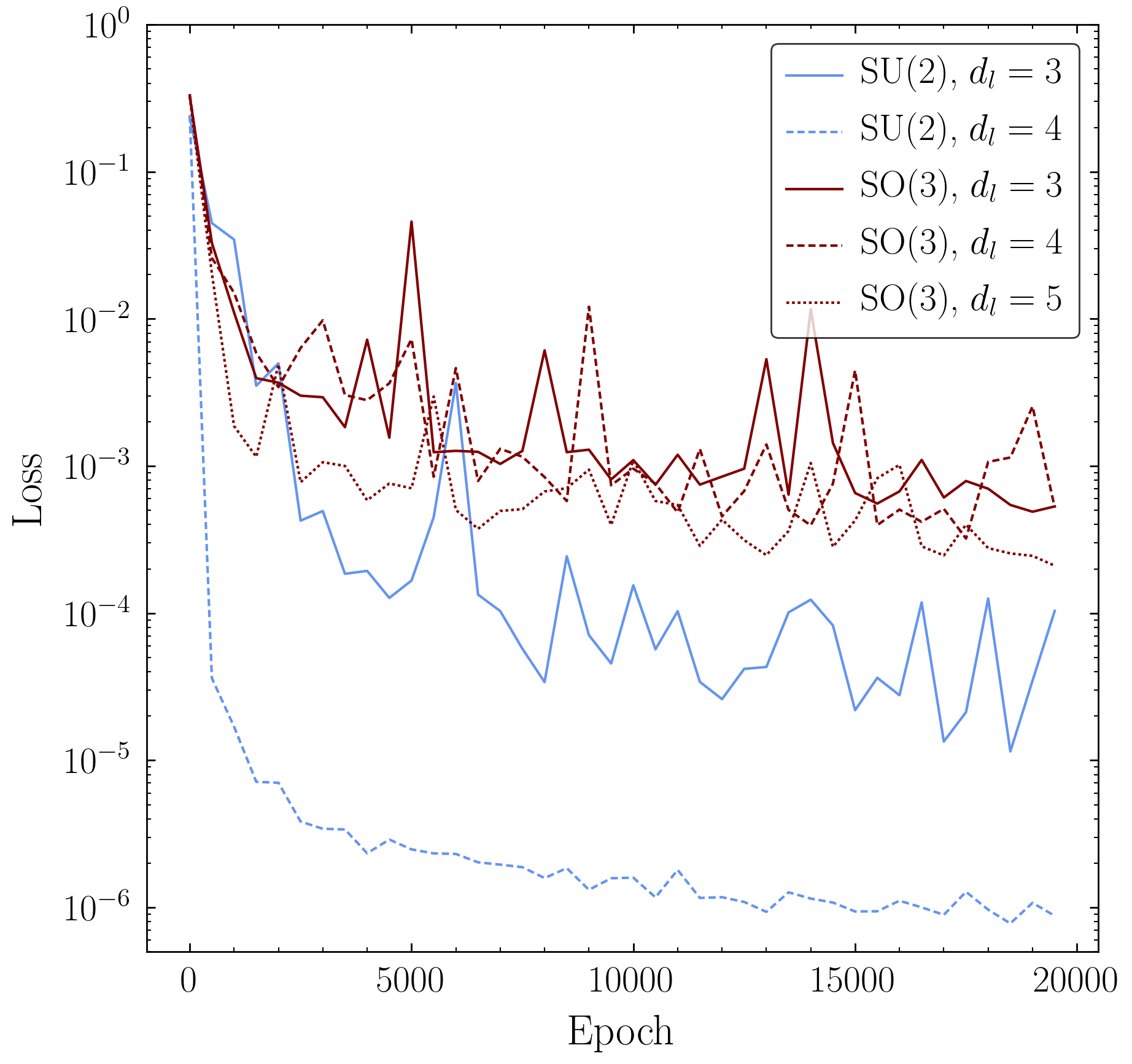}
	\caption{Two Lie groups with the same Lie algebra but different topology. \textbf{Left:} loss-versus-distance plot SO(3) with $d_l = 3$. \textbf{Right:} loss as a function of training for various $d_l$. The SU(2) network approaches perfect reconstruction with $d_l = 4$ because it has the topology of the 3-sphere $S^3$, but SO(3) with the more complicated quotient space topology of $\mathbb{RP}^3$ has similar losses for $d_l = 3, 4, 5$.}
	\label{fig:SU2vsSO3}
\end{figure}

The Lie groups SU(2) and SO(3) have the same local structure with isomorphic Lie algebras, but the global structure of the groups differs in a nontrivial way. Both SU(2) and SO(3) are 3-dimensional, but are topologically distinct, with SU(2) the double cover of SO(3). As both groups can be parametrized with a triplet of Euler angles $(\alpha, \beta, \gamma)$, which can be mapped into a vector of 8 real numbers (the entries of a complex $2 \times 2$ SU(2) matrix $U$ satisfying $U^\dagger U = I_{2 \times 2}$) or 9 real numbers (the entries of a real $3 \times 3$ SO(3) matrix $O$ satisfying $O^T O = I_{3 \times 3}$), looking at the differing behavior of autoencoders trained on these two manifolds can isolate the topological features from the geometric ones. In particular, since SU(2) is diffeomorphic to $S^{3}$, the SU(2) autoencoder will provide an example of a topologically nontrivial manifold embedded in a much higher-dimensional space $\mathbb{R}^8$, which will again be analogous to our phase space example.

Since the geometric structure of these manifolds is difficult to visualize, instead of plotting the output directly, we will evaluate the performance of the autoencoder with the loss-versus-distance plot introduced in Sec.~\ref{sec:dim2}, as well as examining the loss on the test set as a function of training epoch. We generate training sets by uniformly sampling each group according to the Haar measure, the unique invariant measure on Lie groups. The matrices are then flattened row-by-row into an 8-component or 9-component real vector for SU(2) and SO(3), respectively. Fig.~\ref{fig:SU2vsSO3} shows the performance of the deep 7-layer autoencoder trained on these two group samples.\footnote{Note that the MSE loss is computed on the flattened 8- or 9-dimensional vector, which implies a Euclidean metric on those vectors and is different from the natural metric on the group.} Based on the results from the circle and the 2-sphere, it is not surprising that a 3-dimensional latent layer is not able to fully reconstruct the data, while a 4-dimensional latent layer can do so: ${\rm SU(2)} \cong S^{3}$ can be embedded in $\mathbb{R}^4$. On the other hand, for SO(3), the size of the loss after the same amount of training is orders of magnitude larger than for SU(2) and barely improves going from $d_l = 3$ to $d_l = 4$. This is due to topology: SO(3) is diffeomorphic to real projective space $\mathbb{RP}^3$, as it is the quotient of SU(2), a 3-sphere, by a $\mathbb{Z}_2$ action, and a classical theorem of Mahowald states that $\mathbb{RP}^3$ does not embed in $\mathbb{R}^4$ \cite{mahowald1962embeddability}. The loss versus distance plot is suggestive of this same phenomenon, where loss is clearly anti-correlated with distance from the worst point for SU(2) (see Fig.~\ref{fig:Sn}) but not for SO(3): the points maximally distant from the worst point cover 4 orders of magnitude in loss. Even for $d_l = 5$, which is the embedding dimension of $\mathbb{RP}^3$, the loss is of the same order as $d_l = 3$; as was the case for the Clifford torus, the embedding exists but appears to be difficult for the network to find during training. This suggests that there are some topological embeddings which are naively hard for neural networks to untangle.

\end{appendix}

\bibliographystyle{JHEP}
\bibliography{AutoencoderTopologyBib.bib}{}

\end{document}